\documentclass[twocolumn,10pt,aps,nofootinbib,superscriptaddress]{revtex4-2}
\usepackage{amsmath}
\usepackage{amsfonts}
\usepackage{amssymb}
\usepackage{graphicx}
\usepackage{caption}
\usepackage{array}
\usepackage{subcaption}
\usepackage{float}
\usepackage{caption}
\allowdisplaybreaks
\newcommand {\bp}{\begin{pmatrix}}
\newcommand {\ep}{\end{pmatrix}}
\newcommand{\be}{\begin{equation}} \newcommand{\ee}{\end{equation}}
\newcommand{\bea}{\begin{eqnarray}}\newcommand{\eea}{\end{eqnarray}}

\DeclareMathOperator{\sgn}{sgn}

\begin{document}

\title{Flat band and Bulk-Boundary correspondence in a non-Hermitian trimerized lattice model with generic boundary conditions}

\author{Supriyo Ghosh}
\email[]{supriyoghosh711@gmail.com}
\affiliation{Department of Physics, Indian Institute of Technology-Kanpur, Kanpur 208016, India}
\author{Pijush K. Ghosh}
\email[]{pijushkanti.ghosh@visva-bharati.ac.in}
\author{Shreekantha Sil}
\email[]{shreekantha.sil@visva-bharati.ac.in}
\affiliation{ Department of Physics, Siksha-Bhavana, Visva-Bharati,
Santiniketan, PIN 731235, India.}
\date{\today}

\begin{abstract}
We consider a Su-Schrieffer-Heeger(SSH)-type trimer model with next-nearest-neighbor(NNN)
interaction and balanced loss-gain(BLG) to study the combined effect of lattice symmetries,
topology, non-hermiticity and general boundary conditions(GBC) on the existence of flat band
and the nature of Bulk-Boundary correspondence(BBC). We derive the necessary and sufficient
conditions for the existence of an entirely real spectrum under the periodic boundary
condition(PBC). The exact expressions for the compact localized states(CLS) and
energy eigenvalues corresponding to flat bands are derived analytically under the PBC for
${\cal{PT}}$ and/or pseudo-chiral symmetric momentum-space Hamiltonian. We establish topological
phase transitions(TPT) for $\mathcal{PT}$-symmetry and pseudo-chiral symmetry through the computation 
of the Zak phase and sub-lattice Zak phase, respectively.
The Hamiltonian under the open boundary condition(OBC) is studied numerically, and edge states
are observed in the topologically non-trivial phase, thereby establishing the non-hermitian BBC. 
The CLS exists in both bulk and the boundary for systems having only pseudo-chiral symmetry, and 
an additional ${\cal{PT}}$-symmetry destroys the CLS at the boundary. We generalize a known formalism
to study the same Hamiltonian under GBC, and derive analytic expressions for the energy and eigenstates
for a class of boundary conditions in parametric ranges which admit flat band under the PBC. The
edge states for these boundary conditions, including the OBC, are obtained analytically in the
topologically non-trivial phase, thereby establishing BBC. The non-hermitian skin effect(NHSE) is
seen in the model with reciprocal bulk interaction and strongly non-reciprocal boundary terms. 
The winding number based on spectral topology is computed analytically.
The eigen energies of the edge states in the thermodynamic limit is purely imaginary or zero for
non-hermitian and hermitian models, respectively.
\end{abstract}

\maketitle
\tableofcontents{}

\section{Introduction}

Flat band systems have emerged as an important area in condensed matter physics
because of their unique electronic properties and the emergence of strongly correlated quantum phenomena.
In these systems, one or more electronic bands in material are independent of momentum, and hence the energy
remains constant\cite{Sondhi2013CRP,Liu2013IJMP,Derzhko2015IJMP}. The flat band admits CLS\cite{Aoki1996PRB}
which spread only over a few lattice sites, resulting in a high density of states
and enhanced electron–electron interactions. The CLS is different from Anderson localization in the sense that 
it appears in a translation invariant system, and very sensitive to the disorder and interaction. 
These distinctive characteristics have led to the observation of unconventional superconductivity, magnetism,
information storage in quantum networks\cite{Rontgen2019PRL}, algebraic singularity 
in mobility edge curve\cite{Flach2014PRL}, inverse Anderson transition, localization-delocalization-localization
transition\cite{Goda2006PRL, Nishino2007JPS,Shukla2018PRB}, and topological phases in a variety of materials\cite{Dutta2017AP,
Franz2022PRB,Budich2013PRB}. Consequently, understanding the origin and behavior of flat bands has become
an important research direction for the development of novel quantum materials and future electronic technologies.

The first theoretical paper predicting the existence of flat bands date back to 1986 for the dice lattice by 
Sutherland\cite{Sutherland1986PRB}. Later in 1989 Lieb showed certain bipartite lattices provide chiral flat 
bands\cite{Lieb1989PRL}. Subsequently, flat band is observed in periodic electronic networks induced by
Aharonov-Bohm flux\cite{Vidal1998PRL}. Although flat bands have been studied theoretically for many years, their
experimental realization has been achieved only in the last few years\cite{Flach2018APX}.
Recently, two dimensional Lieb lattice that produce flat band was constructed in electronic systems\cite{Qin2016PRB,
Drost2017NP,Slot2017NP}. Flat bands have been experimentally realized in optical Lieb lattice for bosonic cold
atoms \cite{Takahashi2015SA}, photonic systems\cite{Nakata2012PRB,Endo2010PRB,Xu2015SR,Li2008OE,Takeda2004JPCM,
Kajiwara2016PRB,Mukherjee2015PRL}, Moir\'e superlattice\cite{moire}, and exciton-polariton condensate\cite{Jacqmin2014PRL,
Baboux2016PRL,Klembt2017APL, Whittaker2018PRL}.

In the last three decades, $\mathcal{PT}$-symmetry\cite{bender} has evolved from a purely mathematical concept into
a vibrant field of theoretical as well as experimental physics. The $\mathcal{PT}$-symmetric systems
have revealed several interesting phenomena in different domains of physics such as optics\cite{Konotop2016Rev,Musslimani2008PRL},
quantum field theory\cite{Bender2004PRD,Jones2006JPA}, exactly solvable
models\cite{PKG2009JPA,PKG2010JPA,PKG2011IJTP,PKG2019JPA}, quantum phase transition\cite{PKG2009PRE},
level statistics\cite{Deguchi2009PRE, Goldsheid1998PRL,Heinrichs2001PRB,Molinari2009JPA,SKG},
Dirac Hamiltonians of topological insulators\cite{PKG2012JPA}, integrability and chaos \cite{DS, PKG-REV,PR,SKG}, etc.
The studies on $\mathcal{PT}$-symmetric lattice models deserve a special
mention due to its possible applicability in condensed matter and optical systems. The earlier studies 
on non-Hermitian tight binding chain with balanced loss-gain(BLG) type onsite imaginary potential were focused
on $\mathcal{PT}$-symmetric phase transition\cite{Joglekar2010PRA}, exact solvability\cite{Jin2009PRA}, 
and transport properties\cite{Rodriguez2020PLA,Candanedo2015PE}. It was realized afterwards that
the SSH model\cite{Su1979PRL,Fradkin1983PRB,Kivelson1982PRB,Li2014PRB}, which exhibits
topological insulators, may be generalized by including non-hermitian interaction to study the interplay of
the $\mathcal{PT}$-symmetry and topology\cite{Lieu2018PRB,Li2020JPCM,Klett2017PRA,Petrosyan2022PRA}. 
A few such modified SSH models have been studied from the viewpoint of non-Hermitian
skin effect, edge states,  bulk boundary correspondence\cite{Kawabata2019PRX,Supriyo2025PRB,Zhang2022AP}, 
$\mathcal{PT}$-symmetric phase transition\cite{Klett2017PRA,Petrosyan2022PRA,Lu2014PRA}, 
localization and transport properties\cite{Nguyen2016PRA}, and quantum chaos\cite{Obuse2020PRE}. 

With the growing interests on the ${\cal{PT}}$-symmetric lattice systems, flat bands in non-hermitian systems have 
been also studied\cite{Andreanov2021PRB,Flach2017PRB,Zhang2019PRA,Saxena2015OL,Qi2018PRL,Zyuzin2018PRB,Ge2018PR,
Zhou2025PRB,Biesenthal2019PRL,Ramezani2017PRA}. Flat band in non-hermitian system may arise due to any one or more
of the following symmetries \textemdash
$\mathcal{PT}$-symmetry\cite{Flach2017PRB,Ramezani2017PRA}, non-hermitian particle-hole symmetry\cite{Qi2018PRL,Ge2018PR},
or a sub-lattice symmetry\cite{Zhang2019PRA}. In particular, flat band in a photonic lattice with non-hermitian coupling
is obtained due to bipartite sub-lattice symmetry\cite{Flach2017PRB}, whereas real energy flat band is obtained in
non-hermitian Lieb lattice with ${\cal{PT}}$ and chiral symmetries\cite{Zhang2019PRA}. 

The purpose of this article is to study flat band and BBC in a single non-hermitian SSH-type lattice model in
order to study interplay of lattice symmetries, topology, non-hermiticity and generalized boundary conditions.
In particular, a CLS corresponding to a flat band exists in the bulk without requiring a boundary, and its
origin is primarily attributed to the lattice symmetry. With the introduction of non-hermiticity, the strength
of the BLG interaction may be used as an additional parameter to create and control CLS. On the other hand, the
edge states are created at the boundary due to underlying topology of the system. The condition for simultaneous
existence of CLS and edge states in a model under OBC with or without overlap of states is worth exploring.

The standard SSH dimer model admits flat band in the limit of vanishing intercell or intracell interactions leading
to trivial and topological flat bands, respectively. However, in this limit, the chain decouples to individual dimers.
The condition for flat band in SSH model with BLG interaction
remains the same with the strength of the non-hermitian interaction determining whether the energy is real or complex.
The addition of NNN interaction to this  model completely destroys the flat band. It appears that non-trivial flat band
may appear in generalized SSH models for which Bloch Hamiltonians are described by $M \times M$ matrices with $M >2$.
For example, the standard SSH model with non-orientable bulk admits doubly degenerate zero-energy flat bands for which
$M=6$\cite{Mobius}. The multimer or multi-leg SSH models\footnote{A chain of $M$ atoms in a single unit cell is
identified as an $M$-mer or multimer, whereas $M$ parallel chains of SSH models coupled internally correspond to $M$-leg SSH
chain.}  may be a natural choice for having Bloch Hamiltonian with $M \times M$ or $2M \times 2M$ matrices, respectively.
The SSH trimer model for which $M=3$ has been studied previously from the viewpoint of
edge states and topological properties\cite{Alvarez2019PRA,Liu2017SR,Zhang2021OE,Huda2020QM,Guo2015PRB,Diakonos2022PRB},
topology protected oscillatory edge states \cite{Jiang2024OE}, BBC with and without NNN hopping
amplitudes\cite{Verma2024PRB,Diakonos2022PRB}, flat band\cite{Ghuneim2024JPC}, etc. The non-hermitian generalizations
of the trimer lattice has also been studied from the perspective of non-hermitian skin effect and spectral winding
number\cite{He2021JPCM}, topological phase transition and edge state\cite{Jin2017PRA,Du2021OE}. 

In this article, we generalize the trimer lattice by including BLG terms and most general form
of NNN interactions which have not been studied earlier. Further, all previous studies on SSH trimer models with
or without non-hermitian interaction were restricted to periodic and open boundary conditions only. We study the
system under GBC with the following two perspectives \textemdash  the exact solvability of the system beyond PBC, and
effect of modified boundary conditions on edge states, CLS and BBC.  We emphasize that results pertaining
to hermitian SSH trimer chain under GBC, including its solvability, arises as a special case in the limit of
vanishing BLG terms which have not been explored earlier.

We analytically obtain the expression of the spectrum under the PBC and derive the necessary and sufficient
conditions for the existence of an entirely real spectrum. We obtain  necessary conditions for the existence
of flat band with entirely real energies, and  observe that the system admits non-trivial flat band 
for non-vanishing NNN interaction, which is in stark contrast with the standard SSH dimer wherein
the NNN interaction destroys the flat band. In the limit of vanishing NNN interaction, the flat band
exists only when any one of the three couplings corresponding to intracell and intercell interaction
vanishes leading to decoupled trimers. We identify ${\cal{PT}}$-symmetry and pseudo-chiral symmetries of the Bloch
Hamiltonian, and study these symmetric cases in detail. The exact expressions for the
CLS and energy eigen-values are derived analytically under the PBC for Hamiltonian having pseudo-chiral symmetry.
We establish the TPT through the computation of the sub-lattice Zak phase, and obtain edge states in the
topological non-trivial region for the same model under the OBC, thereby showing the BBC. The
bifurcation diagram in the relevant regions are drawn to mark the topologically trivial and non-trivial regions.
The system with ${\cal{PT}}$ symmetry is studied numerically and TPT is shown through the computation of Zak phase.  

We generalize a known formalism to study the same Hamiltonian under GBC, and derive analytic expressions
for energy and the eigenstates for a class of boundary conditions in parametric ranges which admit flat band
under the PBC. We also obtain analytic expressions of edge states under the OBC for a few cases, thereby establishing
non-hermitian BBC. Further, the NHSE is observed in the model for reciprocal bulk Hamiltonian and highly
asymmetric boundary terms \textemdash the boundary term is non-hermitian for this case. An analysis of the
spectral topology ensures the stability of the skin-modes. These results
survive even in the limit of vanishing BLG interaction. The difference between a Hamiltonian with or without
BLG interaction is that energy is purely imaginary or real, respectively, in the thermodynamic
limit. There is no pronounced difference on the eigen functions representing edge states for these two cases.

The plan of this article is the following. We introduce the model in Sec. II along with classifications of various
types of boundary conditions or boundary terms. The Bloch Hamiltonian is obtained for the PBC in Sec. III. The parametric
ranges for which the Hamiltonian admits ${\cal{PT}}$ or pseudo-chiral symmetry are specified in Sec. III.A. In Sec. III.B,
the necessary and sufficient conditions for an entirely real spectrum  are derived. In Sec. III.C, these conditions
are further specialized to obtain the necessary conditions for the existence of a zero-energy flat band.
We classify these flat bands into two distinct types based on the constraints satisfied by the model parameters.
The corresponding conditions are solved analytically. We show TPT in pseudo-chiral symmetric limit
by computing CLS and sub-lattice Zak phase analytically in Sec. IV. For the case of ${\cal{PT}}$ symmetry, the
eigen states and the associated Zak phases are computed numerically. The bifurcation diagram showing topologically trivial
and nontrivial regions in the parameter space are presented.  In Sec. V, the system is considered under the OBC for the
same ranges of parameters for which it was studied under the PBC, and BBC is established by observing edge states in the topologically
non-trivial region. In Sec. VI, a known formalism is generalized to study the Hamiltonian under GBC, and a class of boundary
conditions is identified for which the system is exactly solvable. Further, analytic expressions for edge states are obtained
under the OBC. Finally, in Sec. VII, the results are summarized along with discussions for future directions. The ${\cal{PT}}$-symmetry
along with explicit forms of the parity operators and pseudo-chiral symmetry are discussed in the appendix VIII.A.
The appendix VIII.B includes solutions of the necessary conditions for reality of the entire spectrum. In appendix VIII.C,
general conditions for acceptable solutions of the phases of the momentum-space eigenfunctions under GBC are discussed.
The exact solutions of a particular trigonometric equation appearing in the analysis of the system under GBC are discussed
in Sec. VIII.D.
The necessary formalism for detecting edge states is also given whenever the aforementioned equation is not analytically
solvable.

\section{The Model}
\begin{figure}
	\includegraphics[width=0.7\linewidth]{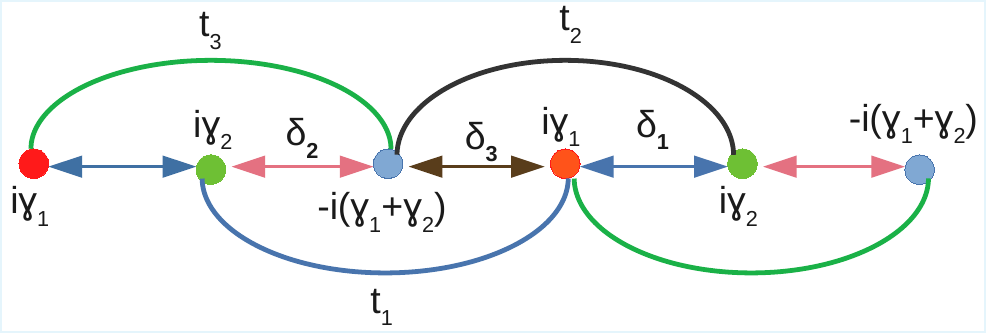}
	\caption{Schematic diagram of the trimerized lattice model}
	\label{model_schematic}
\end{figure}
We consider a one dimensional trimerized tight-binding chain with BLG and NNN interaction that can be described by the Hamiltonian : 
\bea
H = H_{bulk} + H_{B}
\label{main-ham}
\eea
where $H_{bulk}$ indicates the bulk Hamiltonian and $H_{B}$ denotes the boundary term. 
The Schematic diagram of the lattice model represented by $H_{bulk}$ is shown in {\bf Fig. \ref{model_schematic}}.
The bulk Hamiltonian on $N = 3m$ lattice sites is expressed as,
\bea
&& H_{bulk}  = H_{NN} + H_{NNN} + H_{BLG} \nonumber \\ 
&& H_{NN}    =  \left[\sum_{n = 1}^{m} \left( \delta_1 \ a_{n}^{\dagger}b_{n} + \delta_2 \ 
b_{n}^{\dagger}c_{n}\right) + \delta_3 \sum_{n=1}^{m-1} c_{n}^{\dagger}a_{n+1} \right] + \textrm{h.c.}\nonumber \\ 
&& H_{NNN}   =  \left[t_3 \sum_{n=1}^{m} a_{n}^{\dagger} c_{n} + \sum_{n = 1}^{m-1} \left(t_1 b_{n}^{\dagger} a_{n+1} 
+ t_2 c_{n}^{\dagger}b_{n+1} \right)\right] + \textrm{h.c.} \nonumber \\ 
&& H_{BLG}       =  i \sum_{n=1}^{m} \Big[ \gamma_1 \ a_{n}^{\dagger}a_{n} + \gamma_2 \ b_{n}^{\dagger}b_{n} 
- \left(\gamma_1 + \gamma_2\right) \ c_{n}^{\dagger}c_{n} \Big] \nonumber
\eea
\noindent where h.c. denotes hermitian conjugation.
The Hamiltonian \(H_{NN}\) denotes the standard trimerized SSH chain, where \(\delta_1\) and \(\delta_2\) are the
intracell hopping amplitudes, while \(\delta_3\) represents the intercell hopping amplitude. The term \(H_{NNN}\)
accounts for the next-nearest-neighbor interactions, and \(H_{BLG}\) incorporates the BLG contributions.
All the parameters appearing in $H_{NN}$, $H_{NNN}$ and $H_{BLG}$ are considered as real, and hence, the
Hamiltonians $H_{NN}$ and $H_{NNN}$ are hermitian, while $H_{BLG}$ is non-hermitian.

The boundary Hamiltonian is, in general, non-hermitian,
\bea
H_{B} & = & t_{1l} a_{1}^{\dagger}b_{m} + t_{1r} b_{m}^{\dagger}a_1 
+ \delta_{3l} a_{1}^{\dagger}c_m + \delta_{3r} c_{m}^{\dagger}a_{1} \nonumber \\
& + & t_{2l} b_{1}^{\dagger}c_m + t_{2r} c_{m}^{\dagger}b_{1}, \nonumber  
\eea
\noindent and represents various boundary conditions under constraints satisfied by
the coupling parameters. The boundary Hamiltonian $H_B=0$ for the OBC which corresponds
to $$t_{il} = t_{ir} = 0, \delta_{3l} = \delta_{3r} = 0, i=1, 2$$ For no-vanishing
$H_B$, the Hamiltonian is hermitian for specific choices of the parameter,
$$t_{il}^* = t_{ir} \neq t_i, \delta_{3l}^* = \delta_{3r} \neq \delta_3, i= 1, 2$$
where $f^*$ denotes complex conjugation of $f$.
We refer to such boundary condition as hermitian general boundary condition(HGBC).
This condition may arise due to defects at the boundary, and is suitable for studying the effects
of weak or strong defects at the boundary on the bulk Hamiltonian by considering the magnitude
of the boundary parameters to be too small or too large compared to the corresponding bulk parameters, respectively.
The HGBC can be further specialized to twisted boundary condition(TBC) with the following  parametric relations:
$$t_{il} = e^{i\eta_i} t_i, t_{ir} = t_{il}^*, \delta_{3l} = e^{ i \eta_3} \delta_3, \delta_{3r} = \delta_{3l}^*, i=1, 2$$
where $0 \leq \eta_i \leq 2 \pi$ are three independent phases. The PBC and anti-periodic boundary condition(APBC) appear
as special cases for $\eta_i=0 \ \forall \ i$ and $\eta_i=\pi \ \forall \ i$, respectively. 
\begin{enumerate}
\item PBC : $t_{il} = t_{ir} = t_{i}$,
$\delta_{3l} = \delta_{3r} = \delta_3$
\item APBC : $t_{il} = t_{ir} = -t_{i}$,
$\delta_{3l} = \delta_{3r} = -\delta_3$
\end{enumerate}
\noindent For each set of values of $\eta_i(i=1,2,3)$, excluding 0 and $\pi$, independent TBC's are obtained,
for which the boundary parameters become complex.
Among all possible non-hermitian boundary conditions, the anti-hermitian boundary
condition(AHBC) is obtained for 
$$t_{il} = -t_{ir}^* \neq t_{i}, \delta_{3l} = -\delta_{3r}^* \neq \delta_3, i=1, 2$$
Such conditions may be realized through asymmetric defects at the boundary. Further, as
in the case of HGBC, the effect of weak or strong asymmetric perturbations at the boundary
on the bulk Hamiltonian may be studied. 
In this article, unless specified explicitly, we restrict our discussions for real boundary parameters.
We show later in Sec. VI that the Hamiltonian under
GBC is exactly solvable for various limiting values of the bulk parameters, whenever the
condition $\delta_3^2 = \delta_{3l}{\delta_{3r}}$ or $t_1^2 = t_{1l}{t_{1r}}$ is satisfied.
All the boundary conditions as stated above, except the AHBC, fulfil this solvability criteria,
and more details in this regard will be discussed in Sec. VI. 

\section{Periodic Boundary Condition}

We get periodic boundary condition for $t_{1l} = t_{1r} = t_1$, $t_{2l} = t_{2r} = t_2$, and
$\delta_{3l} = \delta_{3r} = \delta_{3}$. In the momentum space, we express the eigen ket $\vert n, a \rangle$ as
$\vert n, a \rangle = e^{ikn} \vert k, a \rangle$ leading to the Bloch Hamiltonian,
\bea
H_{k} = \bp i\gamma_1 && \delta_1 + t_1 e^{-ik} && \delta_3 e^{-ik} + t_3 \\ 
\delta_1 + t_1 e^{ik} && i\gamma_2 && \delta_2  + t_2 e^{-ik} \\ 
\delta_3 e^{ik} + t_3 && \delta_2 + t_2 e^{ik} &&
-i(\gamma_1+\gamma_2) \ep
\eea
\noindent The diagonal terms make the Hamiltonian non-hermitian, and it is not
invariant under the time-reversal symmetry ${\cal{T}}: i \rightarrow -i, k \rightarrow -k$ for non-vanishing
$\gamma_1$ or $\gamma_2$ or both.
The Hamiltonian $H$ and its adjoint $H^{\dagger}$ are related as,
$H_k^{\dagger}(\gamma_1,\gamma_2)=H_k(-\gamma_1,-\gamma_2)= {\cal{T}} H_k$,
which in turn implies that the energy eigenvalues $E_k$ satisfy the condition
$E_k(\gamma_1, \gamma_2)=E_k^*(-\gamma_1, -\gamma_2)$.
Further, for non-degenerate eigenstates,
\bea
|\psi_L(\gamma_1, \gamma_2)\rangle = C |\psi_R(-\gamma_1,-\gamma_2)\rangle,
\eea
where $|\psi_L\rangle$ and  $|\psi_R\rangle$ denote left and right eigenvectors of $H$,
respectively, and $C$ is a constant which may be chosen to be unity for normalized right and left
eigen vectors. The relation between the left and right eigen vectors is useful for computing $|\psi_L\rangle$
from $|\psi_R\rangle$ or the vice versa.

\subsection{Symmetry}
\label{subsec_symmetry}
The parity operator ${\cal{P}}$ is not unique, and we define three ${\cal{P}}_i$'s corresponding to exchange of sub-lattices
$j \leftrightarrow k, j \neq i \neq k \neq j$, and keeping the $i$'th sub-lattice unchanged. In a similar way,
we introduce $\tilde{\cal{P}}_i=R_i {\cal{P}}_i$, where $R_i$ corresponds to rotation around $i$'th axis by an angle $\pi$.
A detail discussion on the ${\cal{PT}}$-symmetry of $H_k$ along with matrix representations of parity operators
is given in Appendix \ref{symmetry}. The Hamiltonian is ${\cal{PT}}$ symmetric, $({\cal{P T}}) H_k ({\cal{P T}})^{-1}=H_k$,
for the following cases:
\bea
&& \textrm{I.(a)}: {\cal{P}} := {\cal{P}}_1, \gamma_1=0, \delta_3=t_1=0, t_3=\delta_1,\nonumber \\
&& \textrm{I.(b)}: {\cal{P}}:= \tilde{\cal{P}}_1, \gamma_1=0, \delta_3=t_1=0, t_3=-\delta_1\nonumber \\
&& \textrm{II.(a)}: {\cal{P}}:= {\cal{P}}_2, \gamma_2=0, t_2=t_1, \delta_2=\delta_1\nonumber \\
&& \textrm{II.(b)}: {\cal{P}}:= \tilde{\cal{P}}_2, \gamma_2=0, t_2=-t_1, \delta_2=-\delta_1\nonumber \\
&& \textrm{III.(a)}: {\cal{P}}:= {\cal{P}}_3, \gamma_2=-\gamma_1, \delta_3=t_2=0, t_3=\delta_2\nonumber \\
&& \textrm{III.(b)}: {\cal{P}}:= \tilde{\cal{P}}_3, \gamma_2=-\gamma_1, \delta_3=t_2=0, t_3=-\delta_2
\eea
\noindent The Hamiltonian has pseudo-chiral symmetry, $\Gamma_{i} H_k^{\dagger} \Gamma_{i}^{-1} = - H_k$ for three
different choices of the parameters and different chiral operators $\Gamma_i$:\\
(i) $\delta_1=t_1=0: \Gamma_1=\textrm{diagonal} (1,1,-1)$, \\ 
(ii)$\delta_2=t_2=0: \Gamma_2=\textrm{diagonal} (1,-1,1)$, \\ 
(iii)$\delta_3=t_3=0: \Gamma_3=\textrm{diagonal} (-1,1,1)$.\\
There are regions in the parameter space for which the Hamiltonian posses both ${\cal{PT}}$ and pseudo-chiral
symmetry. For example, ${\cal{P}}_2 {\cal{T}}$ and $\Gamma_3$ may be realized in the region,
$\gamma_2=0, t_2=t_1, \delta_2=\delta_1, \delta_3=t_3=0$.
The ${\cal{PT}}$, and pseudo-chiral symmetric  $H_k$ will be discussed below in detail. The Hamiltonian
has inversion symmetry  ${\cal{P}} H_k {\cal{P}}^{-1}=H_{-k}$ in certain regions of the parameter space, and
further imposition of reality condition on the entire spectrum reduces it to trivial decoupled systems.
The inversion symmetry is discussed in Appendix VIII.A, and further analysis of the Hamiltonian by utilizing
it will not be discussed in the main text. 

\subsection{Necessary and sufficient conditions for Reality of Spectra}

The necessary condition for $H_k$ to admit three real eigenvalues is that all the co-coefficients
of its characteristic polynomial are real which leads to the following equations,
\bea
&& \gamma_1 \left ( \delta_2^2 + t_{2}^2 - \delta_1^2 -t_1^2 -\gamma_2^2 \right )
=- {\gamma_2}\left(\delta_3^2+t_3^2 -\delta_1^2-t_1^2 -\gamma_1^2 \right)\nonumber \\
&& \gamma_1 \left ( \delta_2 t_2 - \delta_1 t_1 \right ) +
\gamma_2 \left ( \delta_3 t_3 - \delta_1 t_1\right )=0
\label{real-pcondi}
\eea
\noindent The parametric conditions for ${\cal{PT}}$-symmetry are solutions of the above equations.
The necessary conditions arise solely due to the BLG interaction, and are absent for systems without
any loss-gain, i.e. $\gamma_1=\gamma_2=0$. The characteristic polynomial, for the case when
Eq. (\ref{real-pcondi}) holds, is a depressed cubic equation,
\bea
\lambda^{3} + p(k) \lambda + q(k) = 0 
\eea
\noindent with real coefficients $p(k)$ and $q(k)$,
\bea
p(k) & = & \gamma_R^2
- \left [ {\vert \vec{\delta} \vert}^2 +
{\vert \vec{t} \vert}^2 + 2 \vec{\delta} \cdot \vec{t}  \cos k\right ] \nonumber \\
q(k) & = & - 2 \left [ \delta_1 \delta_2 t_3 + \delta_1 \delta_3 t_2 + \delta_2 \delta_3 t_1 
+ t_1t_2t_3 \cos 2k \right .\nonumber \\
& + & \left . \left(   \delta_1 t_2 t_3
+ \delta_2 t_1 t_3 + \delta_3 t_1 t_2 + \delta_1 \delta_2 \delta_3 \right) \cos k\right ]
\eea
where we have defined a quantity $\gamma_R= \sqrt{\gamma_1^2+\gamma_2^2 + \gamma_1 \gamma_2}$, and
two vectors $\vec{\delta}=(\delta_1, \delta_2, \delta_3), \vec{t}=(t_1,t_2,t_3)$. 
It may be noted that the balanced loss-gain
criteria directly casts the cubic equation in its depressed form without any change of variables.
The necessary conditions for reality of the spectra, as given in Eq. (\ref{real-pcondi}), can
be rewritten as,
\bea
\gamma_1 \left (\delta_2 \pm t_2 \right )^2 + \gamma_2 \left (\delta_3 \pm t_3 \right )^2 -
(\gamma_1+\gamma_2) \left (\delta_1 \pm t_1 \right )^2 = \Gamma \nonumber
\eea
\noindent where $\Gamma \equiv \gamma_1 \gamma_2(\gamma_1+\gamma_2)$. The parameter space is
eight dimensional with two constraints for reality of the spectra. For fixed $\gamma_1, \gamma_2$ and
$t_i$'s or $\delta_i$'s, a geometric interpretation of the above equation is allowed. In particular,
the necessary condition for reality of the spectra can be interpreted as the overlapping region of two hyperboloids
in the $\delta$-space centered at $\vec{t}$ and $-\vec{t}$, respectively. A similar interpretation
for fixed $\delta_i, \gamma_i$ and variable $t_i$ is also allowed. The nature of the hyperboloids
depends on the signs of $\gamma_1$ and $\gamma_2$. It may be noted that  $\Gamma$ is invariant
under $\gamma_1 \leftrightarrow \gamma_2$, and 
\begin{itemize}
\item $\Gamma >0$:  (i) $\gamma_i >0 \ \forall \ i$, (ii) $\gamma_i <0\ \forall \ i$,\\
(iii) $0 < \gamma_i > {\vert \gamma_j \vert}$, $i \neq j$ and $ \gamma_j = - {\vert \gamma_j \vert}$
\item $\Gamma=0$: (i) $\gamma_1=0$, (ii) $\gamma_2=0$, (iii) $\gamma_2=-\gamma_1$
\item $\Gamma < 0$: $\gamma_i < {\vert \gamma_j \vert}$ for $i \neq j$ and
$ \gamma_j = - {\vert \gamma_j \vert}$
\end{itemize}
\noindent The general solutions of Eq. (\ref{real-pcondi}) involve fixing any two parameters in terms
of the remaining six parameters, the result is presented in the Appendix \ref{appendix-sol}. There are various limits in
which the hyperboloids reduce to curves, (i) $\gamma_i=0$, (ii) $\gamma_2=-\gamma_1$, (iii) $\delta_i=\pm t_i$.
We will be considering some of these limiting cases while analyzing the reality of the spectra.

The expressions of the eigenvalues are,
\bea
E_i & = & 2 \sqrt{-\frac{p(k)}{3}} \cos\left[ \Theta(k) + \frac{2\pi(i-1)}{3}\right], i=1, 2, 3 \nonumber \\
\Theta(k) & = & \frac{1}{3} \arccos\left(\frac{3q(k)}{2p(k)} \sqrt{-\frac{3}{p(k)}}\right)
\eea
\noindent The reality of the eigenvalues requires,
\bea
p(k) < 0, \ \ -1\leq \left(\frac{3q(k)}{2p(k)} \sqrt{-\frac{3}{p(k)}}\right) \leq 1 \nonumber  
\eea
\noindent which are satisfied by the following inequalities,
\bea
{\vert \vec{\delta}- \vec{t} \vert} >  \gamma_R, \
{\vert q(0) \vert}  \leq \frac{2}{3\sqrt{3}} {\vert p(\pi) \vert}^{\frac{3}{2}}
\label{inequality}
\eea
\noindent where
\bea
p(\pi) = - \left ( {\vert \vec{\delta}- \vec{t} \vert}^2 -\gamma_R^2 \right ), \ \
q(0)=\prod_{i=1}^3 \left ( \delta_i+t_i \right )
\eea
\noindent The first inequality in Eq. (\ref{inequality}), which ensures $p(k) <0$, has a simple geometric interpretation 
\textemdash it describes a region in the parameter space of $(\delta_1, \delta_2, \delta_3)$ which lies 
outside the sphere with radius $\gamma_R$, and centered at $\vec{t}$. A similar interpretation
in the parameter space of $(t_1, t_2, t_3)$ is also allowed where the region lies beyond the
sphere with radius $\gamma_R$, and centered at $\vec{\delta}$. It appears that the second inequality
may not have any simple geometric interpretation either in $\delta$-space or $t$-space. We find
the sufficient condition which ensures reality of the spectra as,
\bea
{\vert \vec{\delta} - \vec{t} \vert}^2 > \frac{4}{2^{\frac{2}{3}}-1} \vec{\delta} \cdot \vec{t}
+\frac{2^{\frac{2}{3}}}{2^{\frac{2}{3}}-1} \gamma_R^2
\eea
\noindent where we have used the identity ${\vert xyz \vert} \leq \frac{1}{3 \sqrt{3}} (x^2+y^2+z^2)^{\frac{3}{2}}$.
It may be noted that this is a stronger condition than ${\vert \vec{\delta} - \vec{t} \vert} > \gamma_R$.
Further, this is only a sufficient condition, not a necessary one.
The inequality is not satisfied when the vectors $\vec{\delta}$ and $\vec{t}$ are parallel or closely aligned
with each other. The inequality is satisfied only when the vectors are well separated, i.e.
${\vert \vec{\delta} -\vec{t} \vert}$ is very large and $\gamma_R$ very small. This can be seen
by  considering a limiting case $s=\frac{\delta_i}{t_i}, t_i \neq 0, \delta_i \neq 0, \ \forall \ i$.
This ansatz leads to the condition,
\bea
\left ( 1-s \right)^2 > \frac{4 s}{2^{\frac{2}{3}}-1} +
\frac{2^{\frac{2}{3}} \gamma_R^2}{(2^{\frac{2}{3}}-1) (\sum_{i=1}^3 \delta_i^2)}.
\eea
Assuming that $(\sum_{i=1}^3 \delta_i^2) \gg \gamma_R^2$ such that the last term on the right hand side
can be neglected,
\bea
s^2-2 s(1+\frac{2}{2^{\frac{2}{3}}-1}) +1  > 0
\eea
There are two solutions, (i) $s < 0.115$, (ii) $s > 8.70$, corresponding
to ${\vert \delta_i \vert} \ll {\vert t_i \vert}$ and
${\vert \delta_i \vert} \gg {\vert t_i \vert}$, respectively.
Since the system has eight parameters, analyzing the spectra of the system with its full generality, either
analytically or numerically, is much more involved, and we study below a few physically interesting limiting
cases. We solve the necessary conditions for reality of the spectra analytically for these limiting cases, and
then obtain the spectra numerically. For the case of flat band, i.e. $q(k)=0 \ \forall \ k$, the transition
from real to complex spectra can be studied analytically.

\subsection{Flat band}

The necessary conditions for reality of the spectrum are further specialized to obtain the necessary conditions
for the existence of a zero-energy flat band. In particular, the condition $q(k)=0 \ \forall \ k$ 
leads to a zero energy flat band in the system with the energy expressions,
\bea
&& E_3(k)=0, \ E_2(k)=-E_1(k),\nonumber \\
&& E_1(k)=\sqrt{\left [ {\vert \vec{\delta} \vert}^2 +
{\vert \vec{t} \vert}^2 + 2 \vec{\delta} \cdot \vec{t}  \cos(k+2\phi)\right ]-\gamma_R^2}
\eea
\noindent The energy eigenvalues are entirely real for ${\vert \vec{\delta} - \vec{t} \vert}
\geq  \gamma_R$ with ${\vert \vec{\delta}- \vec{t} \vert} =  \gamma_R$ defining the exceptional
surface. The zero energy flat band criteria $q(k)=0 \ \forall \ k$ implies the following conditions:
\bea
&& t_1 t_2 t_3 = 0, \ \delta_1 \delta_2 t_3 + \delta_1 \delta_3 t_2 + \delta_2 \delta_3 t_1 = 0 \nonumber \\
&& \delta_1 t_2 t_3 + \delta_2 t_1 t_3 + \delta_3 t_1 t_2 + \delta_1 \delta_2 \delta_3 = 0
\label{flat-condi}
\eea
\noindent A nontrivial flat band does not exist in the absence of NNN interactions ($t_i=0 \ \forall \ i$). In this limit,
Eq. (\ref{flat-condi}) reduces to $\delta_{1}\delta_{2}\delta_{3}=0$, whose solutions correspond to completely disconnected
unit cells rather than a genuine lattice supporting a nontrivial flat band. For the case of a standard
SSH dimer chain, flat bands are not allowed in presence of NNN interaction. The situation for the SSH trimer chain
is different where presence of NNN interaction is essential for the existence of flat bands. At first sight, Eq. (\ref{flat-condi}) appears
to suggest that the existence of the flat band is independent of the loss-gain strengths. This conclusion, however, is misleading
because the expression for $q(k)$ is derived under the constraint imposed by Eq. (\ref{real-pcondi}), whose solutions in terms
of any two parameters from the set ($\delta_i, t_i$) depend explicitly on the BLG strengths. Further, the nature of the upper
and the lower bands corresponding to $E_1$ and $E_2$, particularly reality of the spectra, crucially depends on
them. 

The first condition in Eq. (\ref{flat-condi}) implies that one of the NNN interaction strengths
$(t_1, t_2, t_3)$ is necessarily zero for flat band. Based on particular choice of $t_i$, there
are six possibilities \textemdash two separate cases for each $t_i$:
\begin{itemize}
\item Type-I: $t_i = \delta_i=0, (t_j, \delta_j)$ arbitrary for $j \neq i$
\item Type-II: $t_i=0,t_j=\pm \delta_j, t_k=\mp \delta_k,  i \neq j \neq k \neq i$
\end{itemize}
\noindent The condition for pseudo-chiral symmetry coincides with that for a Type-I flat band. In general,
Type-II flat bands do not correspond to pseudo-chiral-symmetric Hamiltonian. However, for a given $t_i=0$,
pseudo-chiral symmetry can be imposed by introducing an additional condition $\delta_i=0$.
The necessary condition in Eq. (\ref{real-pcondi}) is to be solved for
each of the above six conditions, for uniquely characterizing a system admitting
flat band. Further, for obtaining flat band with entirely real spectrum,  the restriction 
${\vert \vec{\delta} - \vec{t} \vert} \geq  \gamma_R$ is to be worked out in terms
of the parameters satisfying the conditions (\ref{real-pcondi}) and (\ref{flat-condi}).
In the Type-I flat band limit, the two hyperboloids corresponding to the necessary condition
for reality of the eigenvalues reduce to two
identical conic sections except for their centers symmetrically placed at the reflection
points of one another on one of the planes in the $\delta$-space.
The intersection points for real values of the parameters correspond to flat band.\\
(i) $t_1 = 0, \ \delta_1 = 0$: The solutions for $t_3$ and $t_2$ are obtained from
Eq. (\ref{asol1}) and (\ref{asol2}) by putting $t_1=0=\delta_1$:
\bea
t_2^{\pm} = \pm \delta_3 \left [
\frac{\gamma_2(\Gamma - \gamma_1 \delta_2^2 -\gamma_2 \delta_3^2)}{\gamma_1(\gamma_1 \delta_2^2
+\gamma_2 \delta_3^2)}\right ]^{\frac{1}{2}},\
t_3^{\pm}  = -\frac{\gamma_1 \delta_2 }{\gamma_2 \delta_3} t_2^{\pm}\nonumber
\eea
\noindent The solutions are real for $\Gamma > \gamma_1 \delta_2^2 + \gamma_2 \delta_3^2 $,
when both $(\gamma_1, \gamma_2)$ are either positive or negative.
The reality conditions for the case when one of the $\gamma_i$'s is positive and the other one
negative, i.e, $\gamma_i >0, \gamma_j < 0, i \neq j$, are 
$\gamma_i \delta_{i+1}^2 > {\vert \gamma_j \vert} \delta_{j+1}^2$ for 
$\gamma_i \geq {\vert \gamma_j \vert}$, and $\gamma_i \delta_{i+1}^2 <
{\vert \gamma_j \vert} \delta_{j+1}^2$ for $\gamma_i \leq {\vert \gamma_j \vert}$\\
(ii) $t_2 = 0, \ \delta_2 = 0$: The solutions are obtained directly from Eq. (\ref{real-pcondi}),,
\bea
t_3^{\pm} & = & \frac{\delta_1(\gamma_1+\gamma_2)}{\gamma_2 \delta_3} t_1^{\pm},\nonumber \\
t_1^{\pm} & = & \pm \delta_3 \left [ \frac{\gamma_2 \left ( \Gamma + (\gamma_1+\gamma_2)\delta_1^2-
\gamma_2 \delta_3^2 \right )}{(\gamma_1+\gamma_2) \left \{ (\gamma_1+\gamma_2)\delta_1^2-
\gamma_2 \delta_3^2 \right \} } \right ]^{\frac{1}{2}}
\eea
\noindent The solutions are real for $(\gamma_1+\gamma_2)\delta_1^2 > \gamma_2 \delta_3^2 $,
when both $(\gamma_1, \gamma_2)$ are either positive or negative. \\
(iii) $t_3 = 0, \ \delta_3 = 0$: The solutions are real for $(\gamma_1+\gamma_2)\delta_1^2 >
\gamma_1 \delta_2^2 $, when both $(\gamma_1, \gamma_2)$ are either positive or negative:
\bea
t_2 & = &\frac{(\gamma_1+\gamma_2)\delta_1}{\gamma_1 \delta_2} t_1^{\pm},\nonumber \\
t_1^{\pm} & = & \pm \delta_2 \left [ \frac{\gamma_1 \left ( \Gamma + (\gamma_1 +
\gamma_2)\delta_1^2- \gamma_1 \delta_2^2 \right )}{(\gamma_1+\gamma_2)\left \{(\gamma_1 +
\gamma_2)\delta_1^2- \gamma_1 \delta_2^2 \right \}} \right ]^{\frac{1}{2} }
\label{condi1_3a3}
\eea

In the type-II flat band limit, the hyperboloids again reduce to two conic sections which lie
on two different planes of the $\delta$-space. The intersection points for real values of the
parameters correspond to flat band.\\
(i) $t_1=0, t_2=\delta_2, t_3=-\delta_3$:
The solutions for $\delta_2$ and $\delta_3$ are valid only when both the
$\gamma_i$'s are either positive or negative. For other cases, $\delta_2$ becomes imaginary.
\bea
&& \delta_2=\pm \sgn(\delta_3) \sqrt{\frac{\gamma_1+\gamma_2}{4 \gamma_1}} \left [ \gamma_1 \gamma_2 +
\delta_1^2\right]^{\frac{1}{2}}, \nonumber \\
&& \delta_3= \pm \sqrt{\frac{\gamma_1+\gamma_2}{4 {\gamma_2}}} \left [ \gamma_1 \gamma_2 +
\delta_1^2\right]^{\frac{1}{2}}
\eea

(ii) $t_2=0, t_3=\delta_3, t_1=-\delta_1$: Non-zero real solutions for $\delta_3$ and $\delta_1$
exist only if one of the $\gamma_i$'s is positive and the other is negative with the conditions:\\
(a) $\gamma_2 < 0, \gamma_1 > {\vert \gamma_2 \vert}$, (b) $\gamma_1 < 0,
{\vert \gamma_1 \vert} > \gamma_2 >0$\\
The solutions are,
\bea
\delta_1 = \pm \sqrt{\frac{\gamma_1 \delta_2^2- \Gamma}{4 (\gamma_1+\gamma_2)}}, \ \
\delta_3 = \pm \sgn(\delta_1) \sqrt{\frac{\Gamma - \gamma_1 \delta_2^2}{4 \gamma_2}}
\eea

(iii) $t_3=0, t_1=\delta_1, t_2=-\delta_2$: Non-zero real solutions exist when one of the
$\gamma_i$'s is positive and the other is negative. The allowed regions for $\gamma_i$'s
are, (a) $\gamma_2 < 0, 0 < \gamma_1 < {\vert \gamma_2 \vert}$ and (b) $\gamma_1 < 0,
\gamma_2 > {\vert \gamma_1 \vert}$, and $\delta_1, \delta_2$ are determined as,
\bea
\delta_1=\pm\sgn(\delta_2) \sqrt{\frac{\gamma_2 \delta_3^2 - \Gamma}{4 (\gamma_1+\gamma_2)}},
\delta_2 =\pm \sqrt{\frac{\Gamma-\gamma_2 \delta_3^2}{4 \gamma_1}}
\eea

\section{TPT in ${\cal{PT}}$-symmetric and pseudo-chiral limit}

In this section, we discuss topological phase transitions and bulk boundary correspondence 
in the ${\cal{PT}}$-symmetric or/and pseudo-chiral limit subject to the necessary condition
on reality of spectrum in Eq. (\ref{real-pcondi}). The ${\cal{PT}}$-symmetry necessitates the
choice of $\gamma_i$'s as (i) $\gamma_1=0$, (ii) $\gamma_2=0$, and (iii) $\gamma_2=-\gamma_1$
along with additional conditions on $\delta_i$ and $t_i$. For each such choice of $\gamma_i$,
we first impose  necessary condition on the reality of spectrum, followed by restrictions due to
${\cal{PT}}$ and/or pseudo-chiral symmetry which fixes the $\delta_i$ and $t_i$. The Hamiltonian
reduces to that of dimer SSH chain or disconnected trimers for some ${\cal{PT}}$ or
pseudo-chiral symmetric limit, which will not be pursued further.  Moreover, without an loss of
generality, we restrict our attention only for non-negative values of the parameters.  The system
having pseudo-chiral symmetry is exactly solvable, while for the ${\cal{PT}}$-symmetric cases,
the results are presented through numerical analysis.

\subsection{$\gamma_1 = 0$} \label{pbc_1g0}

There are two possible cases which satisfy the necessary conditions in Eq.~(\ref{real-pcondi}) for the 
reality of the energy spectrum, \\
Case - I : $\delta_3 = \pm t_1$, $\delta_1 = \pm t_3$, \\
Case - II : $\delta_3 = \pm \delta_1$, $t_3 = \pm t_1$. \\
There is no restrictions on $\delta_2$, and $t_2$. 

{\bf Pseudo-chiral symmetry}: The system is pseudo-chiral symmetric with respect to 
the chiral operator $\Gamma_2$(defined in Sec. \ref{subsec_symmetry}) for $\delta_2 = t_2 = 0$ for
both the cases. The results for the Case-II may be obtained simply by the replacement
$\delta_1 \leftrightarrow t_1$ in the results for Case-I. This is a duality relation between
two different lattice configurations. The pseudo-chiral symmetry under
$\Gamma_1$ or $\Gamma_3$ reduces the system to a dimer SSH chain, and will not be pursued further.
In the limit $\delta_3 = t_1, \delta_1 = t_3, \delta_2=t_2=0$, the expressions of $\psi_{ja}$,$\psi_{jb}$,$\psi_{jc}$
for compact localized states are
\bea
&& \psi_{n,a} = -\gamma_2, \psi_{n,b} = -i\delta_1, \psi_{n,c} = i\delta_1, \nonumber \\
&& \psi_{n-1,b} = -i\delta_3, \psi_{n-1,c} = i\delta_3 \nonumber
\eea
\noindent and the probability densities at all other sites are zero. The CLS is extended
over only five sites over two unit cells. The Bloch Hamiltonian takes a simple form,
\bea
H_{k} = \bp 0 && W^{*}_{13} && W^{*}_{13} \\ W_{13} && i\gamma_2 && 0 \\ W_{13} && 0 && -i\gamma_2 \ep
\eea
\noindent with its eigenvalues determined as,
\bea
E_0      = 0, \ \ E_{\pm}  =  \pm \sqrt{2 |W_{13}|^2 - \gamma_2^2}
\eea
\noindent where $W_{13} = \delta_1 + \delta_3 e^{ik}$.  
The left and right eigenstates $\langle \phi_{0}|$,$|\psi_{0}\rangle$ corresponding to $E_0$ are,
\bea
\langle \phi_{0} |  = N_0  {\bp \gamma_2 \\ iW^{*}_{13} \\ -iW^{*}_{13} \ep}^{T} \ , \ 
|\psi_{0} \rangle  =  N_0 \bp -\gamma_2 \\ -iW_{13} \\ iW_{13} \ep
\eea
\noindent where the normalization factor $N_0 = \frac{1}{\sqrt{2 |W_{13}|^2 - \gamma_2^{2}}}$ and the superscript
$^T$ denotes transpose of a matrix. The left and right eigenstates $\langle \phi_{\pm}|$,$|\psi_{\pm}\rangle$ 
corresponding to $E_{\pm}$ are 
\bea
\langle \phi_{\pm} | & = & \frac{1}{\sqrt{1 + \cos2\xi_2}} \bp 1 && \pm \frac{1}{\sqrt{2}} e^{-i(\xi_1 \mp \xi_2)} && \pm \frac{1}{\sqrt{2}} e^{-i(\xi_1 \pm \xi_2)} \ep\nonumber \\
|\psi_{\pm} \rangle        & = & \frac{1}{\sqrt{1 + \cos2\xi_2}} \bp 1 \\ \pm \frac{1}{\sqrt{2}} e^{i(\xi_1 \pm \xi_2)} \\ \pm \frac{1}{\sqrt{2}} e^{i(\xi_1 \mp \xi_2)} \ep 
\eea
\noindent where $\xi_1 = arg(W_{13})$,$\xi_2 = arg(E_{\pm} + i\gamma_2)$. The expressions of eigenstates are applicable
for real $E_{\pm}$. It appears that the sub-lattice Zak phase is a proper topological invariant for systems with
pseudo-chiral symmetry, and is defined in terms of the sub-lattice projection operator $P_j=\vert j\rangle\langle
j \vert, j= A, B, C$ as,
\bea
&& Z_{0, \pm}^{j} = \frac{i}{2} \oint \langle \tilde{\phi}_{0, \pm}^{j} \vert \partial_{k} \vert
\tilde{\psi}_{0, \pm}^{j} \rangle \ dk\nonumber \\
&& \vert \tilde{\psi}_{0, \pm}^{j} \rangle = \frac{P_{j} \vert \psi_{0, \pm} \rangle}{\sqrt{\langle \phi_{0,\pm}
\vert P_{j} \vert \psi_{0,\pm}\rangle}}, \
\langle \tilde{\phi}_{0, \pm}^{j} \vert = \frac{\langle \phi_{0, \pm} \vert P_{j}}{\sqrt{\langle \phi_{0,\pm}
\vert P_{j} \vert \psi_{0,\pm}\rangle}}
\label{slzak}
\eea
\noindent The sub-lattice Zak phases corresponding to $B$ and $C$ sub-lattices are,
\bea
Z_{0, \pm}^{B} & = & \begin{cases} -\pi, & \delta_1 < \delta_3 \\
        0, & \delta_1 > \delta_3
\end{cases}\nonumber \\
Z_{0, \pm}^{C} & = & \begin{cases} -\pi, & \delta_1 < \delta_3 \\
        0, & \delta_1 > \delta_3
\end{cases}\nonumber
\eea
\noindent The Zak phase corresponding to $A$-sub-lattice is always zero, $Z_{0, \pm}^{A} = 0$. The system 
undergoes from trivial topology to nontrivial topology when $\delta_{3}$ increases and becomes 
$\delta_3 > \delta_1$. It is seen from {\bf Fig. \ref{spec_obc_1g0_chiral}} that the system, under 
the OBC and the parameter constraints considered in this subsection, admits edge states for 
$\delta_3 > \delta_1$. The analytical results discussed in sec.~\ref{gbc-limit1-obc} also proves 
the existence of edgestate for $\delta_3>\delta_1$. The edge states appear when the system has non-trivial 
topology. So the BBC in the limit of pseudo-chiral symmetry is established through sublattice Zak phase. 

The expressions for the sub-lattice Zak phases for Case-II may be obtained from the results of Case-I
through the duality relation $\delta_1 \leftrightarrow t_1$ as,
\bea
Z_{0, \pm}^{B} & = & \begin{cases} -\pi, & t_1 < \delta_3 \\
        0, & t_1 > \delta_3
\end{cases}\nonumber \\
Z_{0, \pm}^{C} & = & \begin{cases} -\pi, & t_1 < \delta_3 \\
        0, & t_1 > \delta_3
\end{cases}\nonumber
\eea
\noindent The Zak phase corresponding to $A$-sub-lattice is always zero, $Z_{0, \pm}^{A} = 0$.

\begin{figure}
\centering
        \begin{subfigure}{0.48\columnwidth}
                \centering
                \includegraphics[width = 0.98\linewidth]{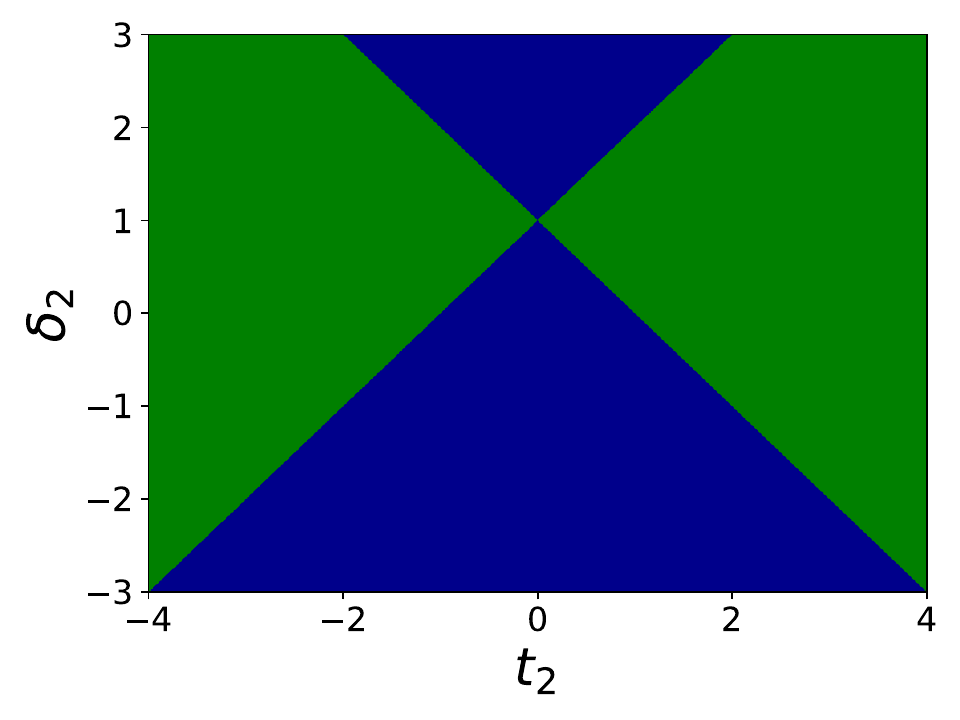}\quad
                \caption{}
                \label{zak_lower_1g0}
        \end{subfigure}
        \hfill
        \begin{subfigure}{0.48\columnwidth}
                \centering
                \includegraphics[width = 0.98\linewidth]{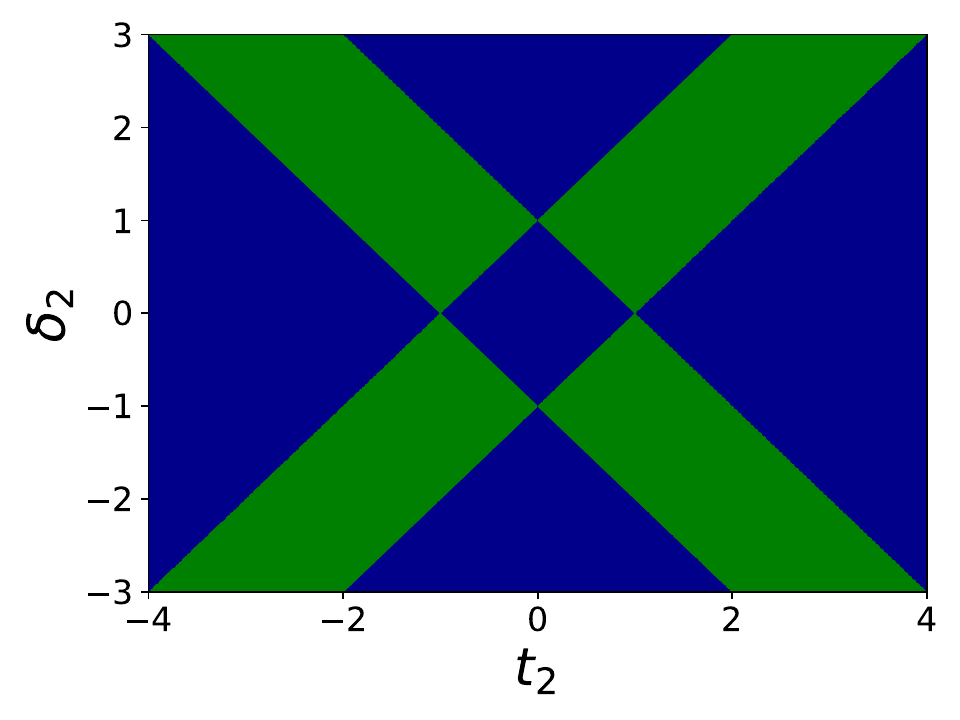}\quad
                \caption{}
                \label{zak_middle_1g0}
        \end{subfigure}
	\vfill
	\begin{subfigure}{0.48\columnwidth}
                \centering
                \includegraphics[width = 0.98\linewidth]{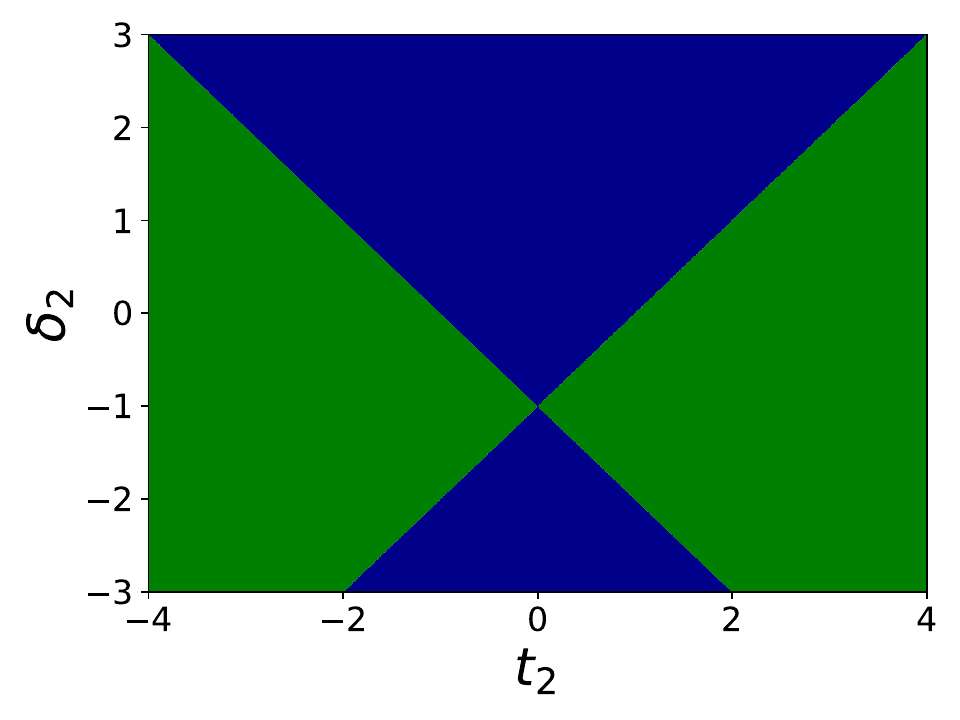}\quad
                \caption{}
                \label{zak_upper_1g0}
        \end{subfigure}
        \hfill
        \begin{subfigure}{0.48\columnwidth}
                \centering
                \includegraphics[width = 0.98\linewidth]{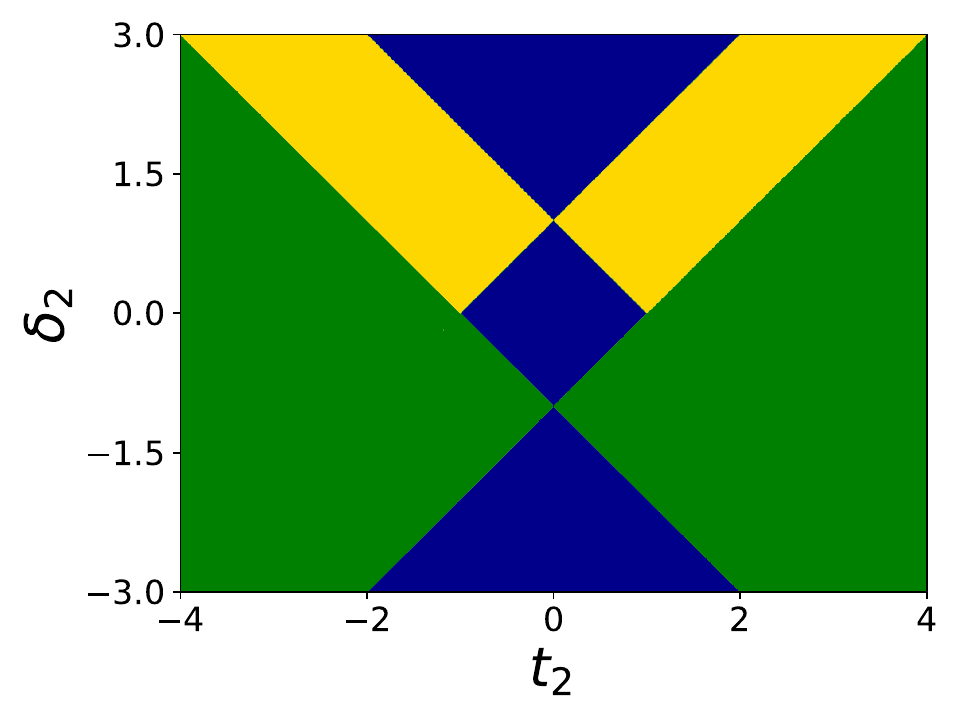}\quad
                \caption{}
                \label{zak_lpm_1g0}
        \end{subfigure}
	\caption{Contour plot of Zak phases for $\gamma_1 = 0$. Blue, green and yellow colors indicates 
	the value of $0, \pi, 2\pi$ respectively. Fig.(a)~ Zak phase corresponding to lower band, 
	Fig.(b)~ Zak phase corresponding to middle band, Fig.(c)~ Zak phase corresponding to upper band, 
	Fig.(d)~ Sum of Zak phases corresponding to lower and middle band. 
        Parameter values : $\gamma_1 = 0$, $\delta_3 = t_1 = 0$, $\delta_1 = t_3 = 1$, $\gamma_2 = 0.3$.}
        \label{zak_1g0_PT}
\end{figure}

{\bf ${\cal{PT}}$ Symmetry}: The Hamiltonian is $\mathcal{PT}$-symmetric with ${\cal{P}}={\cal{P}}_1$
for the Case-I if $\delta_3 = t_1 = 0$, The ${\cal{PT}}$-symmetric limit for Case-II will not be pursued
further, since it reduces the Hamiltonian to a SSH dimer. The system is described in terms of the
parameters $\delta_1 = t_3, \delta_2, t_2, \gamma_2$. We study topological phase transition
in terms of Zak phase which is obtained numerically using right eigenstates. The behavior of Zak
phases corresponding to different bands is shown in the {\bf Fig. \ref{zak_1g0_PT}}. Without loss of
generality, we set $\delta_1 = t_3 = 1$ and express all other parameters in units of $\delta_1$. We
observe through numerical analysis that different values of $\gamma_2$ has no effects on the 
Zak phase. We obtain the Zak phase with the variation of $\delta_2$ and $t_2$ for a fixed value of
$\gamma_2 = 0.3$. 

\subsection{$\gamma_2=0$} 
\label{pbc_2g0}

There are two cases which satisfy the necessary conditions in Eq.~(\ref{real-pcondi}) for reality
of the entire spectrum:\\
Case-I: $t_2 = \pm t_1, \delta_2=\pm \delta_1$,\\
Case-II: $t_2=\pm \delta_1, \delta_2=\pm t_1$\\
with no constraints on $\delta_3, t_3$. The parametric restrictions for these two cases
are related through the duality transformation $t_1 \leftrightarrow \delta_1$. We exclude from  our
discussions parameters with negative values, and those lattice configurations
which are either SSH dimer or disconnected unit cells of trimers. There are three distinct
parametric regions corresponding to different symmetries satisfying the conditions for Case-I and Case-II:
(i) Simultaneous ${\cal{PT}}$ $\&$ pseudo-chiral symmetry,
(ii) Pseudo-chiral symmetry  and (iii) ${\cal{PT}}$-symmetry.

(i) {\bf Simultaneous ${\cal{PT}}$ $\&$ pseudo-chiral symmetry}:
The Hamiltonian $H_k$ possesses simultaneous pseudo-chiral $\&$ ${\cal{PT}}$-symmetry for
$t_2 = t_1, \delta_2= \delta_1, \delta_3 = t_3 = 0$ under $\Gamma_3$ and ${\cal{P}}_2{\cal{T}}$.
This parametric regime contains flat band due to the condition $\delta_3 = t_3 = 0$ corresponding
to pseudo-chiral symmetry. 
The expressions of $\psi_{ja}$,$\psi_{jb}$,$\psi_{jc}$ for CLS are
\bea
&& \psi_{n,a} = i\delta_1, \psi_{n,b} = \gamma_1, \psi_{n,c} = -i\delta_1, \nonumber \\ 
&& \psi_{n+1,a} = it_1, \psi_{n-1,c} = -it_1 \nonumber
\eea
The value of probability densities at all other sites are zero, and CLS is extended 
over only five sites over three unit cells. 
The Bloch Hamiltonian has the form,
\bea
H_{k} = \bp i\gamma_1 && W^{*}_1 &&  0 \\ W_1 && 0 && W^{*}_1 \\ 0 && W_1 && -i\gamma_1 \ep
\nonumber
\eea
with the eigenvalues given by, 
\bea
E_0  =  0, E_{\pm} & = & \pm \sqrt{2|W_1|^{2} - \gamma_1^2} 
\nonumber
\eea
where $W_1 = \delta_1 + t_1 \ e^{ik}$. 
The left and right eigen vectors corresponding to $E_0$ are denoted as, $\langle \phi_{0}|$ and $|\psi_{0}\rangle$, respectively, 
\bea
\langle \phi_0 | = N_0 {\bp -iW_1 \\ -\gamma_1 \\ iW^{*}_1 \ep}^{T}, \ 
|\psi_{0}\rangle = N_0 \bp iW^{*}_1 \\ \gamma_1 \\ -iW_1 \ep
\nonumber
\eea
\noindent where the normalization factor $N_0 = (2|W_1|^{2} - \gamma_1^2)^{-\frac{1}{2}}$.
The left and right eigenstates corresponding to $E_{\pm}$ are obtained as,
\bea
\langle \phi_{\pm} | & = & \frac{1}{\sqrt{1 + \cos2\xi_4}} \bp \pm \frac{1}{\sqrt{2}} e^{i\left(\pm \xi_4 + \xi_3\right)} && 1 && 
\pm \frac{1}{\sqrt{2}} e^{i\left(\mp \xi_4 - \xi_3\right)} \ep \nonumber \\
|\psi_{\pm}\rangle   & = & \frac{1}{\sqrt{1 + \cos2\xi_4}} \bp \pm \frac{1}{\sqrt{2}} e^{i\left(\pm \xi_4 - \xi_3\right)} \\ 1 \\ 
\pm \frac{1}{\sqrt{2}} e^{i\left(\mp \xi_4 + \xi_3\right)} \ep
\nonumber
\eea
where $\xi_4 = arg(E_{+} + i\gamma)$, $\xi_3 = arg(W_1)$. 
We analytically obtain the sub-lattice Zak phase 
utilizing its definition in Eq. (\ref{slzak}) that correctly predicts BBC:
\bea
Z_{0, \pm}^{A} = \begin{cases} \pi, & \delta_1 < t_1 \\
	0, & \delta_1 > t_1 
\end{cases}, \ \
Z_{0, \pm}^{C} =  \begin{cases} -\pi, & \delta_1 < t_1 \\
        0, & \delta_1 > t_1\nonumber 
\end{cases} 
\eea
The Zak phase corresponding to $B$-sub-lattice is always zero, $Z_{0, \pm}^{B} = 0$. We get 
edge states under the OBC when $\delta_1 < t_1$. 

(ii) {\bf Pseudo-chiral symmetry}: The Hamiltonian has Pseudo-chiral symmetry for $t_2= \delta_1, \delta_2= t_1,
t_3=\delta_3=0$. The duality transformation $t_1 \leftrightarrow \delta_1$ may be utilized to obtain
results for this case from that of the system under simultaneous ${\cal{PT}}$ \& pseudo-chiral symmetry
corresponding to $t_2 = t_1, \delta_2= \delta_1, t_3=\delta_3=0$. In particular,
\bea
Z_{0, \pm}^{A} = \begin{cases} \pi, & \delta_1 > t_1 \\
	0, & \delta_1 < t_1 
\end{cases}, \ \
Z_{0, \pm}^{C}  =  \begin{cases} -\pi, & \delta_1 > t_1 \\
        0, & \delta_1 < t_1\nonumber 
\end{cases}
\eea
\noindent The Zak phase corresponding to $B$-sub-lattice is always zero, $Z_{0, \pm}^{B} = 0$.

\begin{figure}
\centering
        \begin{subfigure}{0.48\columnwidth}
                \centering
                \includegraphics[width = 0.98\linewidth]{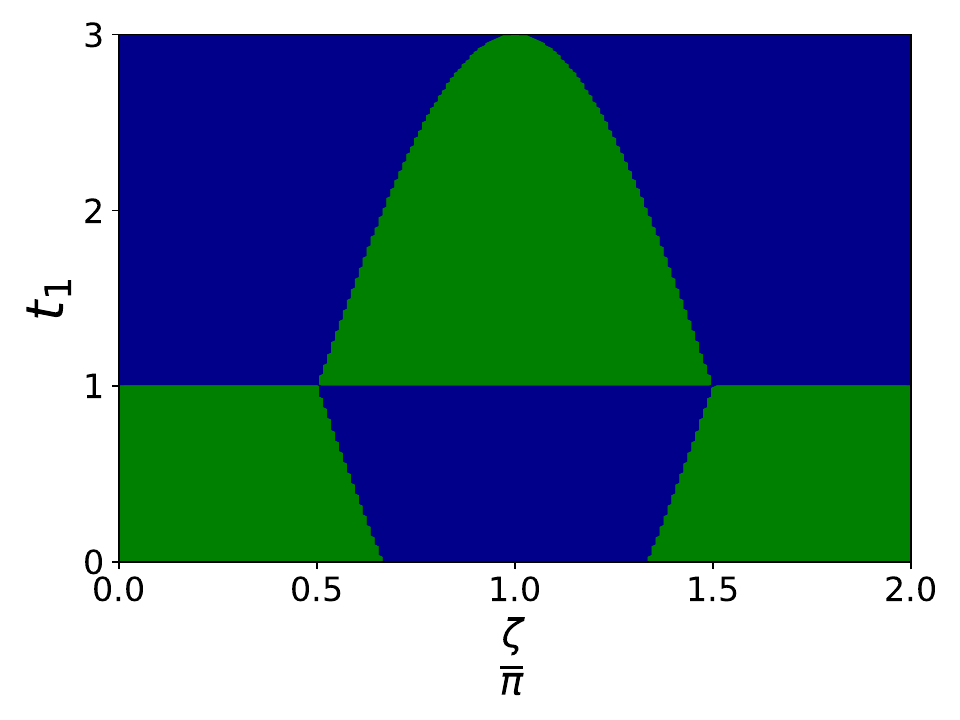}\quad
                \caption{}
                \label{zak_lower_2g0}
        \end{subfigure}
        \hfill
        \begin{subfigure}{0.48\columnwidth}
                \centering
                \includegraphics[width = 0.98\linewidth]{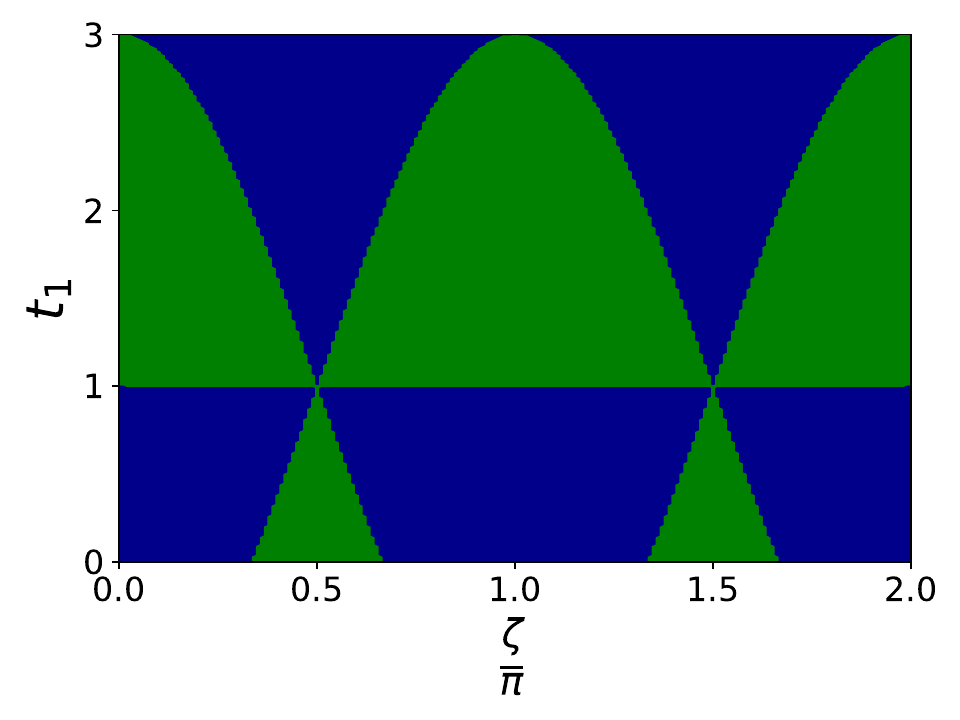}\quad
                \caption{}
                \label{zak_middle_2g0}
        \end{subfigure}
	\vfill
	\begin{subfigure}{0.48\columnwidth}
                \centering
                \includegraphics[width = 0.98\linewidth]{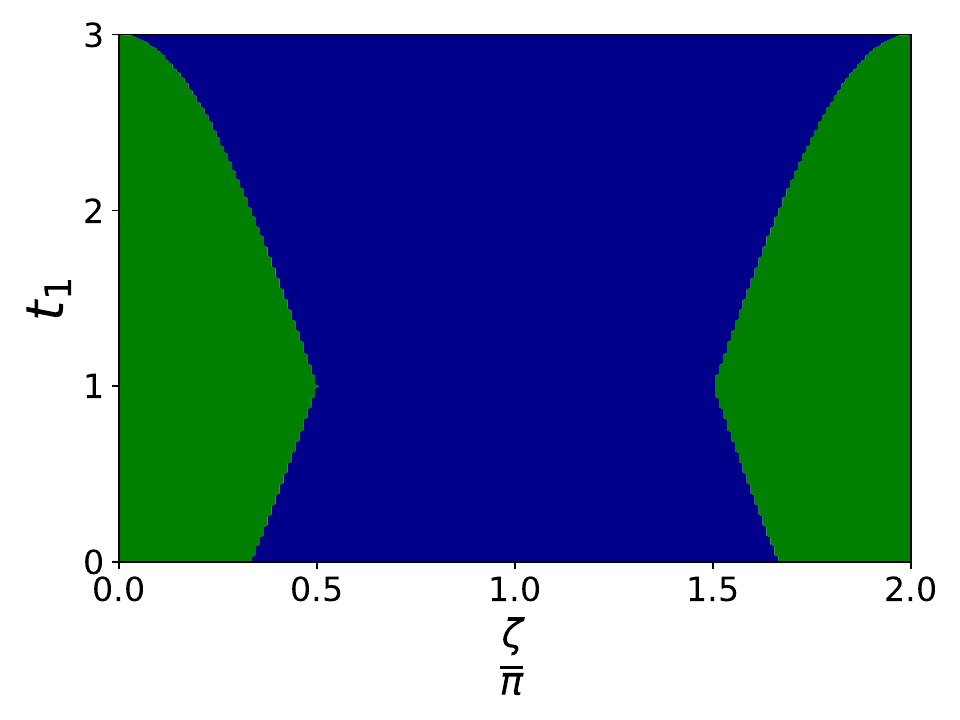}\quad
                \caption{}
                \label{zak_upper_2g0}
        \end{subfigure}
        \hfill
        \begin{subfigure}{0.48\columnwidth}
                \centering
                \includegraphics[width = 0.98\linewidth]{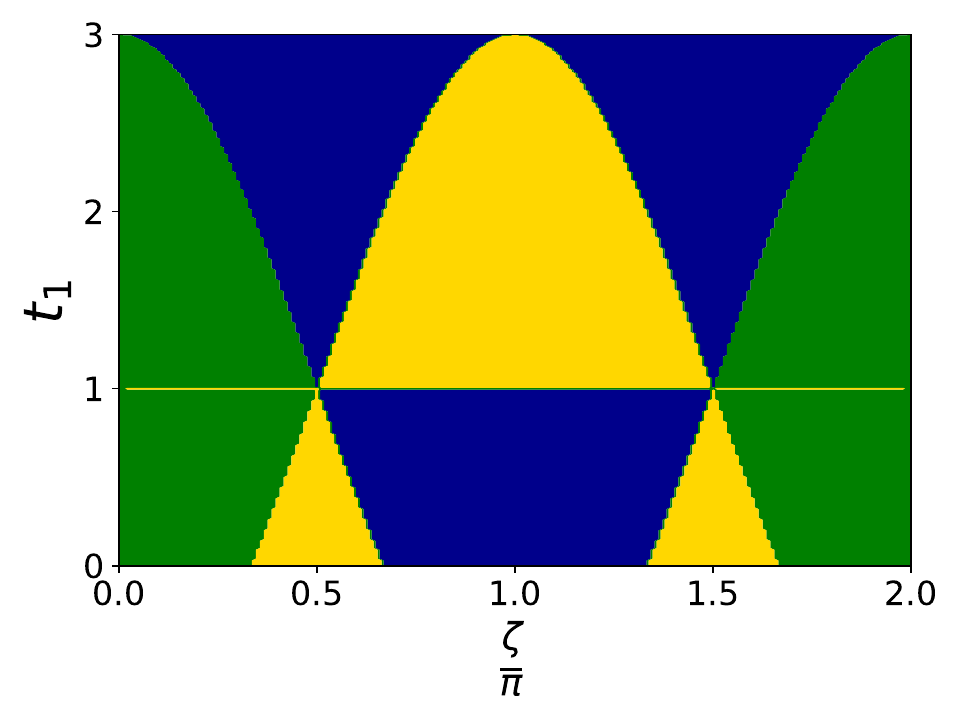}\quad
                \caption{}
                \label{zak_lpm_2g0}
        \end{subfigure}
        \caption{Contour plot of the Zak phase for $\gamma_2 = 0$.
        Parameter values : $\delta_2 = \delta_1 = 1$,$t_2 = t_1$, $\gamma_1 = 0.3$,$\gamma_2 = 0$, 
	$\delta_3 = 1 + \cos(\zeta)$, $t_3 = 1 - \cos(\zeta)$. Blue, green and yellow colors indicates 
	the value of $0, \pi, 2\pi$ respectively. Fig.(a)~ Zak phase corresponding to lower energy band, 
	Fig.(b)~ Zak phase corresponding to middle energy band, Fig.(c)~ Zak phase corresponding to upper 
	energy band, Fig.(d)~Sum of Zak phases corresponding to lower and  middle band}
        \label{zak_2g0_PT}
\end{figure}

(iii) {\bf ${\cal{PT}}$ symmetry}: The Hamiltonian $H_k$ is $\mathcal{PT}$-symmetric 
under ${\cal{P}}_2$ for $t_2 = t_1, \delta_2= \delta_1$ with no restriction on $\delta_3, t_3$.
If $\delta_3=t_3=0$, the Hamiltonian has an additional pseudo-chiral symmetry and has been discussed
earlier. We consider $\delta_3 \neq 0 \neq t_3$ and study the effect of ${\cal{PT}}$-symmetry alone
on the topology.  There are no flat bands in the spectrum. We re-parameterize 
$\delta_3$, $t_3$ as $\delta_3 = 1 + \cos(\zeta), t_3 = 1 - \cos(\zeta), 0 \leq \zeta \leq 2 \pi$. 
The numerically computed Zak phase is shown in {\bf Fig. \ref{zak_2g0_PT}}. The BBC is  established for $t_1 < 1$ in terms 
of Zak phase. The edge states under OBC appear in the band gap between lower and middle bands when the Zak phase
corresponding to lower band has non-trivial value. On the other hand, the edge state for the
OBC appear in the band gap between middle and upper bands when the sum of Zak phases corresponding to lower
and middle band has non trivial value which can be observed from the {\bf Fig. \ref{spec_obc_2g0_PT}}. 

\subsection{$\gamma_2=-\gamma_1$} 
\label{pbc_g2emg1}

The necessary conditions in Eq.~(\ref{real-pcondi}) for reality
of the entire spectrum are fulfilled for the two cases:\\
Case-I: $t_2 = \pm t_3, \delta_2=\pm \delta_3$,\\
Case-II: $t_2=\pm \delta_3, \delta_2=\pm t_3$ \\
with no restrictions on $\delta_1, t_1$. The parametric restrictions for Case-I
and case-II are related through the duality transformation $t_3 \leftrightarrow \delta_3$.

{\bf Pseudo-chiral symmetry:} Both the Case-I and Case-II correspond to pseudo-chiral symmetry for
$\delta_1 = t_1 = 0$ which also ensures the appearance of zero-energy flat bands. We analyze the
system for $t_2 = t_3, \delta_2= \delta_3, \delta_1=t_1=0$ in some detail and utilize
the duality symmetry to obtain TPT for the Case-II.
The expressions of $\psi_{ja}$,$\psi_{jb}$,$\psi_{jc}$ for compact localized states are
\bea
&& \psi_{n,a} = i\delta_2, \psi_{n,b} = -i\delta_2, \psi_{n,c} = \gamma_1, \nonumber \\
&& \psi_{n+1,a} = i\delta_3, \psi_{n+1,b} = -i\delta_3 \nonumber
\eea
\noindent The probability densities at all other sites are zero, and the CLS is extended
over only five sites over two unit cells. The Bloch Hamiltonian and the energy eigenvalues are given by,
\bea
H_{k} = \bp i\gamma_1 & 0 & W^{*} \\ 0 & -i\gamma_1 & W^{*} \\ W & W & 0 \ep\nonumber \\
\eea
\bea
E_{0} =  0, \ E_{\pm} = \pm \sqrt{2|W|^2 - \gamma_1^2},
\eea
where $W = \delta_2 + \delta_3 e^{ik}$. The left and right eigenvectors $\langle \phi_{0} \vert, \vert \psi_{0} \rangle $ 
corresponding to the flat band are,
\bea
\langle \phi_0| = N_0 {\bp -iW \\ iW \\ -\gamma_1 \ep}^{T}, \ |\psi_{0} \rangle & = &  N_0 \bp iW^{*} \\ -iW^{*} \\  \gamma_1 \ep 
\eea
\noindent where the normalization factor $N_0 = \frac{1}{\sqrt{2|W|^{2} - \gamma_1^2}}$.
The left and right eigenstates $\langle \phi_{\pm}|, \ |\psi_{\pm} \rangle$  corresponding to energy $E_{\pm}$ are
\bea
\langle \phi_{\pm} | & = & \frac{1}{\sqrt{1 + \cos2\xi_6}} \bp \pm \frac{1}{\sqrt{2}} e^{i(\xi_5 \pm \xi_6)} &&  \pm \frac{1}{\sqrt{2}} e^{i(\xi_5 \mp \xi_6)}  && 1 \ep \nonumber \\
|\psi_{\pm} \rangle  & = &  \frac{1}{\sqrt{1 + \cos2\xi_6}} \bp \pm \frac{1}{\sqrt{2}} e^{-i\left(\xi_5 \mp \xi_6\right)} \\ 
\pm \frac{1}{\sqrt{2}} e^{-i\left(\xi_5 \pm \xi_6\right)} \\ 1\ep
\eea
\noindent where $\xi_6 = arg(E_{+} + i\gamma_1)$, $\xi_5 = arg(W)$. Following the procedure outlined in previous sections,
we analytically obtain the sub-lattice Zak phase that correctly predicts bulk boundary correspondence,
\bea
Z_{0, \pm}^{A} & = & \begin{cases} \pi, & \delta_2 < \delta_3 \\
        0, & \delta_2 > \delta_3
\end{cases}\nonumber \\
Z_{0, \pm}^{B} & = & \begin{cases} -\pi, & \delta_2 < \delta_3 \\
        0, & \delta_2 > \delta_3\nonumber 
\end{cases}
\eea
The sub-lattice Zak phase corresponding to $C$ sub-lattice is always zero $Z_{0, \pm}^{C} = 0$. We get
edge state in OBC when $\delta_2 < \delta_3$. The Zak phases for the Case-II are obtained through the
duality relation $t_3 \leftrightarrow \delta_3$ as,
\bea
Z_{0, \pm}^{A} & = & \begin{cases} \pi, & \delta_2 < t_3 \\
        0, & \delta_2 > t_3
\end{cases}\nonumber \\
Z_{0, \pm}^{B} & = & \begin{cases} -\pi, & \delta_2 < t_3 \\
        0, & \delta_2 > t_3\nonumber 
\end{cases}
\eea
\noindent and $Z_{0, \pm}^{C} = 0$. The pseudo-chiral symmetry breaks for non-vanishing $\delta_1$ and $t_1$,
and the system no longer admits flat bands. Unlike the case of pseudo-chiral symmetric $H_k$, no analytic expressions
for the eigenstates are available for non-vanishing $\delta_1$ and $t_1$. 

\begin{figure}
\centering
        \begin{subfigure}{0.48\columnwidth}
                \centering
                \includegraphics[width = 0.98\linewidth]{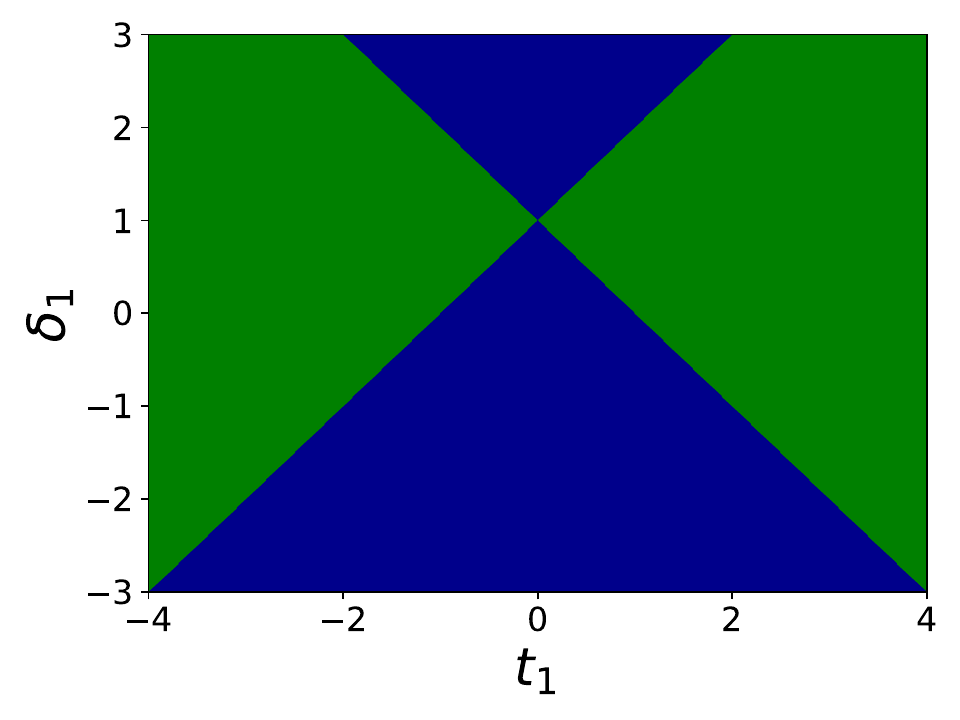}\quad
                \caption{}
                \label{zak_lower_1gem2g}
        \end{subfigure}
        \hfill
        \begin{subfigure}{0.48\columnwidth}
                \centering
                \includegraphics[width = 0.98\linewidth]{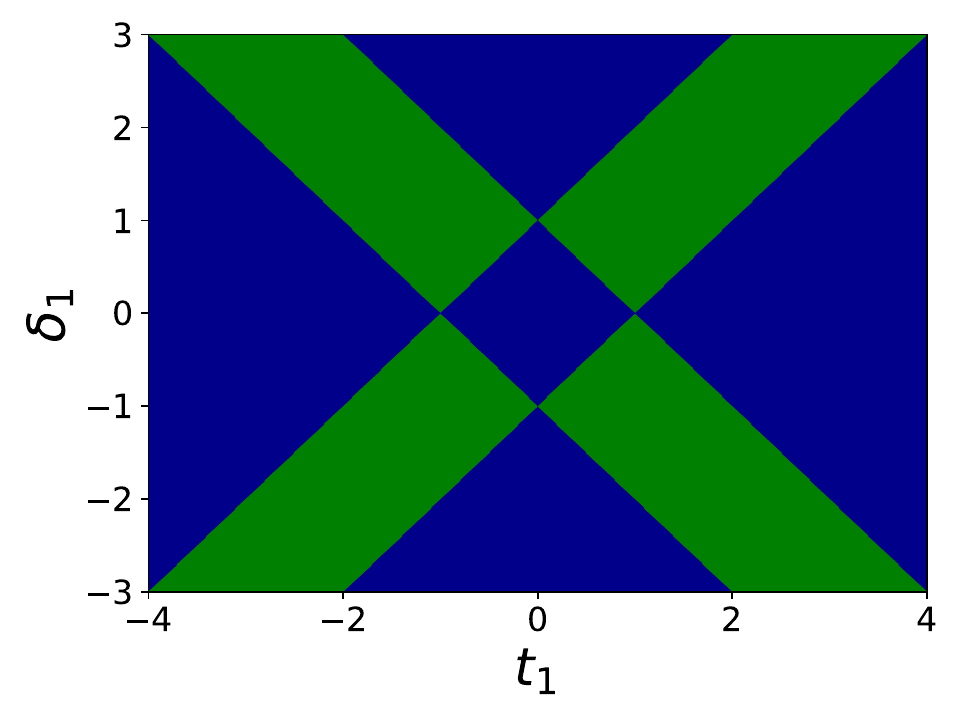}\quad
                \caption{}
                \label{zak_middle_1gem2g}
        \end{subfigure}
        \vfill
        \begin{subfigure}{0.48\columnwidth}
                \centering
                \includegraphics[width = 0.98\linewidth]{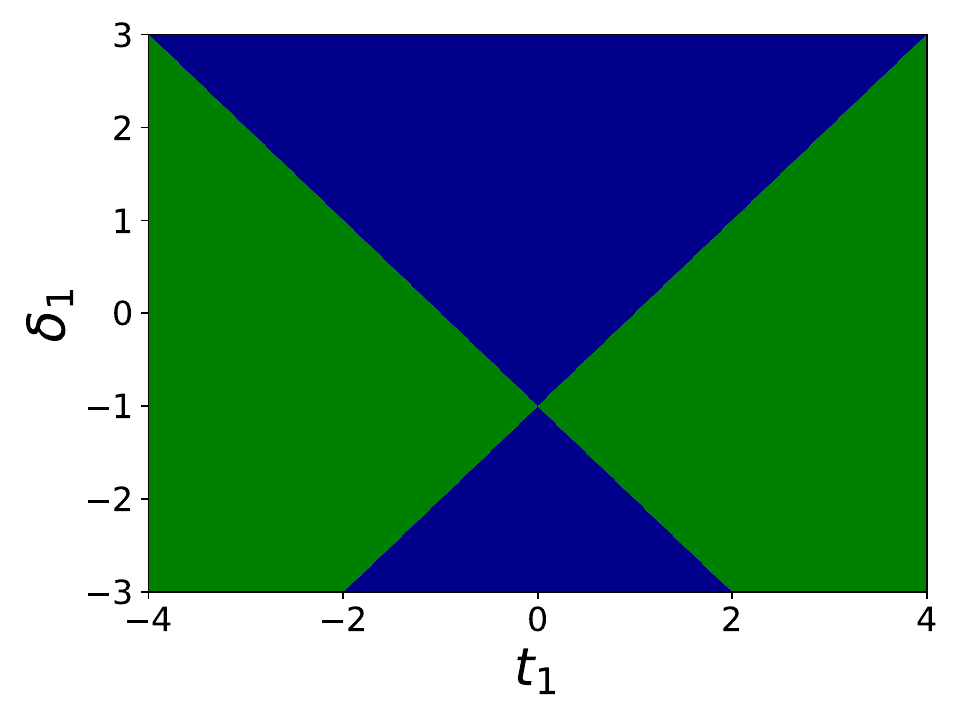}\quad
                \caption{}
                \label{zak_upper_1gem2g}
        \end{subfigure}
        \hfill
        \begin{subfigure}{0.48\columnwidth}
                \centering
                \includegraphics[width = 0.98\linewidth]{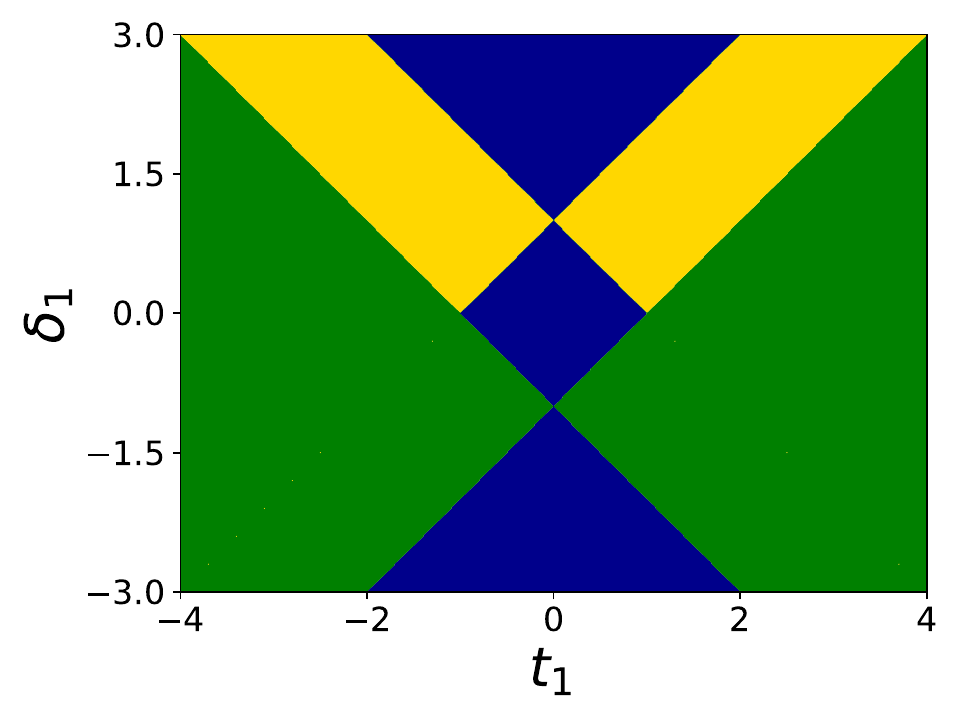}\quad
                \caption{}
                \label{zak_lpm_1gem2g}
        \end{subfigure}
        \caption{Contour plot of Zak phases for $\gamma_2 = -\gamma_1$. Blue, green and yellow colors indicate 
        the value of $0, \pi, 2\pi$ respectively. Fig.(a)~ Zak phase corresponding to lower band,
        Fig.(b)~ Zak phase corresponding to middle band, Fig.(c)~ Zak phase corresponding to upper band,
        Fig.(d)~ Sum of Zak phases corresponding to lower and middle band.
        Parameter values : $\gamma_1 = -\gamma_2 = 0.3$, $\delta_3 = t_2 = 0$, $\delta_2 = t_3 = 1$.}
        \label{zak_1gem2g_PT}
\end{figure}

{\bf ${\mathcal{PT}}$-symmetry}: The system is $\mathcal{PT}$-symmetric for Case-II when $t_2 = \delta_3 = 0$.The Hamiltonian
reduces to SSH dimers for the Case-I, and will not be pursued further.
The topological phase transition is studied for the $\mathcal{PT}$-symmetric system i.e., $\delta_3 = t_2 = 0$ 
and $t_3 = \delta_2$. The Zak phase is computed numerically and its contour plot is shown in {\bf Fig. \ref{zak_1gem2g_PT}}.

\section{Open boundary condition}
\label{obc-sec}

In the previous section, we have studied Zak phases 
of the Bloch Hamiltonian under the PBC, and with pseudo-chiral and/or $\mathcal{PT}$-symmetry. 
The Zak phase is quantized and takes values $0, \pm \pi$. The sudden changes in its value
at some critical points in the parameter space imply topological phase transitions. It is known that
non-trivial topology under the PBC indicates the existence of edge states under the OBC. 
In this section, we numerically study the Hamiltonian under the OBC to identify the parameter regime that admits 
edge states. In the pseudo-chiral limit, the analytical expressions of the edge states and its energy 
are obtainable through the study of GBC, and relevant discussions are included in Sec.~\ref{gbc-limit1-obc}. 
We perform numerical analysis to analyze the entire spectrum, of which only a subset corresponds to the
edge states, and study the nature of the band gap. Further, we employ numerical techniques to study the
system under the $\mathcal{PT}$-symmetry, since it appears that the Hamiltonian may not be amenable for a complete
analytical study. 

\subsection{$\gamma_1 = 0$}
\label{obc-1g0}
\begin{figure}
\centering
        \begin{subfigure}{0.48\columnwidth}
                \centering
                \includegraphics[width = 0.98\linewidth]{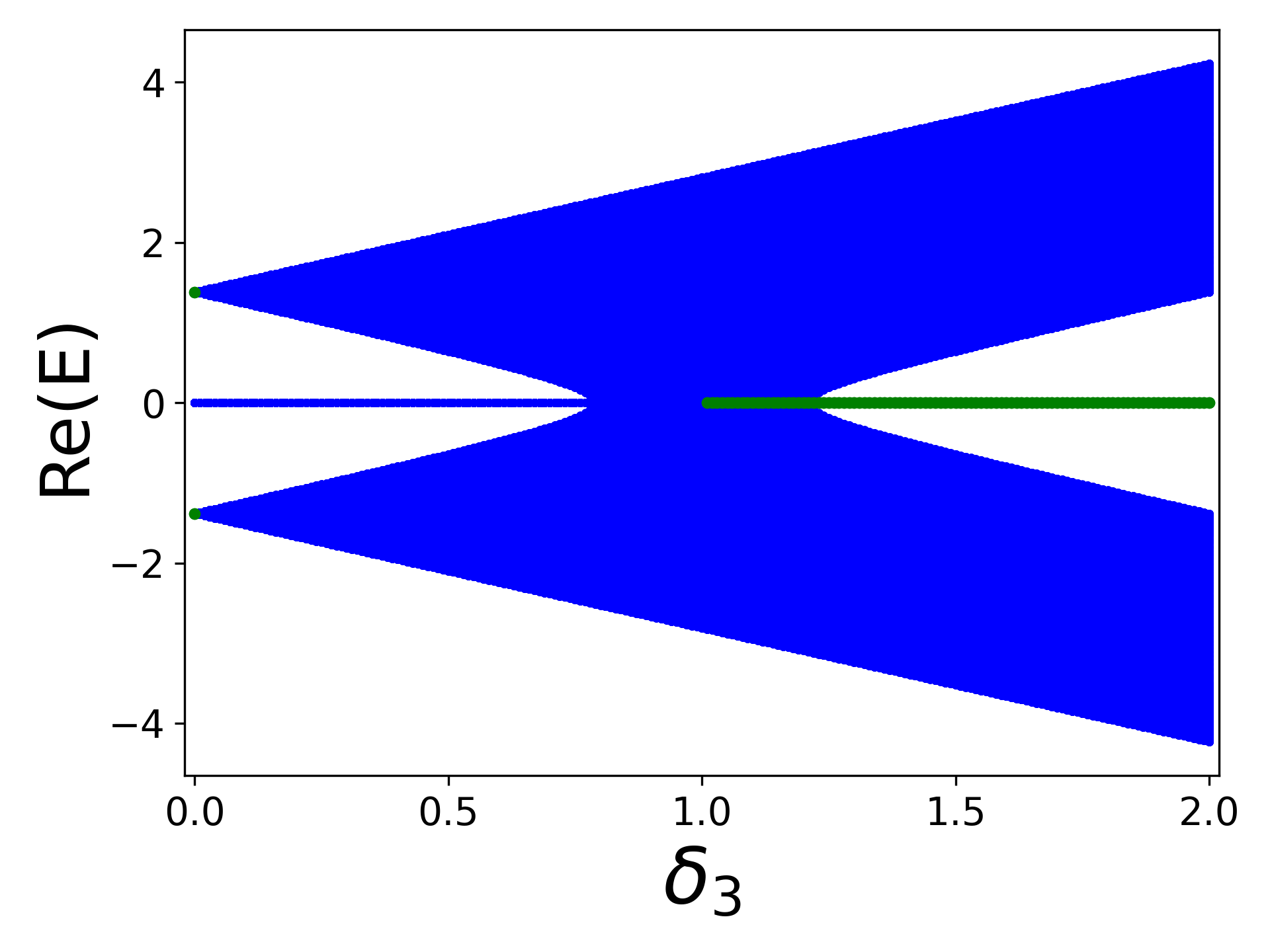}\quad
                \caption{}
                \label{spec_obc_real_1g0_chiral}
        \end{subfigure}
        \hfill
        \begin{subfigure}{0.48\columnwidth}
                \centering
                \includegraphics[width = 0.98\linewidth]{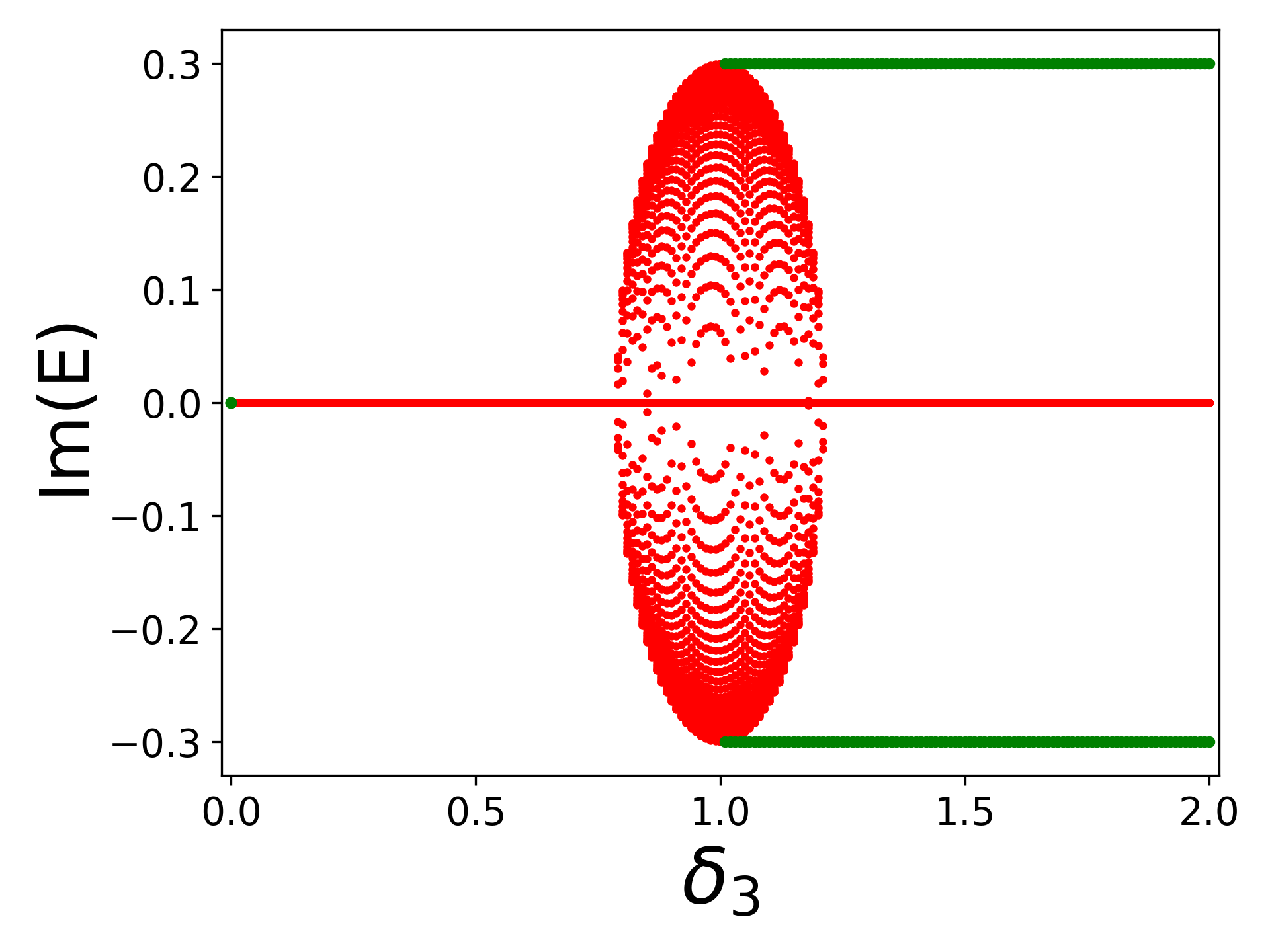}\quad
                \caption{}
                \label{spec_obc_imag_1g0_chiral}
        \end{subfigure}
        \caption{Plot of the spectrum under OBC for $\gamma_1 = 0$.
        Parameter values : $\gamma_1 = 0$,$\gamma_2 = 0.3$, $\delta_2 = t_2 = 0$, $\delta_1 = t_3 =1$, $\delta_3 = t_1$.}
        \label{spec_obc_1g0_chiral}
\end{figure}

\begin{figure}
\centering
        \begin{subfigure}{0.48\columnwidth}
                \centering
                \includegraphics[width = 0.98\linewidth]{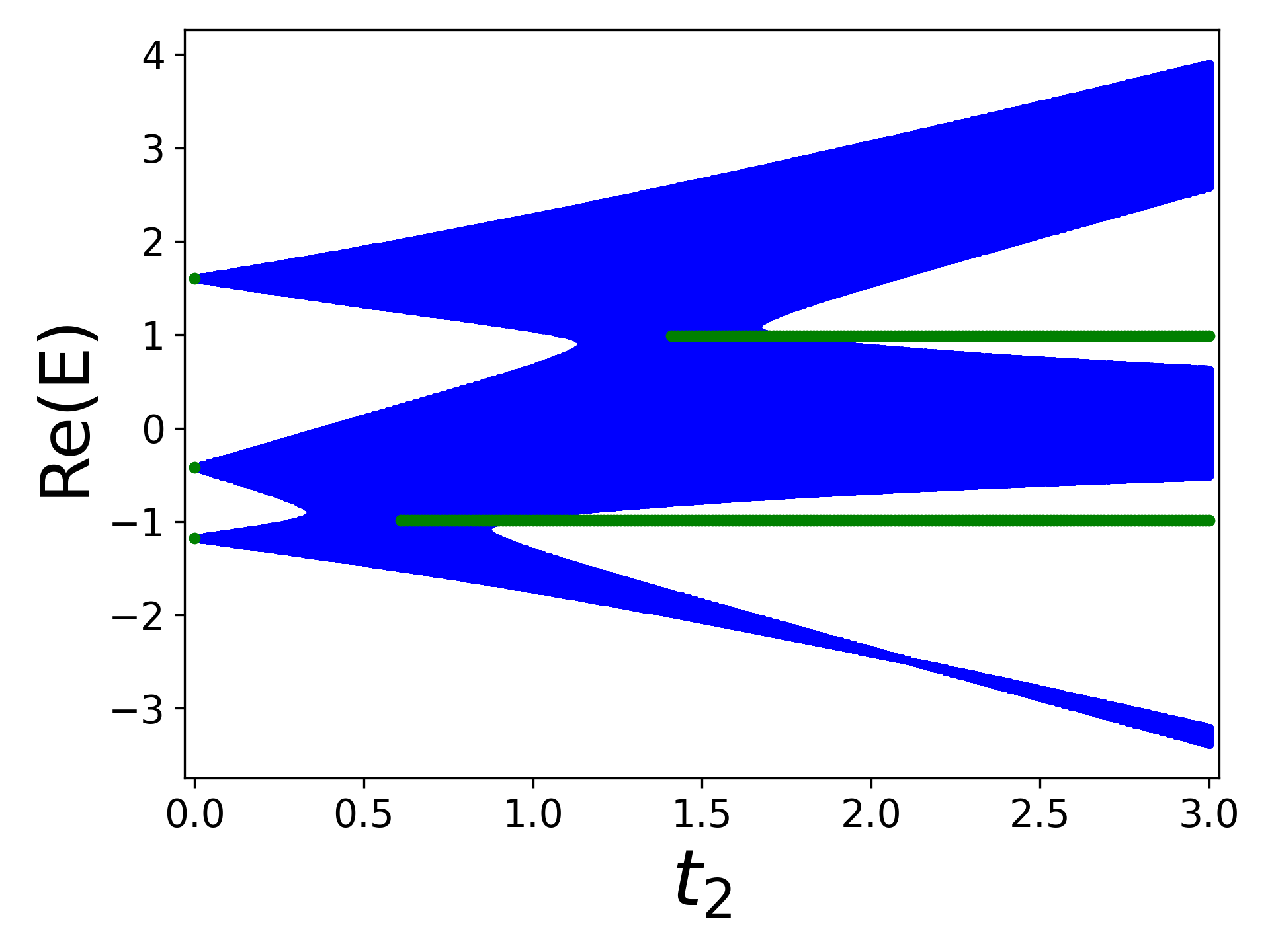}\quad
                \caption{}
                \label{spec_obc_real_1g0_2d04}
        \end{subfigure}
        \hfill
        \begin{subfigure}{0.48\columnwidth}
                \centering
                \includegraphics[width = 0.98\linewidth]{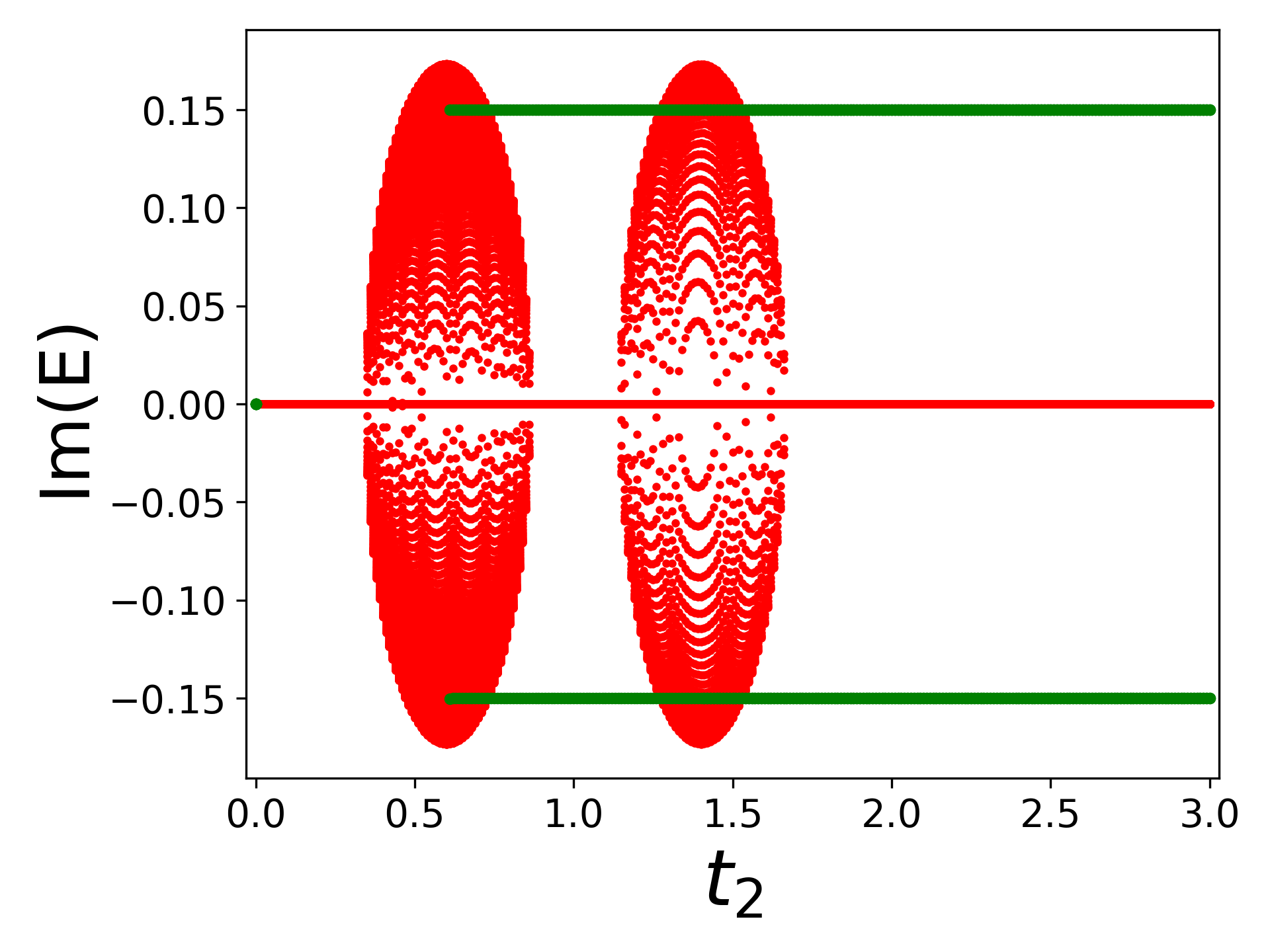}\quad
                \caption{}
                \label{spec_obc_imag_1g0_2d04}
        \end{subfigure}
        \caption{Plot of the spectrum under OBC for $\gamma_1 = 0$.
        Parameter values : $\gamma_1 = 0$,$\gamma_2 = 0.3$, $\delta_1 = t_3 = 1$, $\delta_3 = t_1 = 0$, $\delta_2 = 0.4$.}
        \label{spec_obc_1g0_PT}
\end{figure}

\noindent In this section, we study the edge states corresponding to the parameter regimes as given in Sec.IV.A. 
We first consider the pseudo-chiral limit i.e., $\delta_2 = t_2 = 0$, $\delta_3 = t_1$ and $\delta_1 = t_3$. 
In this limit, the system shows non-trivial topological phase when $\delta_1 < \delta_3$. The spectrum under the OBC 
is shown in {\bf Fig. \ref{spec_obc_1g0_chiral}}. The green color indicates eigenvalues of the edge states. We
observe that edge states appear when the system has non-trivial topology (Sub-lattice Zak phases corresponding
to B,C are $-\pi$). 
In this limit, the system also admits CLS corresponding to flatbands. The expressions of $\psi_{na}$,$\psi_{nb}$ and $\psi_{nc}$ 
for the CLS located around $n = l$ unit cell, are 
\bea
&& \psi_{la} = -\gamma_2, \ \psi_{lb} = -i\delta_1, \ \psi_{lc} = i\delta_1, \nonumber \\
&& \psi_{l-1,a} = 0, \ \psi_{l-1,b} = -i\delta_3, \ \psi_{l-1,c} = i\delta_3,  \nonumber 
\eea
and $\psi_{na} = \psi_{nb} = \psi_{nc} = 0$ for $n \neq l$. This expression is valid for $l=2,3, \dots m$. 
Unlike to the PBC, the expression of the CLS localized at the left boundary is different from CLS located at bulk sites. 
The expressions of CLS located at $n = 1$ unit cell is 
\bea
\psi_{1a} = -\gamma_2, \ \psi_{1b} = -i\delta_1, \ \psi_{1c} = i\delta_1 \nonumber
\eea
and probability density of CLS at other sites is zero. The system in  pseudo-chiral limit admits two types of localized states, 
topology protected edgetstate and CLS. In the $\mathcal{PT}$-symmetric limit i.e., $\delta_3 = t_1 = 0$ and $t_3 =
\delta_1$ , the real and the complex parts of the spectrum under OBC is shown in {\bf Fig. \ref{spec_obc_1g0_PT}}. The green
color indicates the energy of the edge states. We observe that the edge state appear when the system has non-trivial
topology under the PBC. 
The edge states under OBC appear in the band gap between lower and middle bands when the Zak phase corresponding 
to lower band has non-trivial value. On the otherhand, the edge state for the OBC appear in the bandgap between 
middle and upper bands when the sum of Zak phases corresponding to lower and middle band has non trivial value 
which can be observed from the {\bf Fig. \ref{spec_obc_1g0_PT}}. This establishes the BBC of the trimer lattice.
\subsection{$\gamma_2 = 0$} \label{obc-2g0}

\begin{figure}
\centering
        \begin{subfigure}{0.48\columnwidth}
                \centering
                \includegraphics[width = 0.98\linewidth]{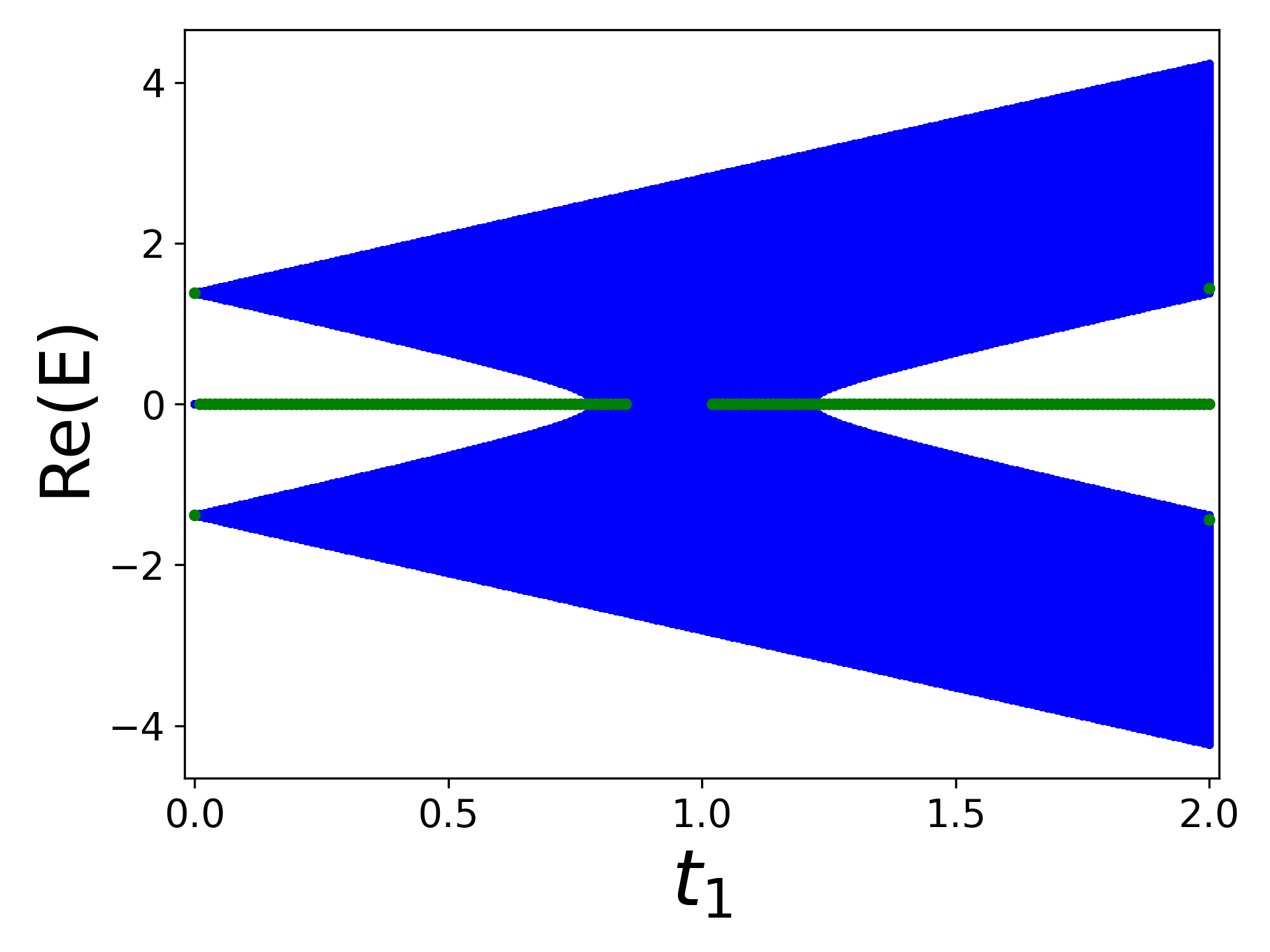}\quad
                \caption{}
                \label{spec_real_obc_2g0_chiral}
        \end{subfigure}
        \hfill
        \begin{subfigure}{0.48\columnwidth}
                \centering
                \includegraphics[width = 0.98\linewidth]{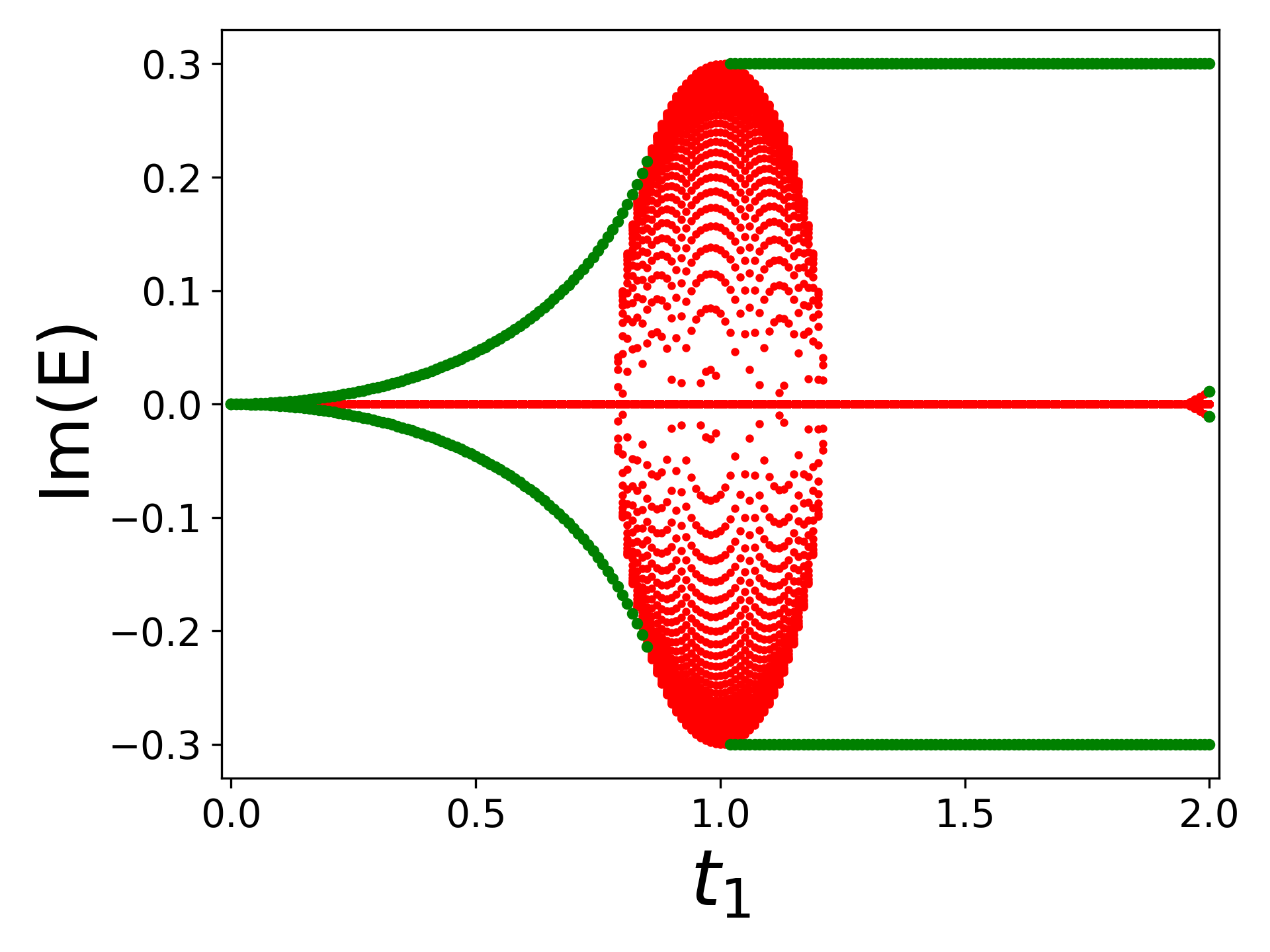}\quad
                \caption{}
                \label{spec_imag_obc_2g0_chiral}
        \end{subfigure}
        \caption{Plot of the spectrum under OBC for $\gamma_2 = 0$.
        Parameter values : $\gamma_2 = 0$, $\gamma_1 = 0.3$, $\delta_1 = \delta_2 = 1$, $t_2 = t_1$, $\delta_3 = t_3 =0$.}
        \label{spec_obc_2g0_chiral}
\end{figure}

\noindent In this section, we study the edge states corresponding to the parameter regimes as stated in Sec.IV.B. 
We first consider the pseudo-chiral limit i.e., $t_3 = \delta_3 = 0$. The spectrum of the system under OBC in 
pseudo-chiral limit is shown in {\bf Fig. \ref{spec_obc_2g0_chiral}}. 
We observe that a pair of eigenvalues corresponding to zero energy flat band under PBC becomes purely imaginary under
OBC and those states show localization at one edges although the sub-lattice Zak phase is trivial. The possible reason
is the interaction between onsite imaginary potential and flat band. This is different from other two cases 
corresponding to $\gamma_1 = 0$ and $\gamma_2 = -\gamma_1$. However, two edgestates appear 
from other two bands when the sub-lattice Zak phase has non-trivial value. In this limit, the system also
admits CLS corresponding to flatbands. 
The expressions of $\psi_{na}$,$\psi_{nb}$ and $\psi_{nc}$ for the CLS located around $n = l$ unit cell, are
\bea
&& \psi_{la} = i\delta_1, \ \psi_{lb} = \gamma_1, \ \psi_{lc} = -i\delta_1, \nonumber \\
&& \psi_{l+1,a} = it_1, \ \psi_{l-1,c} = -it_1, \nonumber
\eea
and probability density is zero at other sites. This expression is valid for $l=2,3, \dots (m-1)$.
There is no CLS located at the boundary for the parameter ranges discussed in this subsection. 

\begin{figure}
\centering
        \begin{subfigure}{0.48\columnwidth}
                \centering
                \includegraphics[width = 0.98\linewidth]{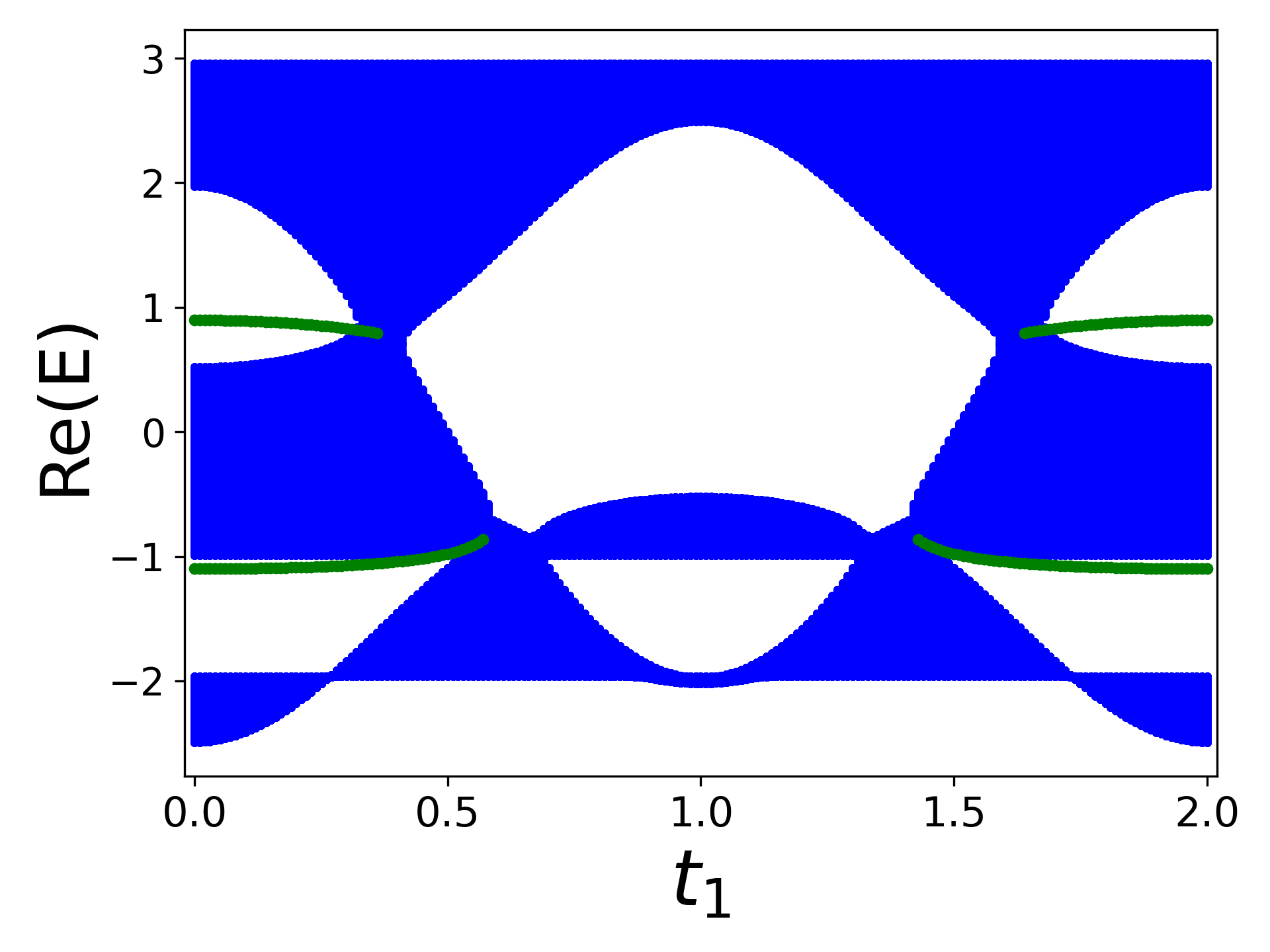}\quad
                \caption{}
                \label{spec_obc_real_2g0_1t02}
        \end{subfigure}
        \hfill
        \begin{subfigure}{0.48\columnwidth}
                \centering
                \includegraphics[width = 0.98\linewidth]{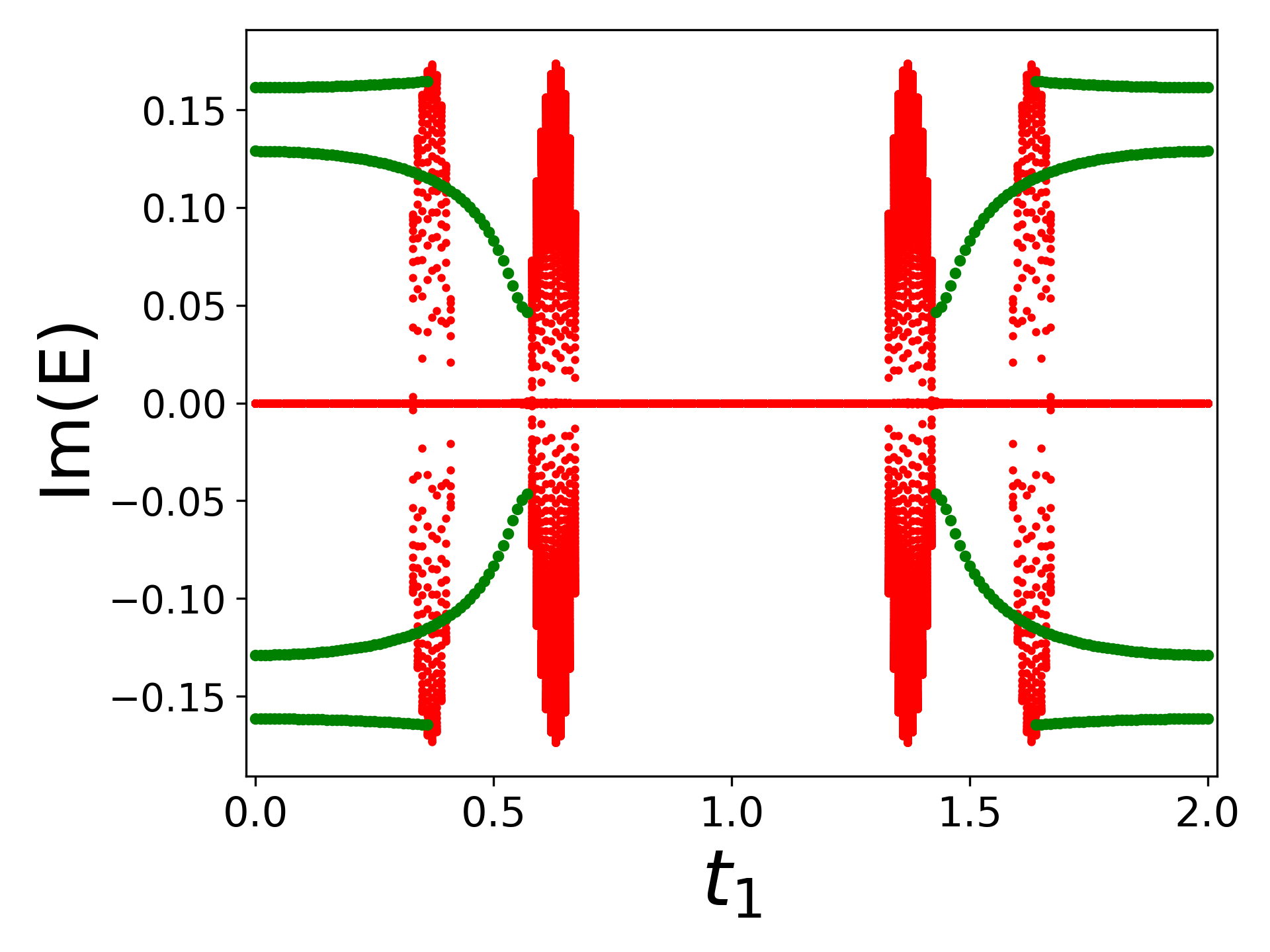}\quad
                \caption{}
                \label{spec_obc_imag_2g0_1t02}
        \end{subfigure}
        \caption{Plot of the spectrum under OBC for $\gamma_2 = 0$.
	Parameter values : $\gamma_2 = 0$, $\gamma_1 = 0.3$, $\delta_1 = \delta_2 = 1$, $t_2 = t_1 = 0.2$, $\delta_3 = 1+ \cos\zeta$, 
	$t_3 = 1 - \cos\zeta$.}
        \label{spec_obc_2g0_PT}
\end{figure}

The real and the complex parts of the spectrum under OBC for $\delta_3 \neq t_3 \neq 0$ is shown 
in {\bf Fig. \ref{spec_obc_2g0_PT}}. The green color indicates the
energy of the edge states. We observe that the edge states appear when the system has non-trivial topology in
PBC for $t_1 < 1$. This establishes the BBC of the trimer lattice for $t_1 < 1$. 
\subsection{$\gamma_2 = -\gamma_1$}
\label{obc-g2emg1}
\begin{figure}
\centering
        \begin{subfigure}{0.48\columnwidth}
                \centering
                \includegraphics[width = 0.98\linewidth]{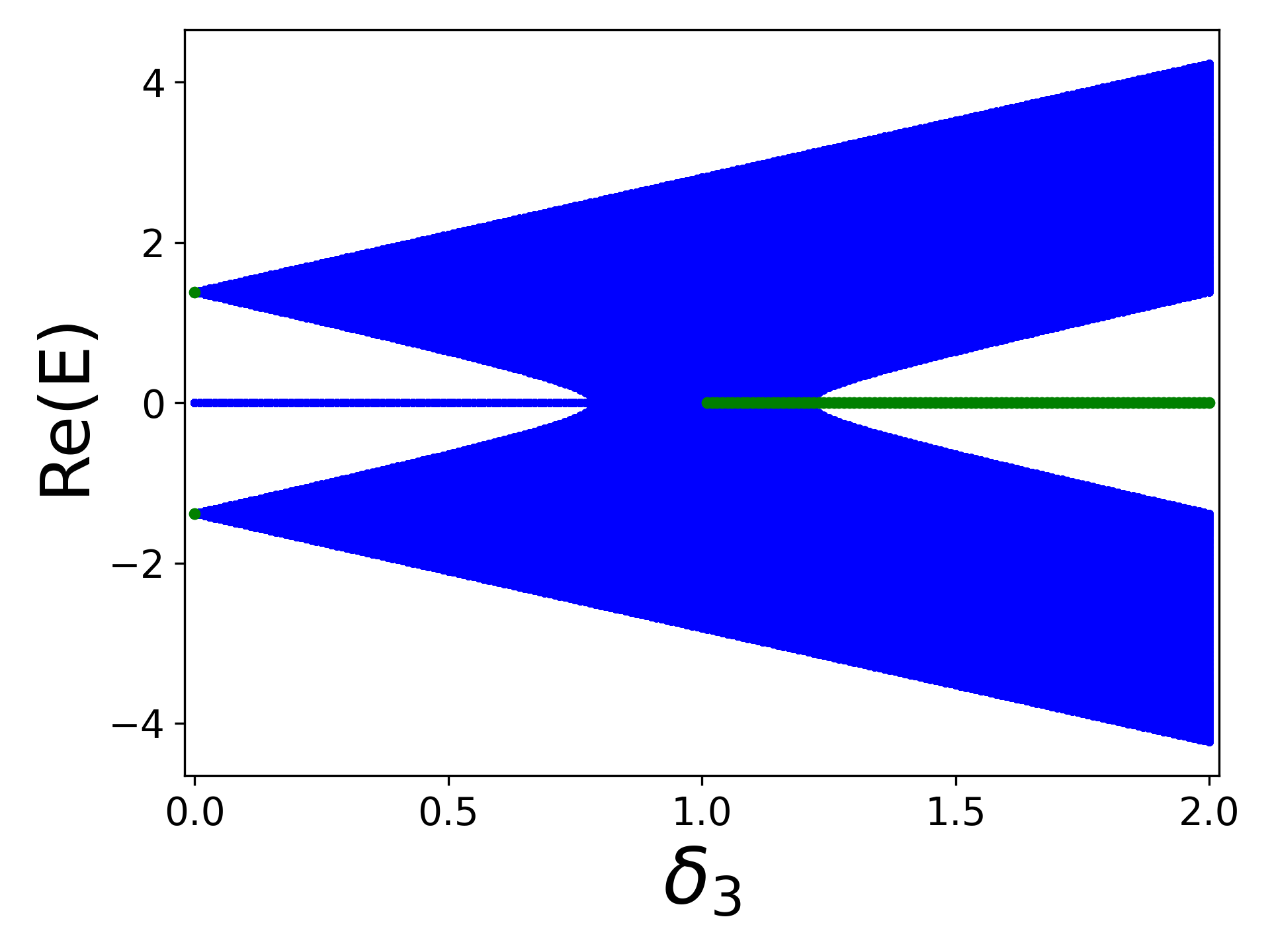}\quad
                \caption{}
                \label{spec_obc_real_1gem2g_chiral}
        \end{subfigure}
        \hfill
        \begin{subfigure}{0.48\columnwidth}
                \centering
                \includegraphics[width = 0.98\linewidth]{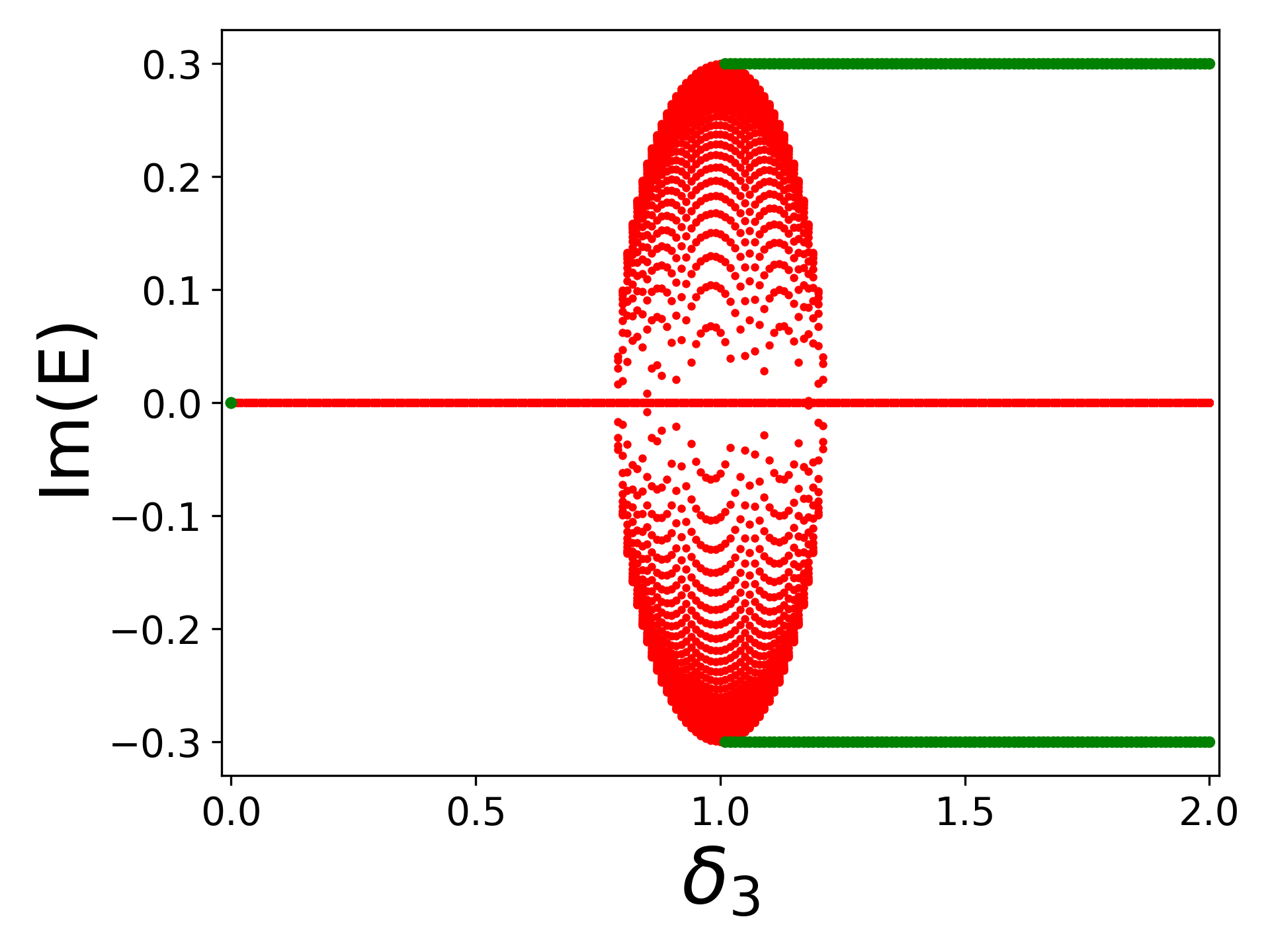}\quad
                \caption{}
                \label{spec_obc_imag_1gem2g_chiral}
        \end{subfigure}
        \caption{Plot of the spectrum under OBC for $\gamma_2 = - \gamma_1$. 
        Parameter values : $\gamma_1 = -\gamma_2 = 0.3$, $\delta_1 = t_1 = 0$, 
	$t_3 = \delta_2 = 1$, $t_2 = \delta_3$.}
        \label{spec_obc_1gem2g_chiral}
\end{figure}

\noindent In this section, we study the edge states corresponding to the parameter regimes given in Sec.IV.C. 
We first consider the pseudo-chiral limit i.e., $\delta_1 = t_1 = 0$, $t_3 = \delta_2$ and $t_2 = \delta_3$. 
The spectrum of the system under the OBC in the pseudo-chiral limit is shown in 
{\bf Fig. \ref{spec_obc_1gem2g_chiral}}. In this limit, 
the system undergoes from trivial topology to non-trivial topology(Non-zero sublattice Zak phase) when $\delta_3$ 
increase and becomes $\delta_3 > \delta_2$. We also observe that edge states appear when $\delta_3 > \delta_2$. 
In this limit, the system also admit CLS corresponding to flatbands. The expressions of
$\psi_{na}$, $\psi_{nb}$ and $\psi_{nc}$ for the CLS located around $n = l$ unit cell, are
\bea
&& \psi_{la} = i\delta_2, \ \psi_{lb} = -i\delta_2, \ \psi_{lc} = i\gamma_1, \nonumber \\
&& \psi_{l+1,a} = i\delta_3, \ \psi_{l+1,b} = -i\delta_3, \ \psi_{l+1,c} = 0,  \nonumber
\eea
and $\psi_{na} = \psi_{nb} = \psi_{nc} = 0$ for $n \neq l, l+1$. This expression is valid for $l=1,2,3, \dots (m-1)$.
Unlike to the PBC, the expression of the CLS localized at the right boundary is different from CLS located at bulk sites.
The expressions of CLS located at $n = m$ unit cell is
\bea
\psi_{ma} = i\delta_2, \ \psi_{mb} = -i\delta_2, \ \psi_{mc} = i\gamma_1 \nonumber
\eea
and probability density of CLS at other sites is zero. So the system in  pseudo-chiral limit admits two types of localized states,
topology protected edgetstate and CLS.

\begin{figure}
\centering
        \begin{subfigure}{0.48\columnwidth}
                \centering
                \includegraphics[width = 0.98\linewidth]{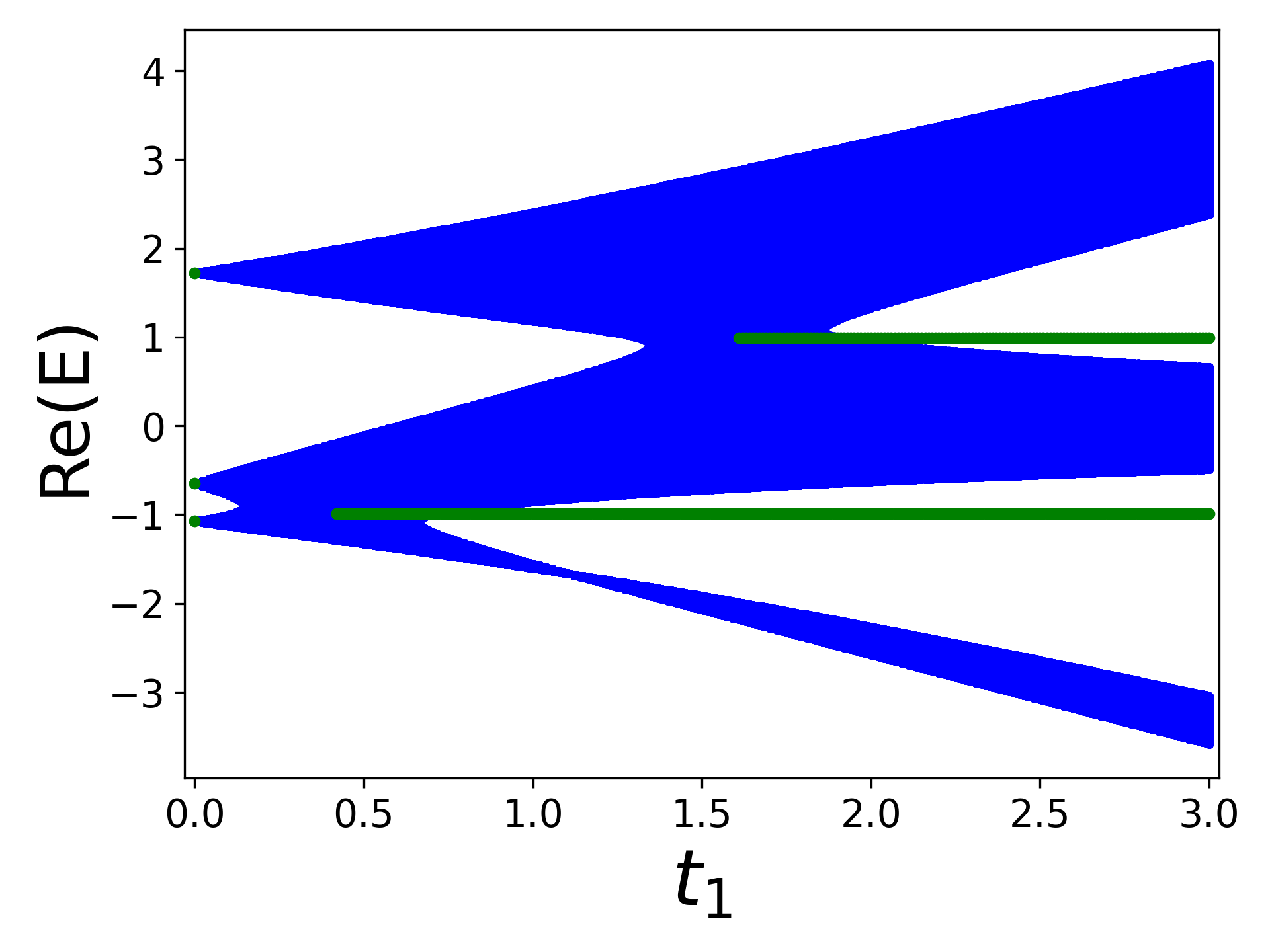}\quad
                \caption{}
                \label{spec_obc_real_1gem2g_1d06}
        \end{subfigure}
        \hfill
        \begin{subfigure}{0.48\columnwidth}
                \centering
                \includegraphics[width = 0.98\linewidth]{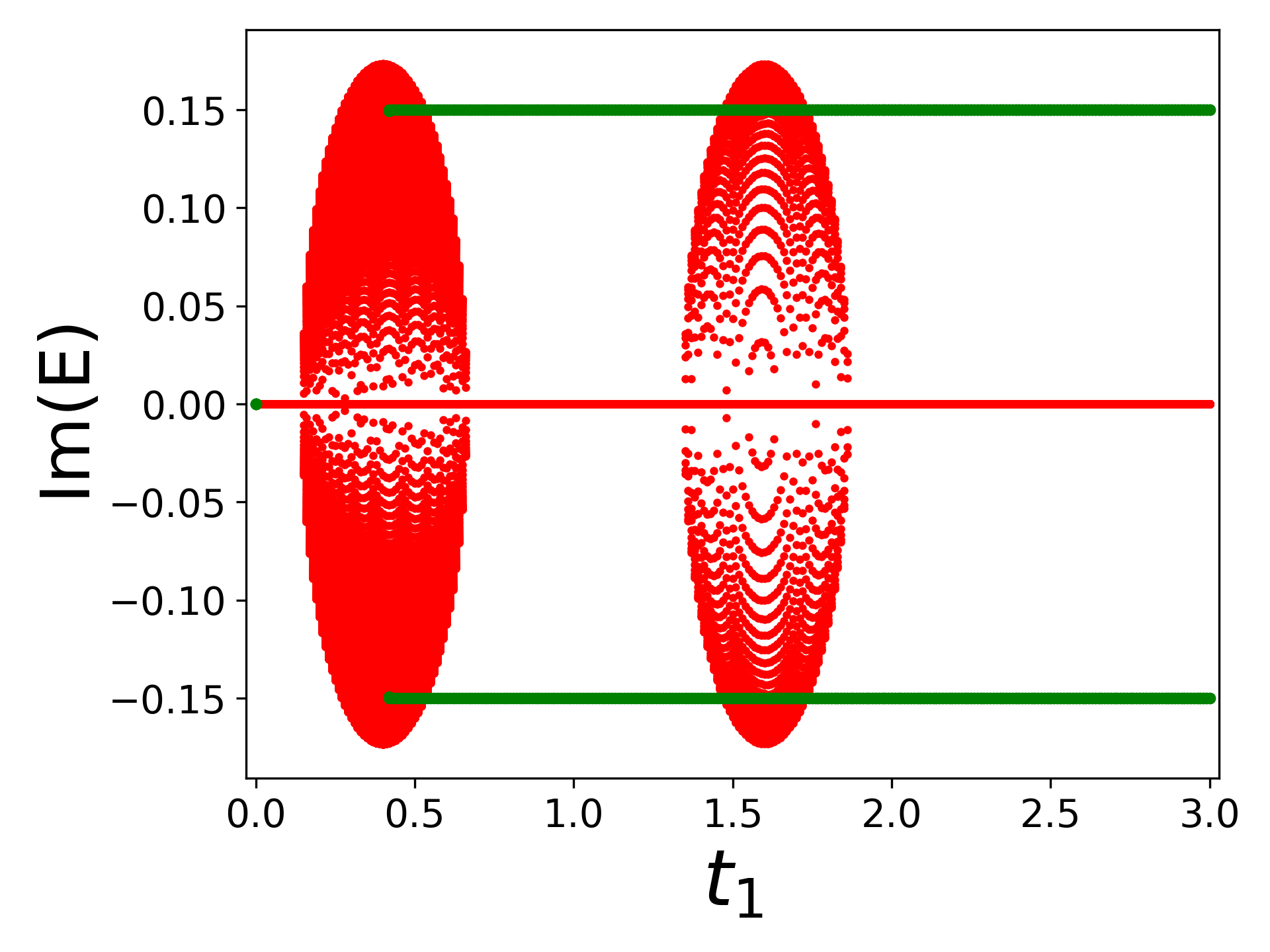}\quad
                \caption{}
                \label{spec_obc_imag_1gem2g_1d06}
        \end{subfigure}
        \caption{Plot of the spectrum under OBC for $\gamma_2 = - \gamma_1$. 
	Parameter values : $\gamma_1 = -\gamma_2 = 0.3$, $\delta_1 = 0.6$, $\delta_3 = t_2 = 0$, $\delta_2 = t_3 = 1$.}
        \label{spec_obc_1gem2g_PT}
\end{figure}

Next we consider the $\mathcal{PT}$-symmetric limit i.e., $t_2 = \delta_3 = 0$, $t_3 = \delta_2$. 
The real and the complex part of the spectrum under the OBC is shown in {\bf Fig. \ref{spec_obc_1gem2g_PT}}.
The green color indicates the energy of the edge states. We observe that the edge state appears when the
system has non-trivial topology under the PBC. This establishes the BBC for the trimer lattice.

We observe that edgestate from non-flat bands appear when the system has non-trivial topology. In case of pseudo-chiral
limit, topological phase transition is shown in terms of sub-lattice Zak phase which is a topological invariant for 
$\mathcal{PT}$-symmetric limits. An interesting phenomena in our model under OBC is that the co-existence of the
two different kinds of the localized states \textemdash{} topology protected edgestates and also CLS. 
In Table~\ref{tab-edge-cls}, we summarize the results obtained under PBC and OBC, highlighting the relevant symmetries
and the existence of CLS and/or edge states across different parameter regimes.

\begin{table*}[ht]
\begin{tabular}{|>{\centering\arraybackslash}m{0.20\textwidth}|
                >{\centering\arraybackslash}m{0.15\textwidth}|
                >{\centering\arraybackslash}m{0.25\textwidth}|
                >{\centering\arraybackslash}m{0.25\textwidth}|}
\hline
	Parameters & Symmetry & Edgestate & CLS \\ \hline
	$\gamma_1 = 0$,$t_1 = \delta_3$,$t_3 = \delta_1$, $\delta_2=t_2 =0$ & Pseudo-chiral symmetry & Topology protected edgestate for $\delta_3 > \delta_1$  & CLS appear in both bulk and boundary sites  \\ \hline
	$\gamma_2 = 0$,$t_2 = t_1$,$\delta_2 = \delta_1$, $\delta_3 = t_3 = 0$ & Pseudo-chiral,$\mathcal{PT}$-symmetry & Topology protected edgestate for $t_1 > \delta_1$ & CLS appear only in the bulk sites not in boundary \\ \hline
	$\gamma_2 = - \gamma_1$, $t_3 = \delta_2$, $t_2 = \delta_3$, $\delta_1 = t_1 = 0$, & Pseudo-chiral symmetry & Topology protected edgestate for $\delta_{3} > \delta_2$ & CLS appear both in the bulk and boundary sites \\ \hline 
\end{tabular}
	\caption{Parameter regimes with both CLS and Edgestates}
\label{tab-edge-cls}
\end{table*}
\section{General Boundary Condition}
\label{gbc-sec}

In this subsection, we investigate the system under GBC from the viewpoint of exact solvability. While tight-binding 
chains and SSH models subject to GBCs have been extensively studied, the corresponding analysis of the trimer-SSH model 
has received comparatively little attention. In this section, we develop an exact treatment of the trimer-SSH model
under GBC in the presence of BLG terms. We emphasize that the formalism and the resulting analytical expressions
remain valid in the limit of vanishing BLG terms which have not been studied earlier.
We choose the eigenvector of the Hamiltonian in Eq. ~(\ref{main-ham}) as, 
\bea
\vert \Psi \rangle = \sum_{n} \left(\psi_{n,a} \vert n,a \rangle + \psi_{n,b} \vert n,b \rangle 
+ \psi_{n,c} \vert n,c \rangle \right)
\eea
The Schr\"odinger equation $H \vert \Psi \rangle = E \vert \Psi \rangle$ gives following set of 
bulk equations, 
\bea
(E - U)\psi_{n} & = & J \psi_{n+1} + J^{\dagger} \psi_{n-1}  
\label{sch-bulk-eqn}
\eea
with $n = 2, 3, \dots , (m-2), (m-1)$ and $\psi_{n} = {\bp \psi_{na} && \psi_{nb} && \psi_{nc}\ep}^{T}$ is a three-component 
column matrix. The expressions of $U$ and $J$ are
\bea
U  =  \bp i\gamma_1 && \delta_1  && t_3  \\ \delta_1  && i\gamma_2 && \delta_2  \\ 
t_3 && \delta_2 && -i(\gamma_1+\gamma_2) \ep, \
J =  \bp 0 && 0 && 0 \\ t_1  && 0 && 0 \\ \delta_3  && t_2  && 0 \ep \nonumber 
\eea
\noindent The matrix $U$ encodes the information about intracell interactions  and onsite imaginary
BLG potentials, whereas $J$ and its adjoint $J^{\dagger}$ keeps tracking of the intercell hopping
strengths. The boundary equations are given as,
\bea
(E - U)\psi_{1} & = & J \psi_{2} + J_{L} \psi_{m} \nonumber \\
(E - U)\psi_{m} & = & J_{R} \psi_{1} + J^{\dagger} \psi_{m-1}
\label{sch-boundary}
\eea
where 
\bea
J_{R}  =  \bp 0 && 0 && 0 \\ t_{1r} && 0 && 0 \\ \delta_{3r} && t_{2r} && 0 \ep \ , \ 
J_{L}  =  \bp 0 && t_{1l} && \delta_{3l} \\ 0 && 0 && t_{2l} \\ 0 && 0 && 0 \ep
\eea
\noindent In general, $J_L$ and $J_R$ are non-hermitian, and it gives rise to  various boundary
conditions under appropriate reductions. Due to the translation symmetry in the bulk Eqs.~(\ref{sch-bulk-eqn}),
the ansatz 
\bea
\psi_{n} = \bp A \\ B \\ D \ep z^{n} \ , \ z \in \mathbb{C} 
\label{ansatz-gbc}
\eea
\noindent reduces it to the equation,
\bea
H_{z} \bp A \\ B \\ D \ep = E \bp A \\ B \\ D \ep
\label{bloch-eqn}
\eea
\noindent where $H_{z}$ is a generalized Bloch matrix
\bea
H_{z} = \bp i\gamma_1 && \delta_1 + \frac{t_1}{z} && t_3 + \frac{\delta_3}{z} \\ 
\delta_1 + t_1 z && i\gamma_2 && \delta_2 + \frac{t_2}{z} \\ 
t_3 + \delta_3 z && \delta_2 + t_2 z && -i(\gamma_1+\gamma_2) \ep \nonumber 
\eea
\noindent The sole purpose of this method is to find an analytic expression of the
complex number $z$ from the boundary conditions. From Eqs.  (\ref{ansatz-gbc}) and (\ref{bloch-eqn}),
the boundary Eq. (\ref{sch-boundary}) reads as,
\bea
J^{\dagger} \psi_{0} = J_{L} \psi_{m} \ , \ J \psi_{m+1} = J_{R} \psi_{1}
\label{sch-boundary1}
\eea
The characteristic polynomial of the Hamiltonian $H_{z}$, subject to the necessary condition
for the reality of the entire spectrum in Eq.~(\ref{real-pcondi}),
is a depressed cubic equation
\bea
\lambda^{3} + p(z) \lambda + q(z) = 0
\eea
\noindent with real coefficients $p(z)$ and $q(z)$, 
\bea
q(z) & = & - \Biggl[ 2\delta_1\delta_2t_3 + 2\delta_1\delta_3t_2 + 2\delta_2\delta_3t_1  \nonumber \\ 
     & + & \left(\delta_1t_3t_2 + \delta_2t_3t_1+\delta_3t_1t_2+\delta_1\delta_2\delta_3\right) 
\left(z + \frac{1}{z} \right) \nonumber \\ 
     & + & \left. t_1t_2t_3 \left(z^2 + \frac{1}{z^2}\right)\right] \nonumber \\ 
p(z) & = & \left[ \gamma_{R}^2 - \sum_{i=1}^{3} \left\{\delta_{i}^2 + t_{i}^{2} 
+  \delta_i t_{i}\left(z + \frac{1}{z}\right) \right\}\right] \nonumber 
\eea
The energy eigenvalues are,
\bea
E_i & = & 2 \sqrt{-\frac{p(z)}{3}} \cos\left[ \Theta(z) + \frac{2\pi(i-1)}{3}\right], i=1, 2, 3 \nonumber \\
\Theta(z) & = & \frac{1}{3} \arccos\left(\frac{3q(z)}{2p(z)} \sqrt{-\frac{3}{p(z)}}\right)
\eea
Although the energy eigenvalues are expressed in terms of $p(z)$ and $q(z)$, $z$ itself is unknown so far.
For a particular value of energy $E$, $z$ satisfies the quartic palindromic polynomial equation, 
\bea
a_0z^{4} + a_1 z^3 + a_2 z^2 + a_1 z + a_0 = 0
\label{z-eqn-quartic}
\eea
where
\bea
a_0 & = & t_1t_2t_3 \nonumber \\
a_1 & = & \left[ E \sum_{i=1}^{3} \delta_{i} t_{i} + \left(\delta_1t_2t_3 + \delta_3 t_1 t_2 
+ \delta_1\delta_2\delta_3\right)\right] \nonumber \\
a_2 & = & -\left[ E^3 + \left(\gamma_R^2 - \sum_{i=1}^{3}\left(\delta_{i}^2 + t_{i}^2\right)\right)
E \right. \nonumber \\
& - & 2\left(\delta_1\delta_2t_3 + \delta_1\delta_3t_2 + \delta_2\delta_3t_1\right)\Biggr] \nonumber  
\eea
\noindent which can always be reduced to a bi-quadratic form. However, the four roots have lengthy 
analytical expressions as shown in Eq.~(\ref{sol-quartic-z-eqn}) in the Appendix VIII.C.
Further analytical study with these expressions of the roots is cumbersome, and an outline
of the general procedure is included in the Appendix VIII.C. 
Note that the roots of the Eq.~(\ref{z-eqn-quartic}) diverges for any $t_{i} = 0, i=1,2,3$. 
The limit $t_{i} = 0$ is singular, and one cannot proceed with this quartic equation (\ref{z-eqn-quartic})
and take the limit at the end. Such cases are to be studied by taking the limits at the very beginning
of the analysis. We analytically study the system under GBC in some limiting conditions
where the Eq.~(\ref{z-eqn-quartic}) reduces to a quadratic equations.
\subsection{$t_2 = \delta_2 = 0$, $t_1 = \delta_3$, $t_3 = \delta_1$, $\gamma_1 = 0$}
\label{gbc-limit1}

In the limiting conditions of $\delta_2 = t_2 = 0$, $t_1 = \delta_3$, $t_3 = \delta_1$, and $\gamma_1 = 0$,
the system under the PBC shows pseudo-chiral symmetry and was discussed in Sec.\ref{pbc_1g0}. In this subsection,
we study the system under GBC with the same limiting conditions of the parameters, and the boundary parameters 
$t_{1l} = \delta_{3l}$, $t_{2l}=t_{2r}=0$, $t_{1r}=\delta_{3r}$. In general, such a choice of the parameters corresponds
to non-hermitian boundary conditions, which can be further specialized to  OBC($t_{1l}=0=t_{1r}$), 
AHBC($t_{1l}=-t_{1r}$, PBC $t_{1l}=t_{1r}=t_1$, etc.
The energy in this limit is determined as,
\bea
E_{0} = 0, \
E_{\pm}  =  \pm \sqrt{2\left(\delta_1 + \delta_3 z\right) \left(\delta_1 + \frac{\delta_3}{z}\right)
- \gamma_{2}^{2}}\nonumber 
\eea
\noindent and Eq.~(\ref{bloch-eqn}), that determines the relations among $A$, $B$ and $D$, reduces to 
\bea
&& EA  =  \left(\delta_1 + \frac{\delta_3}{z}\right) \left(B + D\right) \nonumber \\
&& \left(E - i\gamma_2 \right)B	= \left(\delta_1 + \delta_3 z \right) A \nonumber \\
&& \left(E + i\gamma_2 \right)D = \left(\delta_1 + \delta_3 z \right) A 
\label{bloch-eqn-limit1}
\eea
\noindent For a fixed energy $E_{\pm}$, $z$ is a solution the equation,
\bea
z^{2} + \frac{1}{2\delta_1\delta_3} \left(2\delta_1^2 + 2\delta_3^2 - \gamma_2^2 - E^2 \right) z + 1 = 0
\eea
\noindent The two roots of the equation $z_1$ and $z_2$ satisfy $z_1 z_2 = 1$, and can be parameterized
in terms of a parameter $\theta$ as $z_1 = e^{i\theta}, z_2 = e^{-i\theta}$. The expressions of the
eigenvalues are,
\bea
E_{\pm} & = & \pm \sqrt{2\left(\delta_1^2 + \delta_3^2 + 2\delta_1\delta_3 \cos\theta\right) - \gamma_2^2} \nonumber 
\eea
The general solution of the Schr\"odinger equation may be considered as
\bea
\psi_{n} = \bp c_1A_1 z_1^n + c_2 A_2 z_2^n \\ c_1B_1z_1^n + c_2B_2 z_2^n \\ c_1D_1z_1^n + c_2D_2z_2^n \ep
\eea
\noindent which must satisfy the boundary Eq.~(\ref{sch-boundary1}). 
The boundary Eq.~(\ref{sch-boundary1}) reduces to,
\bea
&&\left(\delta_3 - \delta_{3l} z_1^{m}\right) c_1\left(B_1 + D_1\right) + \left(\delta_3 - \delta_{3l} z_2^{m} \right) c_2\left(B_2 + D_2\right) = 0 \nonumber \\
&&\left(\delta_3 z_1^{m} - \delta_{3r} \right) c_1A_1z_1 + \left(\delta_3 z_{2}^{m} - \delta_{3r} \right) c_2A_2z_2 = 0 \nonumber
\eea
which can be expressed in matrix form as
\bea
H_{B} \bp c_1 \\ c_2 \ep = 0
\label{sch-boundary-limit1}
\eea
where the matrix $H_{B}$ is,
\bea
\bp \left(\delta_3 - \delta_{3l} z_{1}^{m}\right) \left(B_1 + D_1\right) && \left(\delta_3 - \delta_{3l} z_{2}^{m}\right) \left(B_2 + D_2\right) \\
\left(\delta_3 z_1^{m} - \delta_{3r} \right) A_1z_1 && \left(\delta_3 z_2^{m} - \delta_{3r} \right) A_2z_2 \ep \nonumber
\eea
\noindent The condition $Det(H_B) = 0$ ensures non-trivial solution for $c_1$ and $c_2$. 
From Eq.~(\ref{bloch-eqn-limit1}), we can write 
\bea
\frac{B_{i} + D_{i}}{A_{i} z_{i}} = \frac{E}{\delta_1 z_{i} + \delta_3},  \ i=1, 2 \nonumber
\eea
Using the the ratio of $\left(B + D\right)/A$, the condition $Det(H_B) = 0$ leads to,
\bea
\left(z_2^{m+1} -  z_1^{m+1}\right) & + & \frac{\delta_3^2 - \delta_{3l}\delta_{3r}}{\delta_1\delta_3} \left(z_2^m - z_1^m\right)
- \frac{\delta_{3l}\delta_{3r}}{\delta_3^2} \left(z_2^{m-1} \right. \nonumber \\
                                    & - & \left. z_1^{m-1}\right) - \frac{\delta_{3l} + \delta_{3r}}{\delta_3} \left(z_2 - z_1\right) = 0 \nonumber
\eea
The boundary conditions can be rewritten in terms of $\theta$ :
\bea
\hspace*{-2mm}\sin(m+1)\theta +  P \sin m\theta -  Q \sin(m-1)\theta =  R \sin\theta 
\label{boundary-con-limit1}
\eea
where $P = \frac{\delta_3^2 - \delta_{3r}\delta_{3l}}{\delta_1\delta_3}$, $Q =  \frac{\delta_{3l} \delta_{3r}}{\delta_3^2}
 $, $R = \frac{\delta_{3l}+\delta_{3r}}{\delta_3}$. The exact solutions of this equations for specific values of
$P, Q$ and $R$ are discussed in Appendix VIII.D. 
The solution $\theta$ gives the expression of the energies and wavefunctions. 
From the boundary Eq.~(\ref{sch-boundary-limit1}) and bulk Eq.~(\ref{bloch-eqn-limit1}), we can write,
\bea
c_2A_2     & = & -\frac{\delta_{3r}z_1 - \delta_3 z_1^{m-1}}{\delta_{3r} z_2 - \delta_3 z_2^{m-1}} c_1A_1 \nonumber \\
c_{i}B_{i} & = & \frac{\delta_1 +\delta_3 z_{i}}{E - i\gamma_2} c_{i}A_{i} \ ; \ i = 1,2\nonumber \\
c_{i}D_{i} & = & \frac{\delta_1 +\delta_3 z_{i}}{E + i\gamma_2} c_{i}A_{i} \ ; \ i = 1,2
\label{rel-for-wf-limit1}
\eea
We obtain the expression of the $\psi_{na}$,$\psi_{nb}$,$\psi_{nc}$ for the wavevector by using the
relations~(\ref{rel-for-wf-limit1}):
\bea
\psi_{na} & = & c_1A_1 z_1^{n} + c_2A_2 z_2^{n} \nonumber \\
	  & = & \frac{2i c_1A_1 e^{i\theta}}{\delta_3 e^{-i m\theta} - \delta_{3r}} \left[ \sin\left(n-m-1\right)\theta \right. \nonumber \\ 
	  & - & \left. \delta_{3r} \sin(n-1)\theta \right] \nonumber \\
\psi_{nb} & = & c_1B_1 z_1^{n} + c_2B_2 z_2^{n} \nonumber \\
	  & = & \frac{2ic_1A_1 e^{i\theta}}{\left(E - i\gamma_2\right)\left(\delta_3 e^{-i m\theta} - \delta_{3r}\right)} \left[ \delta_1 \delta_3\sin\left(n-m-1\right)\theta \right. \nonumber \\
	  & + & \left. \delta_3^2 \sin\left(n-m\right)\theta - \delta_1\delta_{3r} \sin(n-1)\theta \right. \nonumber \\
	  & - & \left. \delta_3 \delta_{3r} \sin(n\theta) \right] \nonumber \\
\psi_{nc} & = & c_1D_1 z_1^{n} + c_2D_2 z_2^{n} \nonumber \\
          & = & \frac{2ic_1A_1 e^{i\theta}}{\left(E + i\gamma_2\right)\left(\delta_3 e^{-i m\theta} - \delta_{3r}\right)} \left[ \delta_1 \delta_3\sin\left(n-m-1\right)\theta \right. \nonumber \\
          & + & \left. \delta_3^2 \sin\left(n-m\right)\theta - \delta_1\delta_{3r} \sin(n-1)\theta \right. \nonumber \\
          & - & \left. \delta_3 \delta_{3r} \sin(n\theta) \right] 
\label{final-exp-wf}
\eea
Equations analogous to Eq.~(\ref{boundary-con-limit1}) emerge as boundary condition under other limiting cases, as 
discussed in Secs.~\ref{gbc-limit2} and \ref{gbc-limit3}. In the following subsections, we present 
some solvable limits of Eq.~(\ref{boundary-con-limit1}). 

\subsubsection{OBC}
\label{gbc-limit1-obc}
We have numerically analysed the appearance of the edge states in Sec.~\ref{obc-sec} for systems having
${\cal{PT}}$ symmetry and/or pseudo-chiral symmetry. It is instructive, however, to complement this numerical 
analysis with an analytical treatment under OBC. Although deriving the edge-state solutions for the 
$\mathcal{PT}$-symmetric case under OBCs is analytically rather involved, closed-form expressions for the 
edgestate energies and wave functions can be obtained in the pseudo-chiral-symmetric limits. In this subsection, 
our aim is to show the analytical expression of the edgestates for the parameter limits considered in this 
subsection. The OBC correspond to the special case 
$\delta_{3l} = \delta_{3r} = 0$. In case of OBC, the boundary condition~(\ref{boundary-con-limit1}) reduces to, 
\bea
\sin (m+1)\theta + \alpha \sin m\theta = 0
\label{boundary-obc-limit1}
\eea
where $\alpha = \frac{\delta_3}{\delta_1}$. This Eq. is similar to the Eq.~(30) in the article~\cite{Supriyo2025PRB}. 
The Eq.~(\ref{boundary-obc-limit1}) admits $m$ real solutions for $\alpha < \alpha_{c} = 1 + \frac{1}{m}$ and there 
are $m-1$ real solutions and one complex solution for $\alpha > \alpha_{c}$. The complex $\theta$ is associated with 
the edgestates. The critical value of $\alpha_{c}$ is obtained from the relation $f'_{1} (\pi) = f'_{2}(\pi)$, where 
$f_{1}(\theta) = \sin(m + 1)\theta$ , $f_{2}(\theta) = -\alpha_{c} \sin(m\theta)$, and 
$f'_{i}(\pi) = \frac{df_{i}}{d\theta}\vert_{\theta = \pi}$. In case of complex $\theta = \pi + i\zeta$, 
the Eq.~(\ref{boundary-obc-limit1}) becomes, 
\bea
2m\zeta = \ln\left(\frac{e^{-\zeta} - \alpha}{e^{\zeta} - \alpha}\right)
\eea
which has solution only when $e^{\zeta} \approx \alpha$. We set  $e^{\zeta} = \alpha + \mu_{\zeta}\left(\mu_{\zeta} \rightarrow 0\right)$ 
and the expression of $\mu_{\zeta}$ is obtained as,
\bea
\mu_{\zeta} = \frac{\frac{1}{\alpha} - \alpha}{\alpha^{2m} + \frac{1}{\alpha^2}}
\eea
by considering the approximation $\left(\alpha + \mu_{\zeta}\right)^{2m} \approx \alpha^{2m}$ and 
$\frac{1}{\alpha + \mu_{\zeta}} \approx \frac{1}{\alpha} - \frac{\mu_{\zeta}}{\alpha^2}$. The expression for 
$\zeta$ is 
\bea
\zeta = \ln \left[ \alpha \left( 1 + \frac{1 - \alpha^2}{1 + \alpha^{2m+2}}\right)\right] 
\label{expression-zeta-limit1}
\eea
The energy of the edge states are,
\bea
E_{edge} = \pm \sqrt{2\left(\delta_1^2 + \delta_3^2 - 2\delta_1\delta_3\cosh\zeta\right) - \gamma_2^2}
\eea
In the thermodynamics limit, $e^{\zeta} = \alpha$ and $E_{edge} = \pm i\gamma_2$. 
We obtain the expression of the $\psi_{na}$,$\psi_{nb}$,$\psi_{nc}$ for the wavevector under OBC by using the 
relations~(\ref{rel-for-wf-limit1}):
\bea
\psi_{na} & = & 2i c_1A_1 e^{i\left(m+1\right)\theta} \sin\left(n-m-1\right)\theta \nonumber \\
\psi_{nb} & = & \frac{2ic_1A_1}{E - i\gamma_2} e^{i\left(m+1\right)\theta} \left[ \delta_1 \sin\left(n-m-1\right)\theta \right. \nonumber \\
	  & + & \left. \delta_3 \sin\left(n-m\right)\theta \right] \nonumber \\
\psi_{nc} & = & \frac{2ic_1A_1}{E + i\gamma_2} e^{i\left(m+1\right)\theta} \left[ \delta_1 \sin\left(n-m-1\right)\theta \right. \nonumber \\ 
	  & + & \left. \delta_3 \sin\left(n-m\right)\theta \right] 
\eea
The complex $\theta = \pi + i \zeta$ is associated with the edgestate. 
The expression of the $\psi_{na}$,$\psi_{nb}$,$\psi_{nc}$ for edge state is,
\bea
\psi_{na} & = & 2c_1A_1 (-1)^{n+1} e^{-(m+1)\zeta} \sinh(n-m-1)\zeta \nonumber \\
\psi_{nb} & = & \frac{2c_1A_1}{E - i\gamma_2} \ (-1)^{n} e^{-(m+1)\zeta} \left[\delta_3\sinh(n-m)\zeta \right. \nonumber \\ 
          & - & \left. \delta_1 \sinh(n-m-1)\zeta\right]\nonumber \\
\psi_{nc} & = & \frac{2c_1A_1}{E + i\gamma_2} \ (-1)^{n} e^{-(m+1)\zeta} \left[\delta_3\sinh(n-m)\zeta \right. \nonumber \\
          & - & \left. \delta_1 \sinh(n-m-1)\zeta\right]
\eea
where $\zeta$ is given by Eq.~(\ref{expression-zeta-limit1}).
\subsubsection{$P = 0$,\ $Q = 1$}

\begin{figure}
\centering
        \begin{subfigure}{0.48\columnwidth}
                \centering
                \includegraphics[width = 0.98\linewidth]{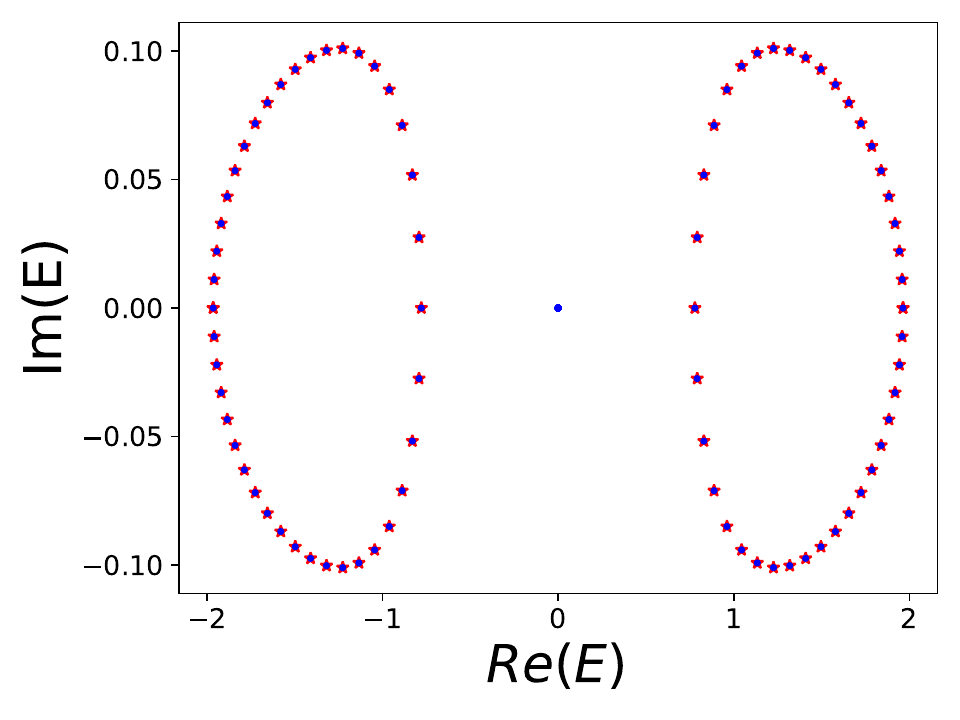}\quad
                \caption{}
                \label{spec-top-gbc1}
        \end{subfigure}
        \hfill
        \begin{subfigure}{0.48\columnwidth}
                \centering
                \includegraphics[width = 0.98\linewidth]{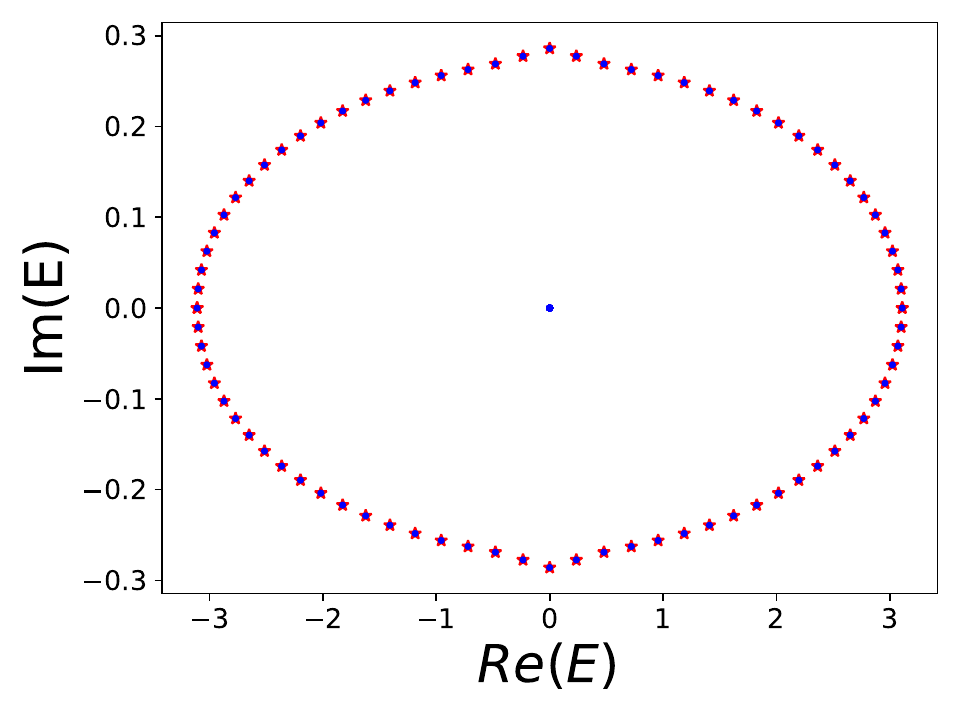}\quad
                \caption{}
                \label{spec_top_gbc2}
        \end{subfigure}
	\caption{Spectrum in the complex plane in the limit of $\delta_{3l}\delta_{3r} = \delta_{3}^2$. The blue and
        red colors indicate the numerical and theoretical values respectively. We have considered the principle branch
        as $[-\pi, \pi]$. So the theoretical points with $Re(E) < 0$ comes from $E_{-}$ and theoretical points with
        $Re(E) > 0$ comes from $E_{+}$. The state in the origin denotes the flatbands. Parameter values :
        $m=40$,$\delta_2 = t_2 = \gamma_1 = 0$, $t_3 = \delta_1 = 1$, $t_{1l} = \delta_{3l} = 1000\delta_3$,
	$t_{1r} = \delta_{3r} = 0.001\delta_{3}$, Fig (a)~ $t_1 = \delta_3 = 0.4$, Fig (b)~ $t_1 = \delta_3 = 1.2$}
        \label{spec-top-gbc}
\end{figure}

The limit $P =0$, \ $Q = 1$ is obtained when $\delta_{3r}\delta_{3l} = \delta_{3}^2$. 
The PBC and APBC falls under this limit. In general, this choice corresponds to non-hermitian boundary
condition. We which reduces to PBC for $\delta_{3r} = \delta_{3l} = \delta_3$, and to APBC for 
$\delta_{3r} = \delta_{3l} = -\delta_3$. If we allow boundary parameters $\delta_{3l}$ and $\delta_{3r}$
to be complex, the TBC can be implemented for various choices of the phases $\eta_1$ and $\eta_2$. We proceed
with real $\delta_{3l}$ and $\delta_{3r}$, and
define $\nu = sgn\left(\frac{\delta_{3r}}{\delta_3}\right) = sgn\left(\frac{\delta_{3l}}{\delta_3}\right)$
so that $\nu = \pm 1$ depending on the values of $\delta_{3r}, \delta_{3l}, \delta_3$. In the limit
$\delta_{3r} \delta_{3l} = \delta_3^2$, the Eq.~(\ref{boundary-con-limit1}) reduces to,
\bea
\Big[ \cos(m\theta) - \nu cosh(u) \Big] \sin\theta = 0
\eea
where $u = \ln\vert\frac{\delta_{3l}}{\delta_3}\vert$
The solution is
\bea
\theta = \frac{2s\pi}{m} \pm \frac{\pi}{2m}(1-\nu) \mp \frac{i}{m} ln\vert\frac{\delta_{3l}}{\delta_3}\vert
\eea
where $s =0, \dots , (m-1)$. The solution corresponding to $\sin\theta = 0$ is not considered, since they lead to trivial 
solution $\Psi = 0$. The phase $\theta$ can be expressed as $\theta = \theta_{R} \mp i\theta_{I}$, where the real part 
$\theta_{R} = \frac{2s\pi}{m} \pm \frac{\pi}{2m}(1-\nu)$ and imaginary part 
$\theta_{I} = \frac{1}{m} \vert \frac{\delta_{3l}}{\delta_3}\vert$. 
Using the expression of $\theta$ into Eq.~(\ref{final-exp-wf}), the expression of the 
wavefunction corresponding to $\delta_{3l}\delta_{3r} = \delta_3^2$ is obtained. 
Due to the factor $\frac{1}{m}$, the imaginary part $\theta_I$ becomes negligible in the large-system-size limit for finite 
$|\frac{\delta_{3l}}{\delta_3}|$. The imaginary contribution becomes appreciable only when 
$\vert\frac{\delta_{3l}}{\delta_3} \vert \approx e^{m}$ i.e., the large asymmetry in the boundary. Therefore, in 
the thermodynamic limit ($m \to \infty$ but $s/m$ finite), a finite imaginary contribution can survive only in the 
limiting regime $\delta_{3l} \to \infty$, $\delta_{3r} \to 0$ while their product $\delta_{3l}\delta_{3r} = \delta_3^2$ 
remains finite. The expression of the energy, excluding $E=0$,  is 
\bea
E_{\pm} = \pm \sqrt{r} \ e^{-\frac{i\phi_r}{2}} \nonumber 
\eea
where,  
\bea
r          & = & \vert  2\left(\delta_1^2 + \delta_3^2\right) - \gamma_2^2 + 4 \delta_1\delta_3 \left(\cos\theta_{R} \cosh\theta_{I} \right. \nonumber \\
	   & - & \left. i \sin\theta_R \sinh\theta_I\right) \vert \nonumber \\
\phi_{r}   & = & \arctan\left(\frac{4\delta_1\delta_3\sin\theta_R\sinh\theta_I}{2\left(\delta_1^2 + \delta_3^2\right) - \gamma_2^2 + 4\delta_1\delta_3\cos\theta_R\cosh\theta_I}\right) \nonumber 
\eea
The spectrum will form closed loops in the complex plane, and it converges from two loops to one loop 
at the transition point $4\delta_1\delta_3 \cos\theta_I = 2\left(\delta_1^2 + \delta_3^2\right)-\gamma_2^2$. 
The spectral winding number $\nu(E_r)$ around a reference point $E_r$ is defined for PBC as,
\bea
\nu(E_r) & = & \frac{1}{2i\pi} \oint dk \frac{d}{dk} \ln \det \left[H(k) - E_r\right] 
\eea
In our case, we can replace $k$ with $\theta_{R}$. The spectral winding number can be rewritten as
\bea
\nu(E_r) = \frac{1}{2\pi} \oint \frac{d}{d\theta_R} arg\left(r e^{-i\phi_r} - E_{r}^2\right) d\theta_{R} \nonumber
\eea
The contribution from the flat bands in spectral winding number will be zero. 
The complex quantity $r e^{-i\phi_r}- E_r^2$ will form an ellipse with semi-major axis $4\delta_1\delta_3\cos\theta_I$ 
,semi-minor axis $4\delta_1\delta_3\sin\theta_I$. In case of spectral topology, $E_r$ can be chosen any point 
with only restriction $E_r$ should not be any eigenvalue. We can choose $E_{r}$ appropriately so that the loop 
includes the  origin and the $arg \left(r e^{-i\phi_r}- E_r^2\right)$ will acquire phase $2\pi$ as $\theta_R$ goes 
from zero to $2\pi$. Hence, we will get non-zero winding number $\nu = 1$. Two representative plots of the spectrum for 
particular values of parameters is shown in {\bf Fig. \ref{spec-top-gbc}}. Due to complex $\theta$, all the
eigenstates are localized at one end of the system which is known as NHSE\cite{Kawabata2019PRX}.
The co-existence of spectral topology and NHSE establishes the non-Hermitian BBC\cite{Kawabata2019PRX}.
\subsubsection{$Q = -1$,\ $R = 0$}

The limit $Q = -1$,\ $R = 0$ can be realized for $\delta_3 = \delta_{3l} = - \delta_{3r}$ that corresponds 
to AHBC. In this limit, the Eq.~(\ref{boundary-con-limit1}) 
reduces to, 
\bea
\sin m\theta \left[ \cos\theta + P  \right] = 0 
\eea
where $P = \frac{2\delta_3}{\delta_1}$. The exact solutions are
\bea 
\theta_{m} = \pm \arccos\left(-\frac{P}{2}\right), \ \theta_{s} = \frac{s\pi}{m} \ ; \ s= 1, 2, \dots, (m-1) \nonumber 
\eea
where $\theta_{m}$ is complex for $|P| > 2$, and corresponds to edgestates. The system has edgestate 
when $|P| > 2$ i.e., $\delta_{3} > \delta_{1}$. 

\subsubsection{$Q = 0$, \ $R = 0$, \ $P = \pm 1$}

The limit $Q = 0$, \ $R = 0$, \ $P = \pm 1$ can be realized for $\delta_{3r} = \delta_{3l} = 0$, $\delta_{3} =
\pm \delta_{1}$. It can also be achieved for non-vanishing $\delta_{3r},  \delta_{3l}$ by taking the approximation
$|\delta_{3}| >> \delta_{3l}$, $\delta_{3r} = - \delta_{3l}$, $\delta_3 = \pm \delta_1$. In this limit, the solution
of Eq.~(\ref{boundary-con-limit1}),
\bea
P = 1 : \ \theta = \frac{2s\pi}{2m + 1} \ ; \ P = -1 : \ \theta = \frac{(2s-1)\pi}{2m + 1} \nonumber 
\eea
\noindent There are no edge states for this case.

\subsection{$\delta_3 = t_3 = 0, \delta_1 = \delta_2, t_1 = t_2, \gamma_2 = 0$} 
\label{gbc-limit2}
We have studied the system under GBC in the limit of $\delta_3 = t_3 = 0$, $\delta_1 = \delta_2$, $t_1 = t_2$, 
$\gamma_2 = 0$. Additionally, we have also considered $\delta_{3l} = \delta_{3r} = 0$, $t_{1l} = t_{2l}$,$t_{2r} = t_{1r}$. 
The method, described in Sec.~(\ref{gbc-limit1}), is also applicable here. From the boundary equation, we get the following
four equations,
\bea
&& \left(t_1 - t_{1l} z_1^{m} \right) c_1 B_1 + \left(t_1 - t_{1l} z_{2}^{m} \right) c_2B_2  =  0 \nonumber \\
&& \left(t_1 - t_{1l} z_1^{m} \right) c_1 D_1 + \left(t_1 - t_{1l} z_{2}^{m} \right) c_2D_2  =  0 \nonumber \\
&& \left(t_1z_1^{m} - t_{1r} \right) c_1 A_1z_1 + \left(t_1 z_2^m - t_{1r} \right) c_2A_2z_2  =  0 \nonumber \\
&& \left(t_1z_1^{m} - t_{1r} \right) c_1 B_1z_1 + \left(t_1 z_2^{m} - t_{1r}\right) c_2B_2z_2  =  0
\eea
These equations can be reduced to two equations as,
\bea
\frac{c_1B_1}{c_2B_2} = \frac{c_1D_1}{c_2D_2} = - \frac{t_1 - t_{1l}z_{2}^m}{t_1 - t_{1l}z_1^m} \nonumber \\
\frac{c_1A_1z_1}{c_2A_2z_2} = \frac{c_1B_1z_1}{c_2B_2z_2} = - \frac{t_1z_2^m - t_{1r}}{t_1z_1^m - t_{1r}} \nonumber 
\eea
Finally, the boundary condition is obtained as,
\bea
\sin(m+1)\theta & + & \frac{t_1^2 - t_{1r}t_{1l}}{\delta_1 t_1} \sin(m\theta) - \frac{t_{1l}t_{1r}}{t_1^2} \sin(m+1)\theta  \nonumber \\
                & - & \frac{t_{1r} + t_{1l}}{t_1} \sin\theta = 0
\eea
This final boundary equation is solvable for $t_{1r}t_{1l} = t_1^2$ and edgestate under OBC as shown in
subsection.~\ref{gbc-limit1}. The general solutions of the $\theta$-equation are discussed in Appendix VIII.D
\subsection{$t_1 = \delta_1 = 0$,$t_2 = \delta_3$,$t_3 = \delta_2$,$\gamma_2 = -\gamma_1$}
\label{gbc-limit3}

In the limiting conditions of $\delta_1 = t_1 = 0$, $t_2 = \delta_3$,$t_3 = \delta_2$, and $\gamma_2 = - \gamma_1$, 
the system under PBC shows pseudo-chiral symmetry and was discussed in Sec.\ref{pbc_g2emg1}. In this subsection, we 
study the system under GBC with same limiting conditions of the parameters. Additionally, we have considered $t_{1l} = t_{1r} = 0$, 
$t_{2l} = \delta_{3l}$,$t_{3r} = \delta_{3r}$. Employing the similar analysis as shown in subsection~\ref{gbc-limit1}, the final
boundary condition is obtained as : 
\bea
\sin(m+1)\theta & + & \frac{\delta_3^2 - \delta_{3r}\delta_{3l}}{\delta_2\delta_3} \sin m\theta 
 -  \frac{\delta_{3l} \delta_{3r}}{\delta_3^2} \sin(m-1)\theta \nonumber \\ 
                & - & \frac{\delta_{3l}+\delta_{3r}}{\delta_3} \sin\theta = 0
\eea
This final boundary equation is solvable for $\delta_{3r}\delta_{3l} = \delta_3^2$ and edgestate under OBC as shown in subsection.~\ref{gbc-limit1}.
\subsection{$t_1=t_2=t_3 = 0$}
\label{gbc-limit4}
The Hamiltonian corresponds to the standard trimer lattice for $t_1=t_2=t_3 = 0$ and 
$t_{1l} = t_{2l} = t_{1r} = t_{2r}$. In this limit, the final boundary condition is obtained 
as, 
\bea
\sin(m+1)\theta & = & \frac{\delta_{3l}\delta_{3r}-\delta_3^2}{\delta_1\delta_2\delta_3}
\left(E - i\gamma_2\right) \sin(m\theta) \nonumber \\
                & + & \frac{\delta_{3l}\delta_{3r}}{\delta_3^2} \sin(m-1)\theta \nonumber \\
                & + & \frac{\delta_{3l} + \delta_{3r}}{\delta_3} \sin\theta
\eea
The equation is exactly solvable in limiting conditions $\delta_{3l}\delta_{3r} = \delta_3^2$ as discussed in 
subsection~\ref{gbc-limit1}. 
\section{Summary and Discussion}

In this work, we have investigated a generalized trimer lattice with BLG and NNN interactions. We have considered
boundary terms which correspond to GBC \textemdash the PBC, OBC, APBC and AHBC appearing as special cases.
We first analyze the system under PBC. Analytical expressions for the energy spectrum are derived,
and the parameter regimes supporting an entirely real spectrum are identified. We further determine
the parameter regions that host a zero-energy flat band. It has been observed that the presence of
NNN interaction is essential for the existence of flat bands. This is to be compared with the SSH
dimer chain where NNN interaction does not allow any flat band.

The symmetries of the system are examined in detail, and the parameter regimes exhibiting $\mathcal{PT}$ symmetry
and/or pseudo-chiral symmetry are identified. In the pseudo-chiral limit, one of the energy bands becomes completely
dispersionless, giving rise to a zero-energy flat band. We derive the corresponding CLS analytically. 

The topological properties of the system are then investigated in both the $\mathcal{PT}$-symmetric and pseudo-chiral limits.
In the $\mathcal{PT}$-symmetric regime, the Zak phase serves as the relevant topological invariant. We compute the Zak phase
numerically and demonstrate the occurrence of TPT. In contrast, the conventional Zak phase becomes inadequate in the
pseudo-chiral limit, where the sub-lattice Zak phase emerges as the appropriate topological invariant. We obtain
analytical expressions for the sub-lattice Zak phase and identify the corresponding topological phase transitions analytically.

We subsequently study the lattice under OBC. The existence of edge states is established through analyses of both
the energy spectrum and the corresponding eigenstates. In the pseudo-chiral parameter regime identified under PBC,
edge states appear whenever the sub-lattice Zak phase assumes a nontrivial value, thereby establishing the bulk-boundary
correspondence in terms of the sub-lattice Zak phase. Similarly, in the $\mathcal{PT}$-symmetric regime, edge states emerge
in the lower band gap when the Zak phase of the lowest band is nontrivial. Edge states also appear in the upper band gap
when the sum of the Zak phases of the lowest and middle bands is nontrivial. Thus, the bulk-boundary correspondence in
the $\mathcal{PT}$-symmetric regime is established through the Zak phase. We also observe that both CLS and edge states
may co-exist at the boundary for the case of systems with pseudo-chiral symmetry. However, an additional ${\cal{PT}}$
symmetry destroys the CLS at the boundary.

Finally, we investigate the system under GBC from the perspective of exact solvability. The trimer SSH model with or
without BLG terms has not been studied earlier under GBC. Using the GBC formalism, we derive
analytical expressions for both the edge-state energies and their corresponding wave functions under OBC. We also obtain
the exact analytical solution of the system under APBC and AHBC. When the product of the nonreciprocal hopping amplitudes
equals the square of the bulk hopping amplitude, the eigenenergy spectrum forms a closed loop in the complex-energy plane
once the boundary-hopping asymmetry becomes comparable to the system-size exponential. The topological phase transition
can be characterized by the corresponding spectral winding numbers. Furthermore, we demonstrate that the system exhibits 
the NHSE. We also show that the system under AHBC admits a pair of edgestate within appropriate parameter ranges. The
existence of NHSE for reciprocal boundary terms and highly asymmetric boundary terms may have significant implications
for future experimental realizations.
\section{Acknowledgments}
SG acknowledges the Department of Physics, Visva-Bharati in India for providing the necessary facilities and support during the 
initial stages of this work. SG also acknowledges Indian Institute of Technology-Kanpur in India for partial financial 
support  through Institute Post-Doctoral Fellowship(PF. No. PDF604).
\section{Appendix}

\subsection{Symmetry of the Bloch Hamiltonian}
\label{symmetry}

The most general parity operator in three dimensions has the form
\bea
{\cal{P}}=\bp 1-2 p^2 &- 2 p q& -2 p r\\
-2 p q & 1-2 q^2 & -2 q r\\ -2 p r & -2 q r & 1-2 r^2\ep
\eea
\noindent where the real parameters $p, q, r$ are constrained to take values
on a unit sphere, i.e. $p^2+q^2+r^2=1$.
The operator satisfies the standard relations $Det({\cal{P}})=-1,
{\cal{P}}^2=I$, where $I$ is the $3 \times 3$
identity matrix. It may be noted that different choices of $(p, q, r)$ satisfying
the aforementioned constraint correspond to choice of
different axes of reflection. For example, $p=r=\frac{1}{\sqrt{2}}, q=0$ correspond to
parity transformation $ x_1 \leftrightarrow x_3, x_2 \leftrightarrow x_2$ in the three
dimensional Euclidean space leading to ${\cal{P}}_{2} = \textrm{anti-diagonal}(1, 1, 1)$.
In the context of lattice systems, this implies that sub-lattices `one' and `three' are swapped,
keeping the sub-lattice `two' invariant. We use the notation that the subscript $i$ in
${\cal{P}}_i$ denotes the axis of inversion(or the sub-lattice which remains unchanged),
and consider the following parity operators:
\bea
{\cal{P}}_{1}=\bp 1 & 0 & 0\\0  & 0 & 1\\ 0 & 1 & 0 \ep,
{\cal{P}}_{2}=\bp 0 & 0 & 1\\ 0 & 1 & 0\\ 1 & 0 & 0 \ep,
{\cal{P}}_{3}=\bp 0 & 1 & 0\\1  & 0 & 0\\ 0 & 0 & 1 \ep\nonumber
\eea
\noindent One may define a more general parity operator $\tilde{{\cal{P}}_i}:= 
R {\cal{P}}_i, R^TR=I, Det(R)=1$, where $R$ is a rotation matrix in 3-dimensional space
and $R^T$ denotes the transpose of $R$. It is to be mentioned that, in general, unlike
the ${\cal{P}}_i$'s, $\tilde{{\cal{P}}}_i^{-1}= {\cal{P}}_i^{-1} R^{-1}
= {\cal{P}}_i^{-1} R^T \neq \tilde{\cal{P}}_i$. We denote the matrix corresponding to the
rotation around $x_i$ by $\pi$ as $R_i$ and redefine $\tilde{{\cal{P}}}_i=R_i {\cal{P}}_i$.
The explicit form of these matrices are,
\bea
&& \tilde{\cal{P}}_{1}=\bp 1 & 0 & 0\\0  & 0 & -1\\ 0 & -1 & 0 \ep,
\tilde{\cal{P}}_{2}=\bp 0 & 0 & -1\\ 0 & 1 & 0\\ -1 & 0 & 0 \ep,\nonumber \\
&& \tilde{\cal{P}}_{3}=\bp 0 & -1 & 0\\-1  & 0 & 0\\ 0 & 0 & 1 \ep\nonumber 
\eea
\noindent Both ${\cal{P}}_i$ and $\tilde{\cal{P}}_i$ are used for finding the symmetry of the
Hamiltonian.

The Hamiltonian $H_k$ is not invariant under the time-reversal symmetry ${\cal{T}}: i \rightarrow
-i, k \rightarrow - k$ alone for non-vanishing loss-gain terms. The Hamiltonian has inversion
symmetry ${\cal{P}} H_k {\cal{P}}^{-1}=H_{-k}$ in the following regions of the parameter space
for different choices of the parity operator:
\bea
&& \textrm{(a)}: {\cal{P}} := {\cal{P}}_1, \gamma_1=-2 \gamma_2, \delta_3=t_1=0, t_3=\delta_1,\nonumber \\
&& \textrm{(b)}: {\cal{P}}:= {\cal{P}}_2, \gamma_2=-2 \gamma_1, t_2=t_1, \delta_2=\delta_1\nonumber \\
&& \textrm{(c)}: {\cal{P}}:= {\cal{P}}_3, \gamma_2=\gamma_1, \delta_3=t_2=0, t_3=\delta_2\nonumber
\eea
\noindent However, it appears that energy spectrum is not entirely real for these cases, and
we exclude discussions on Hamiltonian having inversion symmetry alone. 
We will be looking for ${\cal{P T}}$ symmetry of the system, $({\cal{P T}}) H_k ({\cal{P T}})^{-1}=H_k$: 
\bea
&& \textrm{I.(a)}: {\cal{P}} := {\cal{P}}_1, \gamma_1=0, \delta_3=t_1=0, t_3=\delta_1,\nonumber \\
&& \textrm{I.(b)}: {\cal{P}}:= \tilde{\cal{P}}_1, \gamma_1=0, \delta_3=t_1=0, t_3=-\delta_1\nonumber \\
&& \textrm{II.(a)}: {\cal{P}}:= {\cal{P}}_2, \gamma_2=0, t_2=t_1, \delta_2=\delta_1\nonumber \\
&& \textrm{II.(b)}: {\cal{P}}:= \tilde{\cal{P}}_2, \gamma_2=0, t_2=-t_1, \delta_2=-\delta_1\nonumber \\
&& \textrm{III.(a)}: {\cal{P}}:= {\cal{P}}_3, \gamma_2=-\gamma_1, \delta_3=t_2=0, t_3=\delta_2\nonumber \\
&& \textrm{III.(b)}: {\cal{P}}:= \tilde{\cal{P}}_3, \gamma_2=-\gamma_1, \delta_3=t_2=0, t_3=-\delta_2
\eea
The parametric relations due to ${\cal{PT}}$ symmetry are consistent with the necessary conditions
for the real spectrum in Eq. (\ref{real-pcondi}).

The pseudo-chiral Symmetry is defined as,
\bea
\Gamma H_k^{\dagger} \Gamma^{-1} = - H_k.
\eea
\noindent We find three $\Gamma$'s in different parametric ranges for which the above relation is satisfied:
\bea
&& \Gamma_1=\textrm{diagonal} (1,1,-1) \Rightarrow \delta_1=t_1=0,\nonumber \\
&& \Gamma_2=\textrm{diagonal} (1,-1,1) \Rightarrow \delta_2=t_2=0,\nonumber \\
&& \Gamma_3=\textrm{diagonal} (-1,1,1) \Rightarrow \delta_3=t_3=0,
\eea
\noindent The conditions for pseudo-chirality and the type-I flat bands are the same. In Table-II,
we identify the parametric regions in which the conditions for ${\cal{PT}}$ and/or pseudo-chiral symmetry,
and the necessary conditions for an entirely real spectrum are satisfied simultaneously. We also refer
to the respective sections in the main text where the Hamiltonian in these regions is studied in detail.

\begin{table*}[ht]
\begin{tabular}{|>{\centering\arraybackslash}m{0.085\textwidth}|
                >{\centering\arraybackslash}m{0.17\textwidth}|
                >{\centering\arraybackslash}m{0.20\textwidth}|
                >{\centering\arraybackslash}m{0.13\textwidth}|
		>{\centering\arraybackslash}m{0.25\textwidth}|}
\hline
Symmetry & Restriction due to symmetry & Further restrictions due to reality condition & Free parameters & Remarks\\ \hline
$\mathcal{P}_1 {\cal{T}}$ & $\gamma_1 =0$,$t_1 = \delta_3 = 0$,\newline $t_3 = \delta_1$ & $\gamma_1 =0$,$t_1 = \delta_3$, $t_3 = \delta_1$  & $\gamma_2$,$\delta_1$,$\delta_2$,$t_2$ & Discussed in Sec.~\ref{pbc_1g0}(PBC) and \ref{obc-1g0}(OBC) \\ \hline
$\mathcal{P}_2 {\cal{T}}$ & $\gamma_1 =0$,$t_1 = t_2$, $\delta_2 = \delta_1$ & $\gamma_1 =0$,$t_1 = t_2$, $\delta_2 = \delta_1$ & $\gamma_1$,$\delta_1$,$\delta_3$,$t_1$,$t_3$ & Discussed in Sec.~\ref{pbc_2g0}(PBC) and \ref{obc-2g0}(OBC)\\ \hline
$\mathcal{P}_3 {\cal{T}}$ & $\gamma_2 = - \gamma_1$, $t_3 = \delta_2$, \newline $t_2 = \delta_3=0$, & $\gamma_2 = - \gamma_1$, $t_2 = \delta_3$, $t_3 = \delta_2$ & $\delta_1$,$\delta_2$,$\gamma_1$,$t_1$ & Discussed in Sec.~\ref{pbc_g2emg1}(PBC) and \ref{obc-g2emg1}(OBC)  \\ \hline
$\Gamma_1$ & $\delta_1 = t_1 = 0$ & $\gamma_2 = - \gamma_1$,$t_2 = \delta_3$, $t_3 = \delta_2$ & $\delta_2$,$\delta_3$,$\gamma_1$ & Discussed in Sec.~\ref{pbc_g2emg1}(PBC), \ref{obc-g2emg1}(OBC) and \ref{gbc-limit3}(GBC)\\ \hline
$\Gamma_2$ & $\delta_2 = t_2 = 0$ & $\gamma_1 = 0$,$t_1 = \delta_3$, $t_3 = \delta_1$ & $\delta_1$,$\delta_3$,$\gamma_2$ & Discussed in Sec.~\ref{pbc_1g0}(PBC), \ref{obc-1g0}(OBC) and \ref{gbc-limit1}(GBC)\\ \hline
$\Gamma_3$, $\mathcal{P}_2 {\cal{T}}$ & $\delta_3 = t_3 = 0$,$\gamma_2 = 0$,\newline $\delta_1 = \delta_2$, $t_2 = t_1$ & $\gamma_2 = 0$,$\delta_1 = \delta_2$, $t_2 = t_1$ & $\delta_1$,$t_1$,$\gamma_1$ & Discussed in Sec.~\ref{pbc_2g0}(PBC),\newline \ref{obc-2g0}(OBC) and \ref{gbc-limit2}(GBC)\\ \hline
\end{tabular}
	\caption{Parameter regimes with ${\cal{PT}}$ and/or pseudo-chiral symmetry}
\label{tab-symmetry}
\end{table*}

\subsection{Solutions of necessary conditions for reality of the spectrum, i. e. Eq. (\ref{real-pcondi})}
\label{appendix-sol}

The necessary condition for reality of the eigenvalues of the Bloch matrix is given in Eq.
(\ref{real-pcondi}) in terms of eight independent parameters. One may fix any two parameters
in terms of the remaining six parameters by solving the two constraints.
The following permutation symmetries of Eq. (\ref{real-pcondi}):\\
(i) P1: $ \delta_1 \leftrightarrow t_1 $, and 
$(\delta_2, \delta_3, t_2, t_3, \gamma_1, \gamma_2 )$ unchanged \\
(ii) P2: $\gamma_1 \leftrightarrow \gamma_2, \delta_2 \leftrightarrow \delta_3, t_2 \leftrightarrow t_3$,
$(\delta_1, t_1) \rightarrow (\delta_1, t_1)$\\
(iii) P3: $\gamma_1 \leftrightarrow \gamma_2, \delta_2 \leftrightarrow t_3, \delta_3 \leftrightarrow t_2$,
$(\delta_1, t_1) \rightarrow (\delta_1, t_1)$,\\
becomes useful to extend any pair of solutions to a few other combinations of pair of parameters.
The combined symmetries $P_1P_2$ and  $P_1 P_3$ are equivalent to $P_2$ and $P_3$, respectively.
Before proceeding with this scheme, we first note that the solutions simplify
considerably for $\gamma_2=0$. There are two distinct solutions:
(i) $t_2 = \pm t_1, \delta_2 = \pm \delta_1$, (ii) $t_2 = \pm \delta_1, \delta_2=\pm t_1$ with
no conditions on $t_1, t_3, \delta_1, \delta_3, \gamma_1$.
The solutions for $\gamma_2 \neq 0, \delta_3 \neq 0$ are obtained by fixing $t_3, t_2$ in terms
of the rest six parameters.
\bea
&& t_2^{\pm} = \frac{1}{2 A} \left [ - B \pm \sqrt{B^2 - 4 A C} \right ], B^2 \geq 4 AC, A > 0\nonumber \\
&& t_3^{(\pm)} = \frac{1}{\delta_3} \left [ \left ( 1+ \frac{\gamma_1}{\gamma_2}\right) \delta_1t_1 -
\frac{\gamma_1 \delta_2}{\gamma_2} t_2^{\pm} \right ]
\label{asol1}
\eea
\noindent where the three quantities $A, B, C$ are defined as,
\bea
&& A=\left ( \frac{\gamma_1\delta_2}{\gamma_2 \delta_3} \right )^2 + \frac{\gamma_1}{\gamma_2},
B= -\frac{2\delta_1 \delta_2 t_1\gamma_1}{\gamma_2\delta_3^2} \left ( 1+\frac{\gamma_1}{\gamma_2}\right),
\nonumber \\
&& C = \left ( 1+\frac{\gamma_1}{\gamma_2}\right) \left [\delta_2^2 -t_1^2-\delta_1^2 + \left ( 1+
\frac{\gamma_1}{\gamma_2}\right) \frac{\delta_1^2 t_1^2}{\delta_3^2} \right ]\nonumber \\
&& + \delta_3^2 -\delta_2^2 -\frac{\Gamma}{\gamma_2}
\label{asol2}
\eea
\noindent The reality and the non-singularity conditions for the parameters, i.e. $B^2 \geq 4 AC$ and $A >0$,
respectively, have to be solved separately for specific choices of the parameters. The solutions for
$\delta_3=0$ may be obtained by using the the permutation symmetry $P2$. In particular, we apply
$P2$ to equations (\ref{asol1}) and (\ref{asol2}) and impose the condition $\delta_2 \neq 0,
\gamma_1 \neq 0$, the resulting solutions are non-singular for $\delta_3=0$:
\bea
&& t_3^{\pm} = \frac{1}{2 A_1} \left [ - B_1 \pm \sqrt{B_1^2 - 4 A_1 C_1} \right ], B_1^2 \geq 4 A_1C_1,
A_1 > 0\nonumber \\
&& t_2^{(\pm)} = \frac{1}{\delta_2} \left [ \left ( 1+ \frac{\gamma_2}{\gamma_1}\right) t_1 +
\frac{\gamma_2 \delta_3}{\gamma_1} t_3^{\pm} \right ]
\label{asol3}
\eea
\noindent where the three quantities $A, B, C$ are defined as,
\bea
&& A_1=\left ( \frac{\gamma_2\delta_3}{\gamma_1 \delta_2} \right )^2 + \frac{\gamma_2}{\gamma_1},
B_1=2 \delta_1 \delta_3 t_1 \frac{\gamma_2}{\gamma_1} \left ( 1+\frac{\gamma_2}{\gamma_1}\right),
\nonumber \\
&& C_1 = \left ( 1+\frac{\gamma_2}{\gamma_1}\right) \left [\delta_3^2 -t_1^2-\delta_1^2 + \left ( 1+
\frac{\gamma_2}{\gamma_1}\right) \frac{\delta_1^2 t_1^2}{\delta_2^2} \right ]\nonumber \\
&& + \delta_2^2 -\delta_3^2 -\frac{\Gamma}{\gamma_1}
\label{asol4}
\eea
\noindent The solutions for vanishing as well as non-vanishing values of $\gamma_1, \delta_2$ are contained in
equations (\ref{asol1}) and (\ref{asol2}). 
The permutation symmetry $P3$ may be utilized for directly obtaining $(\delta_2, \delta_3)$
from $(t_2, t_3)$ in Eq. (\ref{asol1}) and (\ref{asol2}):
\bea
&& \delta_3^{\pm} = \frac{1}{2 {A_2}} \left [ - B_2 \pm \sqrt{B_2^2 - 4 A_2 C_2} \right ],
B_2^2 \geq 4 A_2C_2, A_2 > 0\nonumber \\
&& \delta_2^{(\pm)} = \frac{1}{t_2} \left [ \left ( 1+ \frac{\gamma_2}{\gamma_1}\right) t_1 +
\frac{\gamma_2 t_3}{\gamma_1} \delta_3^{\pm} \right ]
\label{asol5}
\eea
\noindent where $\gamma_1 \neq 0, t_2 \neq 0$ and the three quantities $A_1, B_1, C_1$ are defined as,
\bea
&& A_2=\left ( \frac{\gamma_2t_3}{\gamma_1 t_2} \right )^2 + \frac{\gamma_2}{\gamma_1},
B_2=2 \delta_1 t_1 t_3 \frac{\gamma_2}{\gamma_1} \left ( 1+\frac{\gamma_2}{\gamma_1}\right),
\nonumber \\
&& C_2 = \left ( 1+\frac{\gamma_2}{\gamma_1}\right) \left [t_3^2 -t_1^2-\delta_1^2 + \left ( 1+
\frac{\gamma_2}{\gamma_1} \right) \frac{\delta_1^2 t_1^2}{t_2^2} \right ]\nonumber \\
&& + t_2^2 -t_3^2 -\frac{\Gamma}{\gamma_1}
\label{asol6}
\eea
\noindent There are special limits like, $\gamma_1 = - \gamma_2, \gamma_1=0,  \gamma_2=0$, etc., for
which the above expressions take simple form. Some of these limiting cases are discussed in the main
text. It may be noted that the permutation symmetries alone can not generate solutions for all possible
pairs of the parameters from the solutions given above. For example, solutions for the pair $(t_1, t_3)$
or $(t_2, t_3)$ can not be generated  by using permutation symmetries on the solutions given
above. For all such cases, equations in (\ref{real-pcondi}) have to be solved separately.

\def\bad{\section{Appendix: $SU(3)$ \label{su3}}
The Bloch Hamiltonian can be written in terms of the SU(3) generator as,
\bea
H_{k} = \sum_{j=1}^{8} d_{i} \Lambda_{j}
\eea
where $\Lambda_{0}$ is the $3\times3$ identity matrix and 
\bea
d_1 & = & \delta_1 \cos\phi + t_1 \cos(k+2\phi) \nonumber \\
d_2 & = & t_1 \sin(k+2\phi) - \delta_1 \sin\phi \nonumber \\
d_3 & = & \frac{i}{2} \left(\gamma_1 - \gamma_2\right) \nonumber \\
d_4 & = & t_3 \cos(2\phi) + \delta_3 \cos(k+\phi) \nonumber \\
d_5 & = & \delta_3 \sin(k+\phi) - t_3 \sin(2\phi) \nonumber \\
d_6 & = & \delta_2 \cos\phi + t_2 \cos(k+2\phi) \nonumber \\
d_7 & = & -\delta_2 \sin\phi + t_2 \sin(k+2\phi) \nonumber \\
d_8 & = & \frac{\sqrt{3}i}{2} \left(\gamma_1 + \gamma_2\right) \nonumber 
\eea
}
\subsection{Conditions for acceptable solutions of the momentum-space eigenfunctions under the GBC}

The Eq.~(\ref{z-eqn-quartic}) can the written in the bi-quadratic form,
\bea
\left[z^{2} \right. &    +   & \left. \left\{ \frac{a_1}{2} + \frac{1}{2} \sqrt{a_1^2 - 4\left(a_2 - 2\right)}\right\} + 1\right] \nonumber \\
	            & \times & \left[z^{2} + \left\{ \frac{a_1}{2} - \frac{1}{2} \sqrt{a_1^2 - 4\left(a_2 - 2\right)}\right\} + 1\right] = 0 \nonumber \\
\eea
The solutions are
\bea
z_{1,2} & =   & -\frac{a_1}{4} - \frac{1}{4}\sqrt{a_1^2 - 4(a_2 - 2)} \nonumber \\ 
	& \pm & \frac{1}{2} \sqrt{\frac{a_1^2}{2} - a_2 - 2 + \frac{a_1}{2}\sqrt{a_1^2 - 4(a_2 - 2)}} \nonumber \\
z_{3,4} & =   & -\frac{a_1}{4} + \frac{1}{4}\sqrt{a_1^2 - 4(a_2 - 2)} \nonumber \\
        & \pm & \frac{1}{2} \sqrt{\frac{a_1^2}{2} - a_2 - 2 - \frac{a_1}{2}\sqrt{a_1^2 - 4(a_2 - 2)}} 
\label{sol-quartic-z-eqn}
\eea
We can write $z_1 = e^{i\theta_1}$,$z_2 = e^{-i\theta_1}$ and $z_3 = e^{i\theta_2}$,$z_4 = e^{-i\theta_2}$ because 
$z_1z_2 = 1$ and $z_3z_4 = 1$. We have four solutions of $z$ for a fixed energy. Now, the general solution will be 
written as,
\bea
\psi_{n} = \sum_{j=1}^{4} \bp c_{j} A_{j} z_{j}^{n} \\ c_{j}  B_{j} z_{j}^{n} \\ c_{j}  D_{j} z_{j}^{n} \ep
\label{gen-ansatz-gbc}
\eea
Using the general solution (\ref{gen-ansatz-gbc}), we can rewrite the boundary Eq.~(\ref{sch-boundary1}) as
\bea
\sum_{j=1}^{4} \left(t_1 - t_{1l} z_{j}^{m} \right) c_{j}B_{j}      & = & \sum_{j=1}^{4} \left(\delta_{3l} z_{j}^{m} - \delta_3\right) c_{j} D_{j} \nonumber \\
\sum_{j=1}^{4} \left(t_2 - t_{2l} z_{j}^{m} \right) c_{j}D_{j}      & = & 0 \nonumber \\
\sum_{j=1}^{4} \left(t_1 z_{j}^{m} - t_{1r} \right) c_{j}A_{j}z_{j} & = & 0 \nonumber \\
\sum_{j=1}^{4} \left(t_2 z_{j}^{m} - t_{2r} \right) c_{j}B_{j}z_{j} & = & \left(\delta_{3r} - \delta_3 z_{j}^{m} \right) 
\eea
The boundary equations can be written in a matrix form as, $H_{B} C = 0$ where $C = {\bp c_1 && c_2 && c_3 && c_4 \ep}^{T}$ is 
constant column matrix and $H_{B}$ is a $4 \times 4$ matrix with elements $[H_{B}]_{ij}$ , $i,j = 1,2,3,4$ given as
\bea
&& [H_{B}]_{1j} = \left(t_1 - t_{1l} z_{j}^{m} \right)B_{j} + \left(\delta_3 - \delta_{3l} z_{j}^{m} \right) D_{j} \nonumber \\
&& [H_{B}]_{2j} = \left(t_2 - t_{2l} z_{j}^{m} \right)D_{j} \nonumber \\ 
&& [H_{B}]_{3j} = \left(t_1 z_{j}^{m} - t_{1r} \right) A_{j} z_{j}  \nonumber \\
&& [H_{B}]_{4j} = \left(\delta_3 z_{j}^{m} - \delta_{3r} \right) A_{j} z_{j} + \left(t_2 z_{j}^{m} - t_{2r} \right) B_{j} z_{j}
\eea
The condition $Det(H_{B}) = 0$ is imposed for non-trivial solutions of $C$. The constraint $Det(H_B) = 0$ will give
a equation in terms of $\theta_1$ and $\theta_2$ which can be converted 
into an equation in terms of $\theta_1$ as $\cos\theta_1 + \cos\theta_2 = -2a_1/a_0$.
The solution of this equation will give the value of $\theta_1$ and $\theta_2$,
which in turn gives the possible values of the eigenvalues and eigenstates.

\subsection{Solution of phase $\theta$}

We consider the following trigonometric equation which appears in the
description of the model under GBC:
\bea
\sin(m+1)\theta  + A \sin m\theta - B \sin(m-1)\theta  - C \sin\theta = 0\nonumber
\eea
\noindent where $A$, $B$ and $C$ are real constants.
The above equation can be rewritten in terms of the Chebyshev polynomial of second kind
$U_m(x)=\frac{\sin (m+1) \theta}{\sin \theta}, x=\cos\theta$ as,
\bea
U_m(x) + A U_{m-1}(x) - B U_{m-2}(x)-C=0
\label{th_eq2}
\eea
\noindent We will be using both forms of the equations depending on the calculations convenience
to determine exact and approximate solutions, and to find the condition for complex solutions
for $\theta$.

\subsubsection{Exact Solutions}

It may be noted that $\theta=0, \pi$ are exact solutions of the above equation independent of the
constants appearing in it. These are trivial solutions since it leads to constant eigenstates $\psi_n$.
It appears that non-trivial exact solutions of the above equation may not be available. However, exact
solutions may be obtained for specific choices of the constants $A, B, C$.
We discuss a few exact solutions of the above equation for special cases:\\
(I) $A=0, B=1$: The equation reduces to,
\bea
U_m(x) - U_{m-2}(x)-C=0  \Rightarrow 2 T_m(x) - C=0
\eea
\noindent where $T_m(\cos \theta)= \cos m \theta$ is the the Chebyshev polynomyal of the first kind,
and we have used an identity involving Chebyshev polynomials. The exact solution for $m \neq 0$ is,
\bea
\theta_s=\frac{1}{m} \ \cos^{-1} (\frac{C}{2}) + \frac{2 s \pi}{m}, \  s=0, \dots, m-1.
\eea
\noindent Note that $\theta_s \in \mathbb{R} \ \forall \ s$ ${\vert C \vert} \leq 2$,
while  all $\theta_s$'s are complex for ${\vert C \vert} > 2$
leading to edge states. This corresponds to boundary induced NHSE.\\

(II) $B=-1, C=0$: The equation reduces to $(2 x + A) U_{m-1}(x)=0$ with solutions
\bea
\theta_m=\cos^{-1}(-\frac{A}{2}), \
\theta_s=\frac{s \pi}{m}, s=1,\dots, m-1
\eea
\noindent Note that $\theta_m$ is real for ${\vert A \vert} \leq 2$, while  it is complex
for ${\vert A \vert} > 2$ leading to edge states. There can be at most one edge state.\\
(III) $B=0, C=0$: The exact solutions may be obtained for three different cases:
\bea
A=0 & : & \theta_k=\frac{s \pi}{m+1},\nonumber \\
A=1 & : & \theta_k=\frac{2 s \pi}{2 m+1},\nonumber \\
A=-1& : & \theta_k=\frac{(2 s-1) \pi}{2 m+1},\  s=1, \dots, m
\eea
\noindent We can not obtain any closed form expressions for $\theta$ for other values of $A$.

\subsubsection{Realization of solvability limits in terms of system parameters}

We have presented exact solutions of Eq. (\ref{th_eq2}) for (i) $A=0, B=1$
and (ii) $B=0, C=0, A=0, \pm 1$. In this Appendix, we consider explicit expressions
of $A, B, C$ as given in the main text,
\bea
A=\frac{\delta_3^2 - \delta_{3r}\delta_{3l}}{\delta_2\delta_3},
B = \frac{\delta_{3l} \delta_{3r}}{\delta_3^2},
C= \frac{\delta_{3l}+\delta_{3r}}{\delta_3}
\eea
\noindent and find conditions on the systems parameters under which the required
values of $A, B, C$ may be obtained.\\
(I) $A=0, B=1$: This limit is obtained for $\delta_3^2 - \delta_{3r}\delta_{3l}=0$ for
which $C=\pm \left ( \sqrt{\frac{\delta_{3l}}{\delta_{3r}}} + \sqrt{\frac{\delta_{3r}}{\delta_{3l}}} \right )$.
Both $C_{3l}$ and $C_{3r}$ are required to be positive or negative such that $C$ is real.
The limit ${\vert C \vert} > 2 $ for the existence of complex $\theta$ may be achieved
through strongly asymmetric boundary terms $\delta_{3l} > 4 \delta_{3r}$ or
$\delta_{3r} > 4 \delta_{3l}$. The bulk parameter $\delta_3 > \delta_{3l,3r}$
due to the relation $\delta_3^2 - \delta_{3r}\delta_{3l}=0$. This is an example
where strongly asymmetric boundary defects with reciprocal bulk interaction
produces non-hermitian skin effect.\\
(II) $ B=-1, C=0$: This is obtained by $\delta_3=\delta_{3l}=-\delta_{3r}$ for
which $\frac{A}{2}= \frac{\delta_3}{\delta_2}$. The condition for the existence of edge states
is $\delta_3 > \delta_2$.\\
(III) $B=0, C=0, A=\pm 1$: The exact relations $\delta_{3l}=\delta_{3r}=0,
\delta_3= \pm \delta_2$ produce the required values of $A, B, C$. This limit
may also be obtained under the approximation,
${\vert \delta_3 \vert} \gg {\vert \delta_{3l} \vert}, \delta_{3r}=-\delta_{3l},
\delta_3 = \pm \delta_2$ \\
(IV) $B=0, C=0, A=0$: These values for physically nontrivial cases may be
obtained under the following approximations,
${\vert \delta_2 \vert} \gg {\vert \delta_3 \vert} \gg {\vert \delta_{3l} \vert},
\delta_{3r}=-\delta_{3l}$ \\

\subsubsection{Condition for edge states}

In this appendix, we derive conditions under which solutions of $F(x,A)=0$ correspond
to ${\vert x \vert} > 1 $ or equivalently $\theta$ becomes complex, where
\bea
F(x,A) \equiv U_m(x)+AU_{m-1}(x)-BU_{m-2}(x)-C.\nonumber
\eea
\noindent We emphasize that such conditions are required for determining the parametric region in  which
edge states appear. This study includes the regions in which exact analytical solutions for $\theta$
are non-available. We study this transition in terms of  varying $A$ and fixed $B, C$.
The critical values of $A\equiv A_{\pm}^c$ characterizing critical surfaces $F(\pm 1,A_{\pm}^c)=0$
in the parameter space at which $x$ takes its extreme values, i.e. $x=\pm 1$ are:
\bea
A_c^{\pm}= \frac{1}{m} \left [ B (m-1) + (\pm 1)^m C-(m+1)\right ]\nonumber
\eea
\noindent where we have used $ U_m(\pm 1)=(\pm 1)^m (m+1)$. The slope
$\left . \frac{dx(A)}{dA}\right |_{x=1}$, where $x\equiv x(A)$ is an implicit function of $A$, will
determine whether $x >1$ corresponds to $A > A_c^{+}$ or $A < A_c^-$ as $A$ is varied
through the critical surface $F(1,A_{+}^c)=0$. A similar analysis for $x < -1$ is allowed.
We note that
\bea
\frac{dF(x,A)}{dA}  =  \frac{\partial F}{\partial x}\frac{dx}{dA} + \frac{\partial F}{\partial A} =0 \Rightarrow
\frac{dx}{dA}= -\frac{\partial F/\partial A}{\partial F/\partial x},\nonumber
\eea
\noindent and the slope at the critical points is determined in terms of
\bea
F_x(\pm 1,A_c^{\pm}) & = & \frac{(\pm 1)^{m-1}}{3} m \left [ (m+1)(m+2) \pm A_c^{\pm} (m^2-1) \right . \nonumber \\
& - & \left . B (m-2)(m-1) \right ]\nonumber \\
F_A(\pm 1,A_c^{\pm}) & = & U_{m-1}(\pm 1)= (\pm 1)^m m, \nonumber
\eea
\noindent where $F_x(\pm 1,A_c^{\pm})$ and $F_A(\pm 1,A_c^{\pm})$ denote partial derivatives of $F(x,A)$ with respect
to $x$ and $A$, respectively, evaluated at the critical points. We have used the results
$
U_n^{\prime}(\pm 1)=\frac{(\pm 1)^{n-1}}{3} n(n+1)(n+2)$ where $U_n^{\prime}(\pm)$ represent the derivative of $U_n(x)$
evaluated at $x=\pm 1$. We determine the slopes at $x =\pm 1$,
\bea
S_{\pm}=\left . \frac{dx}{dA}\right |_{{\substack{x=\pm 1\\A=A_c^{\pm 1}}}}= (\mp 1)^m \frac{m}{F_x(\pm 1, A_c^{\pm})}
\eea
\noindent The edge states exist for $S_+ < 0$  corresponding to $A< A_c^+$ for $x > 1$. Similarly, for
$S_- < 0$, $x < -1$ is obtained for $A< A_c^-$ .
The sign of the slope may be calculated for specific values of $B, C$ to identify the parametric regions allowing
edge states. A particularly important case is the OBC for which $B=C=0$. In this limit, $A_c^{\pm}=-\frac{m+1}{m},
F_x(1,A_c^{+})=\frac{1}{3} (m+1)(2m+1)$ and the slope $S_+=-\frac{3 m}{(m+1)(2m+1)}$ is always negative. Similarly,
$S_-= \frac{3}{2m^2+4m+1}$ is always positive.


\begin{thebibliography}{10}
\bibitem{Sondhi2013CRP} S. A. Parameswaran, R. Roy, S. L. Sondhi, Fractional quantum Hall physics in topological flat bands, 
	Compt. Rend. Phys. {\bf 14}, 816 (2013).
\bibitem{Liu2013IJMP} E. J. Bergholtz and Z. Liu, Topological flat band models and fractional chern 
	insulators, Int. J. Mod. Phys. B {\bf 27}, 1330017 (2013).
\bibitem{Derzhko2015IJMP} O. Derzhko, J. Richter, and M. Maksymenko, Strongly correlated flat-band systems: The route from
        Heisenberg spins to Hubbard electrons,  Int. J. Mod. Phys. B {\bf 29}, 1530007 (2015).
\bibitem{Aoki1996PRB} H. Aoki, M. Ando and H. Matsumura, Hofstadter butterflies for flat bands, Phys. Rev. B {\bf 54}, R17296 (1996).
\bibitem{Rontgen2019PRL} M. R\"ontgen, C. V. Morfonios, I. Brouzos, F. K. Diakonos, and P. Schmelcher, Quantum network transfer and storage with
        compact localized states induced by local symmetries, Phys. Rev. Lett. {\bf 123}, 080504 (2019).
\bibitem{Flach2014PRL} J.D. Bodyfelt, D. Leykam, C. Danieli, X. Yu and S. Flach, Phys. Rev. Lett. {\bf 113}, 236403 (2014).
\bibitem{Goda2006PRL} M. Goda, S. Nishino and H. Matsuda, Inverse Anderson Transition Caused by Flatbands,
        Phys. Rev. Lett. {\bf 96}, 126401 (2006).
\bibitem{Nishino2007JPS} S. Nishino, H. Matsuda, M. Goda, J. Phys, Soc. Jap. {\bf 76}, 024709 (2007).
\bibitem{Shukla2018PRB} P. Shukla, Disorder perturbed flat bands: Level density and inverse participation
        ratio, Phys. Rev. B {\bf 98}, 054206 (2018).
\bibitem{Dutta2017AP} D. Bercioux, O. Dutta, and E. Rico, Solitons in One-Dimensional Lattices with a Flat Band,
        Ann. Phys. {\bf 529}, 1600262 (2017).
\bibitem{Franz2022PRB} C. Weeks and M. Franz, Flat bands with nontrivial topology in three dimensions, Phys. Rev. B {\bf 85}, 041104 (2012).
\bibitem{Budich2013PRB} J. C. Budich and E. Ardonne, Fractional topological phase in one-dimensional flat bands with
        nontrivial topology, Phys. Rev. B {\bf 88}, 035139 (2013).

\bibitem{Sutherland1986PRB} B. Sutherland, Localization of electronic wave functions due to local topology, Phys. Rev. B {\bf 34}, 5208 (1986).
\bibitem{Lieb1989PRL} E.H. Lieb, Two theorems on the Hubbard model, Phys. Rev. Lett. {\bf 62}, 1201 (1989).
\bibitem{Vidal1998PRL} J. Vidal, R. Mosseri, and B. Doucot, Aharonov-Bohm cages in two-dimensional
        structures, Phys. Rev. Lett. {\bf 81}, 5888 (1998).
\bibitem{Flach2018APX} D. Leykam, A. Andreanov and S. flach, Artificial flat band systems: from lattice 
	models to experiments, Adv. Phy. X {\bf 3}, 1473052 (2018).
\bibitem{Qin2016PRB} W.-X. Qiu, S. Li, J.-H. Gao, Y. Zhou, and F.-C. Zhang, Designing an artificial lieb lattice on a 
metal surface. Phys. Rev. B, {\bf 94}, 241409 (2016).
\bibitem{Drost2017NP} R. Drost, T. Ojanen, A. Harju, and P. Liljeroth, Topological states in engineered atomic lattices, 
	Nat. Phys. {\bf 13}, 668 (2017).
\bibitem{Slot2017NP} M. R. Slot, T. S. Gardenier, P. H. Jacobse, G. C.P. van Miert, S. N. Kempkes, S. J.M. Zevenhuizen, 
	C. M. Smith, D. Vanmaekelbergh, and I. Swart, Experimental realization and characterization of an electronic 
	lieb lattice, Nat. Phys., {\bf 13}, 672 (2017).
\bibitem{Takahashi2015SA} S. Taie, H. Ozawa, T. Ichinose, T. Nishio, S. Nakajima, and Y. Takahashi, Coherent driving and freezing of
bosonic matter wave in an optical lieb lattice, Science Advances, {\bf 1}, 1500854 (2015).
\bibitem{Takeda2004JPCM} H. Takeda, T. Takashima, and K. Yoshino, Flat photonic bands in 
two-dimensional photonic crystals with kagome lattices, J Phys.: Cond. Mat. {\bf 16}, 6317 (2004).
\bibitem{Li2008OE} J. Li, T. P. White, L. O’Faolain, A. G.-Iglesias, and T. F. Krauss, Systematic design of flat band slow light 
in photonic crystal waveguides. Opt. Express {\bf 16}, 6227 (2008).
\bibitem{Xu2015SR} C. Xu, G. Wang, Z. H. Hang, J. Luo, C T Chan, and Y. Lai, Design of full-k-space flat bands in photonic 
	crystals beyond the tight-binding picture, Sci. Rep. {\bf 5}, 18181 (2015).
\bibitem{Endo2010PRB} S. Endo, T. Oka, and H. Aoki, Tight-binding photonic bands in metallophotonic waveguide networks and flat bands 
in kagome lattices, Phys. Rev. B {\bf 81} 113104 (2010).
\bibitem{Nakata2012PRB} Y. Nakata, T. Okada, T. Nakanishi, and M. Kitano, Observation of flat band for terahertz spoof plasmons 
	in a metallic kagom\'e lattice, Phys. Rev. B {\bf 85}, Password205128 (2012).
\bibitem{Kajiwara2016PRB} S. Kajiwara, Y. Urade, Y. Nakata, T. Nakanishi, and M. Kitano, Observation of a nonradiative flat 
	band for spoof surface plasmons in a metallic Lieb lattice. Phys. Rev. B {\bf 93}, 075126 (2016).
\bibitem{Mukherjee2015PRL} S. Mukherjee, A. Spracklen, D. Choudhury, N. Goldman, P. \"Ohberg, E. Andersson, and R. R. Thomson, 
	Observation of a localized flat-band state in a photonic Lieb lattice, Phys. Rev. Lett. {\bf 114}, 245504 (2015).

\bibitem{moire} Y. Cao et. al., Correlated insulator behaviour at half-filling in magic-angle graphene superlattices,
Nature {\bf 56}, 80(2018).

\bibitem{Jacqmin2014PRL} T. Jacqmin, I. Carusotto, I. Sagnes, M. Abbarchi, D. D. Solnyshkov, G. Malpuech, E. Galopin, A. Lemaître, J. Bloch, and A. Amo, 
	Direct observation of dirac cones and a flatband in a honeycomb lattice for polaritons, Phys. Rev. Lett. {\bf 112}, 116402 (2014).
\bibitem{Baboux2016PRL}  F. Baboux, L. Ge, T. Jacqmin, M. Biondi, E. Galopin, A. Lemaître, L. L. Gratiet, I. Sagnes, 
	S. Schmidt, H.E. T\"ureci, A. Amo, and J. Bloch., Bosonic condensation and disorder-induced localization in a 
	flat band, Phys. Rev. Lett. {\bf 116} 066402 (2016).
\bibitem{Klembt2017APL} S. Klembt, T. H. Harder, O. A. Egorov, K. Winkler, H. Suchomel, J. Beierlein, M. Emmerling, C. Schneider, and S. H\"ofling, 
	Polariton condensation in S- and P-flatbands in a two-dimensional Lieb lattice. Applied Phys. Lett. {\bf 111}, 231102 (2017).
\bibitem{Whittaker2018PRL} C. E. Whittaker, E. Cancellieri, P. M. Walker, D. R. Gulevich, H. Schomerus, D. Vaitiekus, B. Royall, D. M. Whittaker, E. Clarke, 
	I. V. Iorsh, I. A. Shelykh, M. S. Skolnick, and D. N. Krizhanovskii, Exciton polaritons in a two-dimensional lieb lattice with 
	spin-orbit coupling. Phys. Rev. Lett. {\bf 120}, 097401 (2018).

\bibitem{bender} C. M. Bender and S. Boettcher, Real spectra in non-Hermitian Hamiltonians having $\mathcal{PT}$ symmetry, 
	Phys. Rev. Lett. {\bf 80}, 5243 (1998); C. M. Bender, $\mathcal{PT}$ Symmetry in Quantum and Classical
Physics (World Scientific, Singapore, 2018
\bibitem{Konotop2016Rev} V. V. Konotop, J. Yang, and D. A. Zezyulin, Nonlinear waves in $\mathcal{PT}$-symmetric systems, 
Rev. Mod. Phys. {\bf 88}, 035002 (2016).
\bibitem{Musslimani2008PRL} Z. H. Musslimani, K. G. Makris, R. El-Ganainy, and D. N. Christodoulides, Optical solitons in 
$\mathcal{PT}$ periodic potentials, Phys. Rev. Lett. {\bf 100}, 030402 (2008).
\bibitem{Bender2004PRD} C. M. Bender, D. C. Brody, and H. F. Jones, Extension of $\mathcal{PT}$-symmetric quantum mechanics to 
quantum field theory with cubic interaction, Phys. Rev. D {\bf 70}, 025001 (2004).
\bibitem{Jones2006JPA} H. F. Jones, Equivalent Hamiltonians for $\mathcal{PT}$-symmetric versions of dual 2D field theories, 
J. Phys. A: Math. Gen. {\bf 39}, 10123 (2006).

\bibitem{PKG2009JPA} T. Deguchi and P. K. Ghosh, Exactly solvable quasi-Hermitian transverse Ising model, 
J. Phys. A: Math. Theor. {\bf 42}, 475208 (2009).

\bibitem{PKG2010JPA} P. K. Ghosh, On the construction of pseudo-Hermitian quantum system with a pre-determined metric
in the Hilbert space, J. Phys. A: Math. Theor. {\bf 43}, 125203 (2010).

\bibitem{PKG2011IJTP} P. K. Ghosh, Constructing exactly solvable pseudo-Hermitian many-particle quantum systems by
isospectral deformation, Int. J. Theor. Phys. {\bf 50}, 1143 (2011).

\bibitem{PKG2019JPA} D. Sinha and P. K. Ghosh, On the bound states and correlation functions of a class of Calogero-type
quantum many-body problems with balanced loss and gain, J. Phys. A: Math. Theor. {\bf 52}, 505203 (2019).

\bibitem{PKG2009PRE} T. Deguchi and P. K. Ghosh, Quantum phase transition in a pseudo-Hermitian Dicke model, 
Phys. Rev. E {\bf 80}, 021107 (2009).

\bibitem{Deguchi2009PRE} T. Deguchi, P. K. Ghosh, and K. Kudo, Level statistics of a pseudo-Hermitian Dicke model, 
Phys. Rev. E {\bf 80}, 026213 (2009).

\bibitem{Goldsheid1998PRL} I. Y. Goldsheid and B. A. Khoruzhenko, Distribution of eigenvalues in non-Hermitian Anderson 
models, Phys. Rev. Lett. {\bf 80}, 2897 (1998).

\bibitem{Heinrichs2001PRB} J. Heinrichs, Eigenvalues in the non-Hermitian Anderson model, Phys. Rev. B {\bf 63}, 165108 (2001).

\bibitem{Molinari2009JPA} L. G. Molinari, Disk-annulus transition and localization in random non-Hermitian tridiagonal 
matrices, J. Phys. A: Math. Theor. {\bf 42}, 265204 (2009).
\bibitem{SKG} S. Ghosh, P.K. Ghosh, Quantum Integrability and Chaos in periodic Toda Lattice with Balanced Loss-Gain,
Chaos, {\bf 34}, 023121 (2024).
\bibitem{PKG2012JPA} P. K. Ghosh, A note on the topological insulator phase in non-Hermitian quantum systems,
J. Phys.: Condens. Matter 24, 145302 (2012).

\bibitem{DS} D. Sinha and P. K. Ghosh, Integrable nonlocal vector nonlinear Schrödinger equation with
self-induced parity-time-symmetric potential, Physics Letters A {\bf 381}, 124(2017).
 
\bibitem{PKG-REV} P KGhosh, Classical Hamiltonian Systems with balanced loss and gain,  J. Phys.: Conf. Ser. {\bf 2038},
012012(2021).
\bibitem{PR} P. Roy and P. K. Ghosh, Balanced loss-gain induced chaos in a periodic Toda lattice,
Phys. Lett. A{\bf 489}, 129156(2023).

\bibitem{Joglekar2010PRA} Y. N. Joglekar, D. Scott, M. Babbey, and A. Saxena, Robust and fragile $\mathcal{PT}$-symmetric phases in 
a tight-binding chain, Phys. Rev. A {\bf 82}, 030103(R) (2010).
\bibitem{Jin2009PRA} L. Jin and Z. Song, Solutions of $\mathcal{PT}$-symmetric tight-binding chain and its equivalent Hermitian 
counterpart, Phys. Rev. A {\bf 80}, 052107 (2009).
\bibitem{Rodriguez2020PLA} L. A. Moreno-Rodríguez, F. M. Izrailev, and J. A. Méndez-Bermúdez, $\mathcal{PT}$-symmetric tight-binding model 
with asymmetric couplings, Phys. Lett. A {\bf 384}, 126494 (2020).
\bibitem{Candanedo2015PE} O. Vázquez-Candanedo, F. M. Izrailev, and D. N. Christodoulides, Spectral and transport properties of the
$\mathcal{PT}$-symmetric dimer model, Physica E {\bf 72}, 7 (2015).
\bibitem{Su1979PRL} W. P. Su, J. R. Schrieffer, and A. J. Heeger, Solitons in polyaeetylene, Phys. Rev. Lett. {\bf 42}, 1698 (1979).
\bibitem{Fradkin1983PRB} E. Fradkin and J. E. Hirsch, Phase diagram of one-dimensional electron–phonon 
	systems. I. The Su–Schrieffer–Heeger model, Phys. Rev. B {\bf 27}, 1680 (1983).
\bibitem{Kivelson1982PRB} S. Kivelson and D. E. Heim, Hubbard versus Peierls and the Su–Schrieffer–Heeger model of polyacetylene, 
	Phys. Rev. B {\bf 26}, 4278 (1982).
\bibitem{Li2014PRB} L. Li, Z. Xu, and S. Chen, Topological phases of generalized Su-Schrieffer-Heeger 
	models, Phys. Rev. B {\bf 89}, 085111 (2014).
\bibitem{Lieu2018PRB} S. Lieu, Topological phases in the non-Hermitian Su-Schrieffer-Heeger model, Phys. Rev. B {\bf 97}, 045106 (2018).
\bibitem{Li2020JPCM} X.-S. Li, Z.-Z. Li, L.-L. Zhang, and W.-J. Gong, PT symmetry of the Su–Schrieffer–Heeger model with imaginary boundary potentials 
and next-nearest-neighboring coupling, J. Phys.: Condens. Matter {\bf 32}, 165401 (2020).
\bibitem{Klett2017PRA} M. Klett, H. Cartarius, D. Dast, J. Main, and G. Wunner, Relation between PT-symmetry breaking and topologically 
	nontrivial phases in the Su-Schrieffer-Heeger and Kitaev models, Phys. Rev. A {\bf 95}, 053626 (2017).
\bibitem{Petrosyan2022PRA} A. F. Tzortzakakis, A. Katsaris, N. E. Palaiodimopoulos, P. A. Kalozoumis, G. Theocharis, F. K. Diakonos, and D. Petrosyan, 
	Topological edge states of the $\mathcal{PT}$-symmetric Su-Schrieffer-Heeger model: An effective two-state description, 
		Phys. Rev. A {\bf 106}, 023513 (2022).
\bibitem{Kawabata2019PRX} K. Kawabata, K. Shiozaki, M. Ueda, and M. Sato, Symmetry and topology in 
non-Hermitian physics, Phys. Rev. X {\bf 9}, 041015 (2019).
\bibitem{Supriyo2025PRB} Supriyo Ghosh, Pijush K. Ghosh, and Shreekantha Sil, Edge states and persistent current in a $\mathcal{PT}$-symmetric 
extended Su-Schrieffer-Heeger model with generic boundary conditions, Phys. Rev. B {\bf 111}, 245428 (2025).
\bibitem{Zhang2022AP} X. Zhang, T. Zhang, M.-H. Lu, and Y.-F. Chen, A review on non-Hermitian skin effect, 
	Adv. Phys. {\bf 7}, 2109431 (2022).
\bibitem{Lu2014PRA} B. Zhu, R. L\"u, and S. Chen, $\mathcal{PT}$-symmetry in the non-Hermitian Su-Schrieffer-Heeger model with complex 
boundary potentials, Phys. Rev. A {\bf 89}, 062102 (2014).
\bibitem{Nguyen2016PRA} B. P. Nguyen and K. Kim, Transport and localization of waves in ladder-shaped lattices with locally PT-symmetric 
	potentials, Phys. Rev. A {\bf 94}, 062122 (2016).
\bibitem{Obuse2020PRE} K. Mochizuki, N. Hatano, J. Feinberg, and H. Obuse, Statistical properties of eigenvalues of the 
	non-Hermitian Su-Schrieffer-Heeger model with random hopping terms, Phys. Rev. E {\bf 102}, 012101 (2020). 

\bibitem{Andreanov2021PRB} W. Maimaiti and A. Andreanov, Non-hermitian flat-band generator in one
        dimension, Phys. Rev. B {\bf 104}, 035115 (2021).
\bibitem{Flach2017PRB} D. Leykam, S. Flach, and Y. D. Chong, Flat bands in lattices with non-Hermitian
        coupling, Phys. Rev. B {\bf 96}, 064305 (2017).
\bibitem{Zhang2019PRA}  S. M. Zhang and L. Jin, Flat band in two-dimensional non-Hermitian optical
        lattices, Phys. Rev. A {\bf 100}, 043808 (2019).
\bibitem{Saxena2015OL} G.-W. Chern and A. Saxena, $\mathcal{PT}$-symmetric phase in kagome-based photonic lattices,
        Opt. Lett. 40, 5806 (2015).
\bibitem{Qi2018PRL} B. Qi, L. Zhang, and L. Ge, Defect States Emerging from a Non-Hermitian Flat Band of Photonic
        Zero Modes, Phys. Rev. Lett. {\bf 120}, 093901 (2018).
\bibitem{Zyuzin2018PRB} A. A. Zyuzin and A. Yu. Zyuzin, Flat band in disorder-driven non-Hermitian Weyl
        semimetals, Phys. Rev. B {\bf 97}, 041203(R) (2018).
\bibitem{Ge2018PR} L. Ge, Non-Hermitian lattices with a flat band and polynomial power increase, Photon. Res. {\bf 6}, A10 (2018).
\bibitem{Zhou2025PRB} K. Zhou, B. Zeng, and Y. Hu, Non-Hermitian Aharonov-Bohm cage in bosonic Bogoliubov-de Gennes
        systems, Phys. Rev. B {\bf 111}, 224308 (2025).
\bibitem{Biesenthal2019PRL} T. Biesenthal, M. Kremer, M. Heinrich, and A. Szameit, Experimental Realization of $\mathcal{PT}$-Symmetric
        Flat Bands, Phys. Rev. Lett. {\bf 123}, 183601 (2019).
\bibitem{Ramezani2017PRA} H. Ramezani, Non-Hermiticity-induced flat band, Phys. Rev. A {\bf 96}, 011802(R) (2017).
\bibitem{Mobius} B. Devanarayanan, Su-Schriffer-Heeger(SSH) model with non-orientable bulk: Union of topology and flat
bands in one dimension, arXiv:2303.00134.
\bibitem{Alvarez2019PRA} V. M. Martinez Alvarez and M. D. Coutinho-Filho, Edge states in trimer lattice, Phys. Review A {\bf 99}, 013833 (2019).
\bibitem{Liu2017SR} X. Liu and G. Agarwal, The new phases due to symmetry protected piecewise berry phases; enhanced pumping and non-reciprocity 
	in trimer lattices, Sci. Rep. {\bf 7}, 45015 (2017).
\bibitem{Zhang2021OE} Y. Zhang, B. Ren, Y. Li, and F. Ye, Topological states in the super-ssh model, Opt. Express {\bf 29}, 42827 (2021).
\bibitem{Huda2020QM} M. N. Huda, S. Kezilebieke, T. Ojanen, R. Drost and P. Liljeroth, Tuneable topological domain wall states in 
	engineered atomic chains, npj Quantum Mater. {\bf 5}, 17 (2020).
\bibitem{Guo2015PRB} H. Guo and S. Chen, Kaleidoscope of symmetry-protected topological phases in one-dimensional 
	periodically modulated lattices, Phys. Rev. B {\bf 91}, 041402 (2015).
\bibitem{Diakonos2022PRB} A. Anastasiadis, G. Styliaris, R. Chaunsali, G. Theocharis, and F. K. Diakonos, Bulk-edge correspondence in the
        trimer su-schrieffer-heeger model, Phys. Rev. B {\bf 106}, 085109 (2022).
\bibitem{Jiang2024OE} Tao Jiang, Jin Zhang, Guoguo Xin, Yu Dang, Anli Xiang, Xinyuan Qi, Wenjing Zhang, and Zhanying Yang, Topological oscillated edge states in 
	trimer lattices, Opt. Express {\bf 32}, 18605 (2024). 
\bibitem{Verma2024PRB} S. Verma and T. K. Ghosh, Bulk-boundary correspondence in extended trimer Su-Schrieffer-Heeger
model, Phys. Rev. B {\bf 110}, 125424 (2024).
\bibitem{Ghuneim2024JPC} M. Ghuneim and R. W. Bomantara, Topological phases of tight-binding trimer lattice in the BDI
symmetry class, J. Phys.: Condens. Matter {\bf 36}, 495402 (2024).
\bibitem{He2021JPCM} Y. He and C.-C. Chien, Non-Hermitian generalizations of extended Su–Schrieffer–Heeger models, 
J. Phys.: Condens. Matter {\bf 33} 085501 (2021).
\bibitem{Jin2017PRA} L. Jin, Topological phases and edge states in a non-Hermitian trimerized optical
lattice, Phys. Rev. A {\bf 96}, 032103 (2017).
\bibitem{Du2021OE} Chuan-Xun Du, Nan Xu, Lei Du, Yan Zhang, and Jin-Hui Wu, Topological edge states controlled by next-nearest-neighbor 
	coupling and Peierls phase in a $\mathcal{PT}$-symmetric trimerized lattice, Opt. Express {\bf 29}, 37722 (2021).
\end{thebibliography}
\end{document}